\documentclass[12pt, a4paper]{report}

\usepackage[T1]{fontenc}
\usepackage[utf8]{inputenc}
\usepackage[english]{babel}
\usepackage{siunitx}
\usepackage{graphicx}
\usepackage{tipa} 
\usepackage{graphics} 
\usepackage{times} 
\usepackage{amsmath}
\usepackage{latexsym}
\usepackage{amscd}
\usepackage{mathrsfs} 
\usepackage{relsize}
\usepackage{delarray}
\usepackage{pstricks}
\usepackage{theorem}
\usepackage{changepage}
\usepackage{euscript}
\usepackage{textcomp}
\usepackage{esvect}
\usepackage{placeins}
\usepackage{subfigure}
\usepackage{array}
\usepackage{mathtools}
\usepackage{delarray}
\usepackage{stmaryrd}
\usepackage{fancyhdr}
\usepackage{graphpap}
\usepackage{makeidx}
\usepackage{enumerate}
\usepackage{esint}
\usepackage{datetime}
\usepackage{caption}
\usepackage{smartdiagram}
\usesmartdiagramlibrary{additions}
\usepackage{abstract}
\usepackage{amsfonts}

\usepackage{color}
\definecolor{igreen}{rgb}{0.0, 0.56, 0.0}
\usepackage{xcolor, colortbl}
\colorlet{gred}{-red!75!green!65!}
\colorlet{mamber}{-red!75!green!15!blue!50!}
\colorlet{grown}{-red!75!blue!20!green}
\colorlet{bled}{-red!85!blue!40!green!45!}
\colorlet{waters}{cyan!25} 
\colorlet{water}{cyan!25!green!20!} 
\definecolor{grin}{HTML}{00F9DE}
\usepackage{rotating}

\usepackage{arydshln}
\usepackage{booktabs}
\usepackage{graphicx}
\usepackage{tikz}
\usepackage{tikz-3dplot}
\usetikzlibrary{
arrows,
arrows.meta,
automata,
backgrounds,
calc,
decorations,
decorations.pathmorphing,
decorations.pathreplacing,
decorations.fractals,
external,
fit,
matrix,
petri,
positioning,
shadows,
shapes,
shapes.multipart,
topaths,
intersections
}
\usepackage{eso-pic}
\def\ba{\begin{array}}
\def\ea{\end{array}}
\def\beann{\begin{eqnarray*}}
\def\eeann{\end{eqnarray*}}
\def\bea{\begin{eqnarray}}
\def\eea{\end{eqnarray}}

\makeatletter
\newlength\qvec@height
\newlength\qvec@depth
\newlength\qvec@width
\newcommand{\qvec}[2][]{
    \settoheight{\qvec@height}{$#2$}
    \settodepth{\qvec@depth}{$#2$}
    \settowidth{\qvec@width}{$#2$}
  \def\qvec@arg{#1}
  \raisebox{.2ex}{\raisebox{\qvec@height}{\rlap{%
    \kern.05em
    \begin{tikzpicture}[scale=1,shorten >=-3pt,shorten <=-3pt]
    \pgfsetroundcap
    \coordinate (Stx) at (.05em,0) ;
		\coordinate (Arx) at (\qvec@width-.05em,0) ;
    \draw[->](Stx) to[bend left] (Arx);
    \end{tikzpicture}
  }}}
  #2
}
\makeatother
\makeatletter
\newlength\pvec@height
\newlength\pvec@depth
\newlength\pvec@width
\newcommand{\pvec}[2][]{
    \settoheight{\pvec@height}{$#2$}
    \settodepth{\pvec@depth}{$#2$}
    \settowidth{\pvec@width}{$#2$}
  \def\pvec@arg{#1}
  \raisebox{.2ex}{\raisebox{\pvec@height}{\rlap{%
    \kern.05em
    \begin{tikzpicture}[scale=1,shorten >=-3pt,shorten <=-3pt]
    \pgfsetroundcap
    \coordinate (Stx) at (.05em,0) ;
		\coordinate (Arx) at (\pvec@width-.05em,0) ;
    \draw[->](Stx) to[bend right] (Arx);
    \end{tikzpicture}
  }}}
  #2
}
\makeatother
\makeatletter
\newlength\vvec@height%
\newlength\vvec@depth%
\newlength\vvec@width%
\newcommand{\vvec}[2][]{%
  \ifmmode%
    \settoheight{\vvec@height}{$#2$}%
    \settodepth{\vvec@depth}{$#2$}%
    \settowidth{\vvec@width}{$#2$}%
  \else 
    \settoheight{\vvec@height}{#2}%
    \settodepth{\vvec@depth}{#2}%
    \settowidth{\vvec@width}{#2}%
  \fi%
  \def\vvec@arg{#1}%
  \def\vvec@dd{:}%
  \def\vvec@d{.}%
  \raisebox{.2ex}{\raisebox{\vvec@height}{\rlap{%
    \kern.05em%
    \begin{tikzpicture}[scale=1]
    \pgfsetroundcap
    \draw (.05em,0)--(\vvec@width-.05em,0);
    \draw (\vvec@width-.05em,0)--(\vvec@width-.15em, .075em);
    \draw (\vvec@width-.05em,0)--(\vvec@width-.15em,-.075em);
    \ifx\vvec@arg\vvec@d%
      \fill(\vvec@width*.45,.5ex) circle (.5pt);%
    \else\ifx\vvec@arg\vvec@dd%
      \fill(\vvec@width*.30,.5ex) circle (.5pt);%
      \fill(\vvec@width*.65,.5ex) circle (.5pt);%
    \fi\fi%
    \end{tikzpicture}%
  }}}%
  #2%
}
\makeatother
\def\ba{\begin{array}}
\def\ea{\end{array}}
\def\beann{\begin{eqnarray*}}
\def\eeann{\end{eqnarray*}}
\def\bea{\begin{eqnarray}}
\def\eea{\end{eqnarray}}

\usepackage{titlesec}
\usepackage{multirow}
\usepackage{hyperref}
\usepackage{apacite}
\usepackage{lipsum}
\usepackage{tikz-cd}
\usepackage{float}
\usepackage{titling}
\usepackage{epigraph}
\usepackage[title, titletoc]{appendix}
\titleformat{\chapter}{\normalfont\LARGE}{\thechapter\,$\vert$}{20pt}{\LARGE}{\setcounter{chapter}{0}}
\titlespacing*{\chapter}{0pt}{-70pt}{40pt} 
\makeatletter
\newcommand\BackgroundPicturea[3]{
	\setlength{\unitlength}{1pt}
	\put(0,\strip@pt\paperheight){
		\parbox[t]{\paperwidth}{
			\vspace{#2}\hspace{#3}
			\mbox{\includegraphics[scale=0.5]{#1}}
}}}
\newcommand\BackgroundPictureb[3]{
	\setlength{\unitlength}{1pt}
	\put(0,\strip@pt\paperheight){
		\parbox[t]{\paperwidth}{
			\vspace{#2}\hspace{#3}
			\mbox{\includegraphics[scale=0.3]{#1}}
}}}
\makeatother

\usepackage{setspace}
\addto\captionsenglish{
	\renewcommand{\contentsname}%
	{Table of Contents}
	\setcounter{tocdepth}{3}
	\setcounter{secnumdepth}{3}
	}

\hypersetup{pdftitle = Thesis, pdfauthor = {First Last}, pdfstartview=FitH, pdfkeywords = essay, pdfpagemode=FullScreen, colorlinks, anchorcolor = red, citecolor = blue, urlcolor=blue, filecolor=green, linkcolor=red, plainpages=false}
\fancyheadoffset{9pt}
\fancyfootoffset{9pt}
\date{}

\title{String Compactification, Effective Field Theory And Holography Swampland}
\author{\\ \Large{Sirui Ning}
\\ St Hugh's College
\\
\\
\\
\\ University of Oxford
\\
A thesis presented for the degree of \\ \textit{Doctor of Philosophy}
\\ \\
Trinity 2024
}
\begin{document}
\AddToShipoutPicture*{\BackgroundPicturea{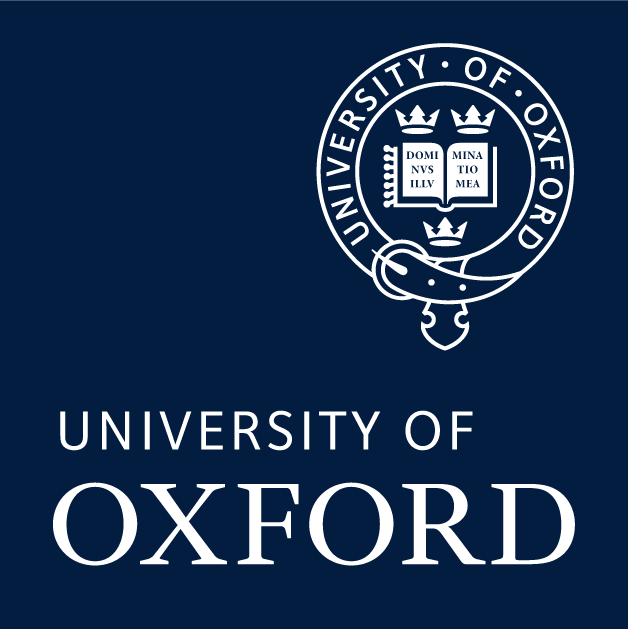}{0.7in}{5.8in}}
\thispagestyle{headings}
	\maketitle
\FloatBarrier
\pagenumbering{arabic}

\newpage`
\thispagestyle{empty}
\vspace*{\fill}
\begin{center}
Copyright \copyright  \thinspace 2024 by Sirui Ning \\ All Rights Reserved
\end{center}
\vspace*{\fill}
\newpage
\thispagestyle{empty}
\epigraph{The Book of Nature is written in the language of mathematics.}{--- \textup{Galileo Galilei}}

\thispagestyle{empty}
\chapter*{Acknowledgements}
First and foremost, I am deeply thankful to my supervisor, Joseph Conlon, for his guidance, expertise, and unwavering support. His insightful feedback and constructive suggestions have significantly contributed to the refinement of this work. 

I am also very grateful to Sean Hartnoll, for sharing his idea with me, providing intensive feedback on my research and many inspring discussions. During my toughest times, you gave me the confidence to keep going. I will never forget your help in my life.

I extend my appreciation to John Wheater, John March Russel, Georges Obied, Andrei Starinets, Alejandra Castro, Fernando Quevedo, Aron Wall, Nele Callebaut for their help and useful discussions during my DPhil, and to John Chalker, Steve Simons, Ben Simons, Francesco Mori, Siddharth Parameswaran and Shivaji Sondhi for useful chat in condensed matter physics and soft matter. 

I want to thank all the other people in Rudolf Peierls Centre for Theoretical Physics, all my fanstic collaborators for the help and discusssions, and all my friends in Oxford. You are all great people and I feel so lucky to be able to collaborate with you, learn from you and spend time with you.

I want to thank Ian Shipsey for his encouragement and suggestions, and Helen Smith for her help and support. I also want to thank Rosanna Wells, Mark Robinson and Antonia Chan. Special thanks to Steven Allan Kivelson and Pamela Davis Kivelson for everything you have done for me.
 
Lastly, I would like to express my heartfelt gratitude to my beloved Helena and my family, it is your love and support have made this journey even more meaningful.

\thispagestyle{empty}
\chapter*{Declaration}
This thesis represents the sole effort of the author, except where explicitly indicated in the text. None of the content in this thesis has been presented for any other academic qualification, either at the University of Oxford or any other institution. Chapter 2 is based on \cite{Conlon:2021cjk}, Chapter 3 is based on \cite{Ning:2022zqx} and Chapter 4 is based on \cite{Blacker:2023ezy}.

I express my sincere gratitude for the valuable contributions of all my coauthors. In particular, Section \ref{1} in \cite{Blacker:2023ezy}   is based on
calculations mainly performed by Matthew Blacker. It is included for the coherence of the presentation.

\vspace{3cm}
\noindent\begin{tabular}{ll}
\makebox[2.5in]{\hrulefill} & \makebox[2.5in]{\hrulefill}\\
\textit{Signature}: Sirui Ning & \textit{Date} 29th Feb 2024\\
\end{tabular}

\thispagestyle{empty}
\begin{abstract}
This thesis primarily dives into investigating the details of four-dimensional vacua within String Theory using the AdS/CFT correspondence. In the first and second part of the thesis, we study the fibred Calabi-Yau and M-theory moduli stabilization scenario. We consider both flux-stabilized models and non-perturbative stabilization methods. We perform
a holographic analysis to determine the spectrum of the assumed dual $CFT_3$ to understand its
AdS/CFT implications. For the flux stabilization, which relies on a large complex Chern-Simons invariant, moduli have integer dimensions similar to the DGKT flux-stabilized
model in type IIA. For the non-perturbative stabilization, the results are similar to race-track models in type IIB.

In the last part of the thesis, we solve the Wheeler DeWitt equation for the planar Reissner-Nordstrom-AdS black hole in a minisuperspace approximation. We construct semiclassical 
Wheeler-DeWitt states from Gaussian wavepackets that are peaked on classical black hole interior solutions. By using the metric component $g_{xx}$ as a clock, these states are evolved
through both the exterior and interior horizons. Close to the singularity, we show that
quantum fluctuations in the wavepacket become important, and therefore the classicality of the minisuperspace approximation breaks down. Towards the AdS boundary,
the Wheeler-DeWitt states are used to recover the Lorentzian partition function of the
dual theory living on this boundary. This partition function is specified by an energy
and a charge. Finally, we show that the Wheeler-DeWitt states know about the black
hole thermodynamics, recovering the grand canonical thermodynamic potential after
an appropriate averaging at the black hole horizon.
\end{abstract}
\tableofcontents
\thispagestyle{plain}
\listoffigures
\listoftables

\chapter{Introduction}
\section{Origins Of String Theory}

String theory originated from the need to describe the strong interaction force, responsible for binding quarks within neutrons and protons, and further, holding these particles together to form atomic nuclei. As exploration continued, it became evident that the internal structure of string theory is exceptionally intricate, giving rise to various types of particles. Over time, string theory has evolved into a prominent contender for a quantum gravity theory. In this theoretical framework, the smallest fundamental unit of matter is no longer a particle but a string. The diverse vibrational modes of these strings give rise to a spectrum of particles, encompassing gravitons, gauge bosons, and chiral fermions.

The earliest bud of string theory can date back to Kaluza-Klein theory \cite{Kaluza:1921tu}. Kaluza-Klein theory is a gravitational theory in five dimensions compactified on a one-dimensional circle. Through dimensional reduction, the five-dimensional metric transforms into a graviton, graviphoton, and dilaton within four-dimensional spacetime. The isometry group of the circle corresponds to the gauge transformation of the graviphoton, unifying gravity and electromagnetism.

However, the Kaluza-Klein mechanism encounters several challenges. Firstly, the semiclassical instability of the vacua in the Kaluza-Klein mechanism was highlighted by Witten \cite{Witten:1981gj}, which connects to the idea that the Kaluza-Klein theory doesn't satisfy the positive energy conjecture. Secondly, the theory lacks the capability to accommodate chiral fermions. These difficulties prompt the exploration of a more suitable candidate for quantum gravity that aligns with the intricacies of our real world: string theory.

Within the Standard Model, there exist 19 free parameters encompassing the masses of quarks and other fermions, the Higgs boson mass derived from the Higgs potential, three gauge coupling constants (U(1), SU(2), SU(3)), three CKM angles, one CP-violating phase, and one QCD vacua angle. In contrast, string theory presents a unique framework where no free parameters exist. Each constant finds expression in the values of moduli, serving as labels for string vacua. While the journey to explaining Standard Model parameters within string theory is ongoing, the prospect of deriving these parameters rather than imposing them by hand offers promise.

On the other hand, from the persepctive of quantum field theory (QFT), a consistent coupling of QFT with gravity faces challenges due to nonrenormalizability. Although pure gravity exhibits one-loop divergences, which can be removed through a field redefinition \cite{tHooft:1974toh}, coupling gravity with a scalar field introduces persistent one-loop divergences. String theory emerges as a solution to this issue, allowing for a harmonious interaction between gravity and quantum fields. Particles are point like, strings are extended; this blurs the interaction point.

The application of string theory spread in nearly all corners of theoretical physics:
from interpreting Bekenstein's black hole entropy formula by counting microstates  \cite{Strominger:1996sh} to giving an argument to the cosmological constant problem  \cite{Bousso:2000xa}.

The cosmological constant problem arises from the theoretical prediction that the cosmological constant should be $10^{120}$ times larger than the observed value. If we consider the cosmological constant as the energy of empty space and use the Planck energy as a cutoff, the calculated vacuum energy corresponding to the cosmological constant is approximately $(10^{18}GeV)^4$. However, astronomical observations reveal that the vacua energy of dark energy is around $(10^{-12}GeV)^4$. To address this issue, Weinberg established an anthropic upper bound for the cosmological constant   \cite{Weinberg:1988cp}, closely aligning with observational results. However, achieving concordance with the theoretical prediction requires the theory to encompass a vast number of vacua.

Bousso and Polchinski proposed a string theory-based argument to tackle this challenge. By strategically manipulating the topology of the internal manifold and introducing fluxes subject to Dirac quantization, the fluxes become quantized and assume discrete values. If the topology is suitably complex, the permissible number of fluxes can easily reach $10^{120}$.

In this thesis, we discuss two different approaches to quantum gravity: top-down from string theory and Wheeler DeWitt equation.

In the first part of the thesis, we determine the holographic properties (assuming a holographic description exists) of two different moduli stabilization scenarios: Fibred Large Volume Scenario and M theory scenario stabilized by flux and non perturbative effects.

The fibred Large Volume Scenario example extends the mixing anomalous dimension conjecture proposed by \cite{Conlon:2020wmc}. The conjecture claims that if the mixing anomalous dimension for any of two different moduli is always negative, then the theory is in the landscape, otherwise it is in the swampland. This relates to the CFT version of Axionic Weak Gravity Conjecture \cite{Arkani-Hamed:2006emk}. However, in the fibred LVS example, the anomalous dimension between two different axions is positive, which means that in the case of multiple axions, the conjecture should be modified. In the M theory scenario stabilized by flux, we show that the conformal dimensions for the volume and axion moduli are integers. This is a very interesting aspect of this scale separated theory, similar to that observed in \cite{Conlon:2021cjk, Apers:2022tfm} for DGKT type IIA models. 

In the second part of the thesis, we look at quantum gravity from a different prospective, give Wheeler-DeWitt state descriptions of charged AdS black holes and associate them with a class of boundary partition functions specified by an energy and a charge. We also study the quantum fluctuation near the timelike singularity of the charged AdS black hole, which shows a different behavior with spacelike singularity in the approximation of mini superspace.

Wheeler DeWitt equation is a canonical quantum gravity theory (unlike string theory which starts with string equation of motion). We introduce the Hamiltonian formalism of general relativity, for a global hyperbolic spacetime. Starting from a Cauchy surface $\Sigma$, we have the following ansatz:
\begin{equation}
    ds^2=-N(x,t)^2dt^2+h_{ij}(x,t)(dx^i+N^i(x,t)dt)(dx^j+N^j(x,t)dt),
\end{equation}
where $N(x,t)$ is the lapse function, $N_i(x,t)$ is the shift vector, $h_{ij}(x,t)$ is the induced metric on the Cauchy slice $\Sigma$, and $t$ is a global defined time function.

In classical general relativity, the diffeomorphism invariance of spacetime coordinates leads to Hamiltonian and momentum constraints:
\begin{equation}
    \begin{aligned}
        &\sqrt{h}(R^{3}-2\Lambda+K^2-K_{ab}K^{ab})=\sqrt{h}(R^{(3)}-2\Lambda)-\frac{1}{\sqrt{h}}(\pi^{ab}\pi_{ab}-\frac{1}{2}\pi^2)=0,\\
        &\nabla^a K_{ab}-\nabla^a K=\nabla^a\pi_{ab}=0,\label{HamiltonianMomentumConstraints}
    \end{aligned}
\end{equation}
where $R^{3}$ is the curvature for Cauchy slice $\Sigma$ and the canonical momentum $\pi_{ab}$ is related to extrinsic curvature as $\pi_{ab}=-\sqrt{h}(K_{ab}-h_{ab}K)$. Eq \ref{HamiltonianMomentumConstraints} is equivalent to the Einstein equation. Promoting the canonical momenta at each point of spacetime to operators and imposing the commutation relation:
\begin{equation}
    \pi^{ij}(x)\rightarrow -i\frac{\delta}{\delta h_{ij}(x)},\quad [h_{ij}(x),\pi^{mn}(y)]=i\delta_{mn}^{ij}\delta^{4}(x-y),
\end{equation}
we reach the Wheeler DeWitt equation:
\begin{equation}
   \begin{aligned}
       &\left(\sqrt{h(x)}(R^{3}-2\Lambda)-\frac{1}{\sqrt{h(x)}}:(\pi_{ab}(x)\pi^{ab}(x)-\frac{1}{2}\pi(x)^2):\right)\Psi[h]=0,\\
       &\nabla^a\pi_{ab}(x)\Psi[h]=0.
   \end{aligned} 
\end{equation}
Given that the Wheeler DeWitt equation arises from canonical quantum gravity, it would be very interesting if we could derive the Wheeler DeWitt equation from string theory. In string theory, the dynamic fields are spacetime coordinates $X^\mu(\sigma,t)$, where $\sigma,t$ are coordinates on the worldsheet. The string action can be written as:
\begin{equation}
    S\propto \int d\sigma dt \sqrt{g}g^{\alpha\beta}\partial_\alpha X^\mu\partial_\beta X^\nu G_{uv}(X),
\end{equation}
where $g^{\alpha\beta}$ is the metric for the worldsheet and $G_{\mu\nu}$ is the metric for spacetime. So in string theory, the spacetime metric is no longer a dynamical field, but a coupling constant. One can check that the beta function for the spacetime metric is exactly the Einstein equation. A possible line for future research would be to derive Wheeler DeWitt equation from beta function, perhaps with same reformulation of the string action.

This thesis comprises several key chapters. Chapter 1 serves as an introduction, providing insights into both string theory and the fundamental principles of conformal field theory. In Chapter 2, we give a detailed calculation involving the fibred Large Volume Scenario (LVS), utilizing it as an illustrative example within the context of holography in the swampland. Moving forward to Chapter 3, a succinct exploration of M-theory compactification is presented, accompanied by the derivation of properties associated with the presumed holographic dual. Lastly, in Chapter 4, our focus shifts to the examination of the Wheeler-DeWitt state of a charged black hole.
The thesis is based on my papers \cite{Conlon:2021cjk,Ning:2022zqx, Blacker:2023ezy}. I have also written other papers during my DPhil which cover a wide range in theoretical high energy physics, from string theory \cite{Apers:2022tfm}, supergravity \cite{Bomans:2023ouw}, mathematical physics \cite{Hu:2023xog} to quantum chaos \cite{Ning:2023pmz}, cosmology \cite{Ning:2023ybc} and extra dimension \cite{Cui:2023wzo}. These are not included for word limit and for coherence. 
\section{Moduli Stabilization}
\subsection{Fundamental Concepts}
Achieving compatibility between string theory and empirical reality necessitates the process of string compactification. In the realm of string theory, the spacetime is either 10-dimensional for superstring theory or 11-dimensional for M theory. The 26-dimensional bosonic string is ruled out as a viable candidate due to the presence of tachyons and the absence of fermions. At a minimum, the vacuum configuration must manifest itself as effectively four-dimensional to align with the observable dimensions of our physical universe.

Tracing the evolution of string compactification, we adopt the assumption that the 10-dimensional spacetime can be expressed as the direct product of a 4-dimensional real-world space and a 6-dimensional internal manifold: $M_{10}=M_{4}\times M_{6}$. Here, $M_4$ represents a maximally symmetric Riemannian manifold, while $M_6$ denotes a compact 6-dimensional manifold. The selection of $M_6$ is crucial to ensure the existence of $\mathcal{N}=1$ supersymmetry in four dimensions, a condition favorable for stability. It has been established that $M_6$ must be a Calabi-Yau manifold \cite{Candelas:1985en}. A more generalized assumption, maintaining maximal symmetry and Poincare invariance for $M_4$, is:
\begin{equation}
    ds^{2}=\Delta(y)^{-1}g_{uv}dx^{u}dx^{v}+\Delta(y)g_{mn}(y)dy^{m}dy^{n},\\
\end{equation}
where $u,v,=1,2,...4$ and $m,n=5,6,...,10$. The warp factor $\Delta(y)$ only includes the internal manifold coordinate to preserve the Poincare invariance of $M_{4}$. The spacetime is no longer a direct product because of the warp factor.

In order to define what a Calabi-Yau manifold is mathematically, we start with the notation of an almost complex manifold. The almost complex structure is defined as $J_{p}: T_{p}M\rightarrow T_{p}M$ such that $J_{p}^{2}=-\text{Id}_{T_{p}M}$, where $T_pM$ is the tangent space for each point p on manifold $M$. If there exists an almost complex structure at each point $p$ on a 2n-dimensional real manifold, then the manifold is said to be an almost complex manifold.

The existence of an almost complex structure necessitates that the manifold be oriented and have an even dimension. However, a detailed exploration of these conditions lies outside the scope of this thesis.

In the case where the corresponding Nijenhuis tensor is null, the almost complex manifold achieves integrability. This, in turn, implies the presence of a globally defined complex structure tensor:
\begin{equation}
    J= idz^{i}\otimes\frac{\partial}{\partial z^{i}}-id\overline{z}^{j}\otimes \frac{\partial}{\partial \overline{z}^{j}},
\end{equation}
which satisfies $J^{2}=-Id$.
A 2n-dimensional real manifold equipped with a complex structure is said to be a complex manifold.

We can also define a complex metric on a complex manifold. A metric is said to be Hermitian if in local coordinates:
\begin{equation}
    g_{ij}=g_{\overline{i}\overline{j}}=0.
\end{equation}

A hermitian metric allows us to build a (1,1) form: $J=ig_{i\overline{j}}dz^{i}\wedge d\overline{z}^{\overline{j}}$. The K\"ahler manifold is defined as a hermitian manifold with a closed (1,1) form J.

Using $dJ=0$, the metric can be written locally as:
\begin{equation}
    g_{i\overline{j}}=\frac{\partial^{2}K}{\partial z^{i}\partial \overline{z}^{\overline{j}}},
\end{equation}
which implies $J=i\partial\overline{\partial} K$, where $
K$ is known as K\"ahler potential.

We now have all the ingredients to define the Calabi-Yau manifold. A renowned theorem, proved by Yau, gives the following definition:

\textbf{Yau's Theorem}: Let X be a compact K\"ahler manifold with vanishing first Chern class, then there exists a unique Ricci flat metric on X up in the cohomology class of the K\"ahler form and complex structure.

For a K\"ahler manifold $X$, the Ricci tensor only has non zero mixed components. Similar to the K\"ahler form, the Ricci form is defined as:
\begin{equation}
    R_{1,1}=R_{i\overline{j}}dz^{i}\wedge d\overline{z}^{\overline{j}}.
\end{equation}

One can prove that the Ricci form is closed and therefore it belongs to $H^{1,1}(X)$. The first Chern class is $c_{1}=\frac{ R_{1,1}}{2\pi}$. So the vanishing of first Chern class implies that the Ricci form $R_{1,1}$ is also exact, which also leads to the Ricci flat condition.

In general relativity, the Ricci flat condition corresponds to the vacuum solution of Einstein equation with vanishing cosmological constant:
\begin{equation}
    R_{ij}-\frac{1}{2}Rg_{ij}=0.
\end{equation}

We give two examples of Ricci flat solutions:

\textbf{Minkowski metric}:
\begin{equation}
    ds^{2}=-c^{2}dt^{2}+dx^{2}+dy^{2}+dz^{2},
\end{equation}
which is a flat spacetime.

\textbf{Schwarzschild metric}:
\begin{equation}
    ds^{2}=-\left(1-\frac{2GM}{r}\right)dt^{2}+\left(1-\frac{2GM}{r}\right)^{-1}dr^{2}+r^{2}\left(d\theta^{2}+sin^2\theta d\phi^{2}\right),
\end{equation}

This depicts a spherically symmetric vacuum solution to the Einstein equation, with the mass M located at $r=0$. The solution does not cover the entire spacetime due to the exclusion of the point $r=0$. Nevertheless, one can achieve a comprehensive coverage by employing Kruskal coordinates.

Given that K\"ahler manifolds are Riemannian, it is always possible to locally choose normal coordinates that satisfy the Ricci flatness condition. However, ensuring the global existence of the Ricci flat condition becomes non-trivial, for reference see \cite{Acharya:2004nq}.

The topological properties of a Calabi-Yau three fold X is characterized by its Hodge numbers $h^{p,q}$, which are the dimension of the Dolbeault cohomology groups $H^{p,q}(X)$. $h^{p,q}$ counts the number of independent harmonic (p,q) form on the Calabi-Yau manifold:

\begin{equation}
    w=w_{i_{1}i_{2}...i_{p}\overline{j}_{1}...\overline{j}_{q}}dz^{i_{1}}\wedge...dz^{i_{p}}\wedge d\overline{z}^{\overline{j}_{1}}\wedge...d\overline{z}^{\overline{j}_{q}}.
\end{equation}

Hodge duality and complex conjugation implies the following relations:
\begin{equation}
    h^{p,0}=h^{n-p,0},\quad h^{p,q}=h^{q,p}.
\end{equation}

Poincare duality states that:
\begin{equation}
    h^{p,q}=h^{n-p,n-q}.
\end{equation}

Following \cite{joyce2002lectures}, $h^{0,0}=1$, $h^{1,0}=h^{0,1}=0$ for all Calabi-Yau 3-folds. So the only non-specified Hodge numbers of the Calabi-Yau 3-fold are $h^{1,1}$ and $h^{2,1}$, which gives the Euler characteristic of Calabi-Yau 3 fold: $\chi=2(h^{1,1}-h^{2,1})$.
The condition that $h^{3,0}=1$ implies a non vanishing harmonic (3,0) form:
\begin{equation}
\Omega(z^{1},z^{2},z^{3})=f(z^{1},z^{2},z^{3})dz^{1}\wedge dz^{2}\wedge dz^{3},
\end{equation}
which is equivalent to the condition that the first Chern class of the K\"ahler manifold vanishes. A good introduction to Calabi-Yau manifolds is \cite{Greene:1996cy}.
\subsection{Moduli}
During the process of string compactification, novel massless scalar fields emerge as 4-dimensional remnants of the extra-dimensional geometry. These scalar fields, known as moduli, initially do not manifest in the 4-dimensional effective potential, rendering them unconstrained. Mathematically, within a given vacuum solution, the moduli space parametrizes a continuous family of vacua in its vicinity, sharing topological equivalence due to the preservation of topological properties under continuous deformations.

The presence of moduli, from a physics standpoint, is a natural consequence of extra dimensions. Considering gravitational waves in $(D+1)$ spacetime, we encounter $\frac{D(D-1)}{2}-1$ polarizations. Consequently, for a 10-dimensional spacetime, there exist 35 polarizations. These novel gravitational wave patterns propagate solely within the extra dimensions, behaving as massless quantum fields in 4-dimensional spacetime, elucidating the physical origin of moduli.

On a different note, in accordance with Einstein's equations, spacetime geometry is dynamic. Moduli can be interpreted as the massless excitations of the vacuum spacetime, exerting an influence on the spacetime geometry as the zero mode of metric deformation. Given their massless nature, distinct values of moduli serve as labels for different vacuum states.

Considering the metric deformation for Calabi-Yau three-folds, Ricci flatness needs to be satisfied by both $g_{uv}$ and $g_{uv}+\delta g_{uv}$:
\begin{equation}
    R_{mn}(g)=R_{mn}(g+\delta g)=0.\label{ricciflatness}
\end{equation}

Expanding Eq \ref{ricciflatness} at the leading order leads to the Lichnerowicz equation :
\begin{equation}
    \nabla^{k}\nabla_{k}\delta g_{mn}+2R_{m\ n}^{ p\ q}\delta g_{pq}=0.
\end{equation}

For K\"ahler manifolds, the equations for the mixed $(\delta g_{a\overline{b}})$ and pure $(\delta g_{ab})$ are independent and do not couple, which corresponds to K\"ahler and complex structure deformation.

Nevertheless, the existence of massless moduli contradicts experimental and observational results, highlighting the imperative to stabilize these moduli.

Moduli stabilization relies on the incorporation of fluxes and non-perturbative effects, such as world sheet instantons and gaugino condensation. These elements introduce new terms in the 4-dimensional effective potential, facilitating the stabilization of moduli.

Viewed as a higher-dimensional generalization of Maxwell theory, fluxes play a crucial role. In a $D$-dimensional warped spacetime, if the vacuum solution of a certain $n$-form $A_n$ is non-zero, two distinct fluxes associated with the n-form $A_n$ can be defined, each coupled with an $(n+1)$-form field strength $F_{n+1}=dA_n$:
\begin{equation}
\begin{aligned}
    &\textbf{Magnetic Flux:}\int_{\gamma_{n+1}}F_{n+1},\\
     &\textbf{Electric Flux:}\int_{\gamma_{D-n-1}}*F_{n+1}.\\  
\end{aligned}   
\end{equation}

The flux value is influenced by both the background field values and the homology class of $\gamma_{n+1}$. Adhering to the generalized Dirac quantization condition, the associated vacua are labelled by discrete integers. The collective set of vacua, arising from various flux choices, constitutes what is commonly referred to as the "landscape."

For type IIA and type IIB, there are two possible gauge fields: One is the NS-NS two form $H_{3}=dB_{2}$, which comes from the product of two vector representations of SO(8) arising both in type IIA and type IIB. The other is the R-R p-form. In type IIA, there are odd forms $F_{2}=dC_{1}, F_{4}=dC_{3}$, $F_{10}=dC_{9}$ (Romans mass for massive IIA). In type IIB, there are even form $F_{1}=dC_{0}, F_{3}=dC_{2}, F_{5}=dC_{4}$. For a p-brane, it can couple to both $(p+1)$ form gauge field electrically and $(D-p-3)$ form gauge field magnetically .

Merely relying on flux stabilization is insufficient without accounting for quantum corrections. For illustrative purposes, consider M-theory compacted on a 8-dimensional internal manifold $X_8$. The bosonic supergravity action of M-theory can be expressed as:
\begin{equation}
    S=\frac{1}{2k_{11}^{2}}\int d^{11}x\sqrt{-G}(R-\frac{1}{2}|F_{4}|^{2})-\frac{1}{6}\int A_{3}\wedge F_{4}\wedge F_{4},
\end{equation}
where $A_{3}$ is the only gauge field in M-theory and $F_{4}=d A_{3}$. The appearance of Chern-Simons term is for anomaly cancellation. The equation of motion and supersymmetry constraint leads to the following equation relating the warp factor and field strength:
\begin{equation}
    d \ast d\;\text{ln}\Delta=\frac{1}{3}F\wedge F.
\end{equation}

The integration over the entire internal manifold $X_8$ results in the flux vanishing, as the left-hand side of the equation becomes a boundary term. Consequently, it becomes imperative to incorporate quantum corrections into the supergravity theory. This involves considering one-loop scattering terms in string theory and introducing sources to render the flux compactification non-trivial. The introduction of these terms modifies the equation and leads to the emergence of the so-called tadpole condition:
\begin{equation}
    \frac{1}{4k_{11}^{2}T_{M2}}\int_{X_8}F\wedge F=\frac{\chi}{24},
\end{equation}
where $\chi$ is the Euler characteristic of the internal manifold $X_8$. Although we discuss M-theory here, the result is also general in other superstring theories, like type IIA and type IIB. 
\subsection{Moduli stabilization: An Overview }
Examples of scenarios of moduli stabilization include the type IIA string compact-
ifications like DGKT \cite{DeWolfe:2005uu}  and type IIB string compactifications like the Large Volume
Scenario(LVS) \cite{Conlon:2005ki,Conlon:2006gv,Conlon:2006wz}, KKLT \cite{Kachru:2003aw} and Racetrack \cite{Blanco-Pillado:2004aap}. In DGKT all geometric moduli are
stabilized at SUSY AdS vacua in the large flux limit. In the Large Volume Scenario the
moduli are stabilized at non-SUSY AdS vacua with an exponentially large volume. KKLT
stabilizes the moduli with both fluxes and non-perturbative effects at SUSY vacua in the
limit of small $W_{0}$. The Racetrack model uses two different non-perturbative effects to
stabilize the moduli at a SUSY vacuum without flux.

For explanatory purposes, we provide a brief overview of how moduli stabilization operates in type IIB superstring theory, examining it from both the 10-dimensional and 4-dimensional perspectives.

In the 10-dimensional scenario, it is essential to couple the supergravity theory with localized sources. The introduction of O-planes, contributing negatively to the 6-dimensional curvature, serves to offset the positive contributions from fluxes and the warped factor. Without this compensation, a no-go theorem implies that the flux would vanish, and the warped factor would be a constant without negative tension objects. Thus, moduli stabilization becomes unattainable without coupling to localized sources. Mathematically, an O-plane is defined as the fixed points of the following action:
\begin{equation}
    O=\Omega(-1)^{F_{L}}\sigma,
\end{equation}
where $\Omega$ is the worldsheet parity  transformation which inverts the orientation of the string: $\sigma\rightarrow 2\pi-\sigma$. $(-1)^{F_{L}}$ counts for the number of left moving fermions, which acts the same on  both R-NS and R-R sector. $\sigma$ is an involution operator, which means $\sigma(\sigma(x))=x$. $\sigma$ acts on holomorphic 3-form $\Omega$ and complex structure (1,1) form J differently in type IIA and type IIB theory:
\begin{equation}
    \begin{aligned}
    &\sigma(J)=-J,\quad \sigma(\Omega)=\overline{\Omega}\quad\text{for type IIA},\\
     &\sigma(J)=J,\quad \sigma(\Omega)=\Omega  \,\text{or}\,  \sigma(\Omega)=-\Omega \quad\text{for type IIB}.\\
     \end{aligned}
\end{equation}

We initiate our exploration with the GKP construction, an orientifold compactification that encompasses both geometric background fluxes and non-dynamical entities such as O-planes. The spacetime characteristics are encapsulated by a warped metric:
\begin{equation}
ds^{2}=e^{2A(y)}\eta_{uv}dx^{u}dx^{v}+e^{-2A(y)}g_{mn}(y)dy^{m}dy^{n}.
\end{equation}

In the context of our investigation, we consider a 4-dimensional spacetime that adheres to the Minkowski metric, with the inclusion of an internal conformally Calabi-Yau manifold. Within the framework of the Einstein frame, the action of Type IIB supergravity is expressed as follows
\begin{equation}
    \begin{aligned}
        &S_{IIB}=\frac{1}{2k_{10}^{2}}\int dx^{10}\sqrt{-g}\left(R-2\frac{\partial_{u}\tau\partial^{u}\overline{\tau}}{2(\text{Im}\tau)^{2}}-\frac{G_{3}\cdot \overline{G}_{3}}{12\text{Im}\tau}-\frac{\tilde{F}_{5}^{2}}{4\cdot 5!}\right)\\
        &+\frac{1}{8i k_{10}^{2}}\int C_{4}\wedge G_{3}\wedge \overline{G}_{3}+S_{loc},\\
    \end{aligned}
\end{equation}
where the coupling constant is written as $2k_{10}^{2}=\frac{1}{2\pi}(2\pi l_{s})^{8}$ and $l_{s}=2\pi\sqrt{\alpha'}$, which is known as the string scale.

The fluxes are defined as:
\begin{equation}
    \begin{aligned}
        &F_{1}=dC_{0},\quad F_{3}=dC_{2},\quad F_{5}=dC_{4},\\
        &H_{3}=dB_{2},\\
        &\Tilde{G_{3}}=F_{3}-\tau H_{3},\\
        &\tau= C_{0}+ie^{-\phi}\label{axio},\\
        &\Tilde{F_{5}}=F_{5}-\frac{1}{2}C_{2}\wedge H_{3}+\frac{1}{2}B_{2}\wedge F_{3},\\
    \end{aligned}
\end{equation}
where $C_{0}, C_{2}, C_{4}$ are RR fields and $B_{2}$ is NS-NS field. $\phi$ is the dilaton. The five-form $\Tilde{F_{5}}$ must satisfy the self-duality constraint due to equation of motion, which is related to the warp factor $\alpha=e^{4A(y)}$:
\begin{equation}
    \Tilde{F}_{5}=(1+*)d\alpha \wedge dx^{0}\wedge dx^{1}\wedge dx^{2}\wedge dx^{3}.
\end{equation}

The action governing localized sources such as $Dp$-branes and $Op$-planes wrapping a $(p-3)$-cycle $\Sigma$ in the extra dimensions is expressed by:
\begin{equation}
    S_{loc}=-T_{p}\int_{R_{4}\times \Sigma}d^{p+1}\zeta \sqrt{-g}+u_{p}\int_{R_{4}\times \Sigma} C_{p+1},
\end{equation}
where an Op-plane has minus tension $T_{p}$ and RR charge $u_{p}$ in contrast with Dp brane.  For positive tension objects, $T_{p}=|u_{p}|e^{(p-3)\phi/4}$ in Einstein frame.

The equation of motion for $F_{5}$ is:
\begin{equation}
    d F_{5}=H_{3}\wedge F_{3}+2k_{10}^{2}T_{3}\rho_{3}^{local},
\end{equation}
where $\rho_{3}^{local}$ is the charge density for O3 plane and D7 brane. Integrating over the whole extra dimension gives the following tadpole condition:
\begin{equation}
    \frac{1}{2k_{10}^{2}T_{3}}\int H_{3}\wedge F_{3}+Q^{loc}=0,
\end{equation}
which means the total D3 charge vanishes.

The 3-form flux satisfy the imaginary self-dual condition following equation of motion and Bianchi identities:
\begin{equation}
    G_{3}=i *_{6}G_{3},
\end{equation}
which fixes the complex structure moduli and dilaton, as the dilaton is included in the definition of $G_{3}$, and the Hodge dual on the internal space is related to the complex structure.

In the large volume limit, the warp factor will vanish:
\begin{equation}
    \alpha\sim 1+\frac{1}{\mathcal{V}^{\frac{2}{3}}}.
\end{equation}
The solution describe the leading order behavior of  $\mathcal{N}=1$ 4d supergravity coupled to O3/O7 plane and D3/D7 branes expanding around $g_{s}$ and $\alpha'$.

From a 4-dimensional standpoint, at low energy, the aforementioned compactification is effectively captured by a 4-dimensional $\mathcal{N}=1$ supergravity theory. In this context, "low energy" denotes an energy scale significantly lower than the Kaluza-Klein scale. To proceed with our analysis, we initially introduce chiral coordinates as a means of organizing the scalar fields. For K\"ahler moduli, we have:
\begin{equation}
T_{u}=\tau_{u}+ib_{u},\quad u=1,2,...,h^{1,1},    
\end{equation}
where $\tau_{u}$ is the volume of the 4-cycle $X_{u}$ in the Einstein frame, and $b_{u}$ is the integration of the RR 4-form $C_{4}$ along the 4-cycle $X_{u}$: $b_{u}=\int_{X_{u}}C_{4}.$

For complex structure moduli, we have:
\begin{equation}
    \Pi_{i}=\int_{\sigma_{i}}\Omega,
\end{equation}
where $\Omega$ is the holomorphic three form and $\Sigma_{i}$ is an arbitrary 3-cycle. $\Pi_{i}$,  the period vector, is invariant under rescalings. Finally, the axio-dilaton is defined as $\tau=C_{0}+ie^{-\phi}$.

The leading order K\"ahler potential is:
\begin{equation}
    K=-2\text{log}(\mathcal{V})-\text{log}\left(\int\Omega\wedge\overline{\Omega}\right)-\text{log}(\tau-\overline{\tau}).
\end{equation}

The superpotential is given by:
\begin{equation}
    W=\frac{1}{2\pi}\int_{M}\Omega\wedge G_{3},
\end{equation}
which depends on all of the complex structure moduli and dilaton.

Combining them together, we have the $\mathcal{N}=1$ F term supergravity scalar potential:
\begin{equation}
    V=e^{K}\left(\sum K^{i\overline{j}}D_{i}WD_{\overline{j}}\overline{W}-3|W|^{2}\right),
\end{equation}
where the sum is over all  K\"ahler and complex structre moduli. A special feature of type IIB theory is that the sum over all K\"ahler moduli cancels with the $-3|W|^{2}$ term and leaves a sum limited to complex moduli space:
\begin{equation}
    V_{\text{no scale}}=e^{K}K^{a\overline{b}}D_{a}WD_{\overline{b}}W,
\end{equation}
where $a,b$ run over complex structure moduli and dilaton.

In this particular example, only the stabilization of complex structure moduli and the dilaton is achieved, attributed to the presence of a no-scale structure. To extend stabilization to K\"ahler moduli, it becomes necessary to incorporate perturbative corrections to the K\"ahler potential and non-perturbative corrections to the superpotential.
\section{Review of Swampland}
 The landscape of string theory vacua relates to the idea that string theory contains a vast configurations of consistent vacua. One example of this would be the many choices of flux in flux compactification that all appear consistent in the low-energy effective supergravity framework. In the pursuit of explaining the observable universe, string theory posits that only solutions adhering to the constraints observed in our real-world context constitute a consistent quantum gravity theory, discarding others. The viable solutions form what is commonly referred to as the Landscape, standing in contrast to the Swampland, a realm deemed incompatible with the principles of quantum gravity. This distinction arises from a top-down perspective.

Alternatively, adopting a bottom-up approach, the Landscape can be viewed as the set of low-energy effective theories possessing a complete UV description within the framework of string theory. This perspective seeks to leverage fundamental principles in quantum gravity to systematically eliminate ineffective field theories, thereby delineating the necessary properties that a low-energy effective field theory must satisfy within the landscape. This section provides a succinct overview of several conjectures pertinent to constructing a low-energy effective theory within the landscape.

\textbf{No Global Symmetries}:

Symmetry holds a prominent position in contemporary quantum field theory, delineated into two distinct categories: global symmetry and gauge symmetry.

Among the earliest conjectures within the swampland paradigm is the proposition that global symmetry finds no place in the realm of quantum gravity \cite{Banks:1988yz,Banks:2010zn,Harlow:2018tng}. This conjecture asserts that global symmetries are either gauged or spontaneously broken, applying to both continuous and discrete instances.

The motivation behind this conjecture stems from considerations related to black holes. In accordance with the no-hair theorem, black holes possess only three conserved quantities: mass, charge, and angular momentum. A pertinent example involves the amalgamation of numerous particles carrying baryon number to form a black hole. Due to Hawking radiation, the black hole undergoes evaporation, progressively diminishing in mass. However, since the mass of baryon is much heavier than the energy scale of Hawking radiation, during Hawking radiation, light particles, such as photons, are more prone to radiation. Imagine an initial state of a black hole consisting entirely of baryons, with the final state being entirely photons, which leads to the non-conservation of baryon numbers. Consequently, the global baryon number symmetry breaks in quantum gravity.

\textbf{Weak Gravity Conjecture}:

A natural question is if quantum gravity restricts gauge symmetries \cite{Arkani-Hamed:2006emk, Harlow:2022ich}. Consider a static, stationary black hole solution with U(1) gauge charge as an example. The geometry is given by Reissner-Nordstrom metric:
\begin{equation}
    ds^{2}=-\left(1-\frac{2M}{r}+\frac{2Q^{2}}{r^{2}}\right)dt^{2}+\frac{dr^{2}}{\left(1-\frac{2M}{r}+\frac{2Q^{2}}{r^{2}}\right)}+r^{2}d\Omega^{2},
\end{equation}
which is a symmetric solution of the following Lagrangian:
\begin{equation}
    S=\int d^{4}x \sqrt{-g}\left(R-\frac{1}{2}F_{uv}F^{uv}\right).
\end{equation}

The criterion ensuring the absence of a naked singularity for the black hole is expressed as $M \geq \sqrt{2}Q$. When $M > \sqrt{2}Q$, the black hole manifests both inner and outer horizons. In the case where $M = \sqrt{2}Q$, the black hole is termed extremal, possessing only a singular horizon. This critical threshold implies a finite count of black holes with gauge charge below a fixed mass, thereby permitting the coexistence of gauge symmetry within the realm of quantum gravity.

In the extremal scenario, the charged black hole undergoes continual decay through Hawking radiation. Initially characterized by a charge $Q$ and mass $M$, the black hole gradually disintegrates into numerous lighter components denoted by $m_i$ and $q_i$. Following the principles of energy and charge conservation, this decay process unfolds.:
\begin{equation}
    M=\mathop{\sum}\limits_{i}m_{i}+E_{kinetic},\quad Q=\mathop{\sum}\limits_{i}q_{i}.
\end{equation}
which leads to the inequality:
\begin{equation}
    \frac{M}{Q}\geq \frac{1}{Q}\mathop{\sum}\limits_{i}m_{i}=\frac{1}{Q}\mathop{\sum}\limits_{i}\frac{m_{i}}{q_{i}}q_{i}\geq \frac{1}{Q}\left(\frac{m}{q}\right)|_{min}\mathop{\sum}\limits_{i}q_{i}=\left(\frac{m}{q}\right)|_{min}.
\end{equation}

Hence, in the analysis of the decay of a nearly extremal black hole, it follows that there must exist at least one particle in the theoretical framework characterized by a charge $q$ and a mass $m$, according to the governing principles:
\begin{equation}
    m\leq \sqrt{2}q.
\end{equation}
This is known as the electric weak gravity conjecture.

\textbf{Distance Conjecture}

The distance conjecture is intimately tied to the moduli space's geometry, positing that traversing towards infinity along a direction within the moduli space results in the emergence of an infinite tower of light states, thereby disrupting the original effective field theory (EFT) description of the theory. The mass scale associated with these novel light states follows a proportionality of $e^{-aT}$, where $a$ is a constant of order 1, and $T$ denotes the distance in the moduli space.

Several instances within string theory substantiate this conjecture. A straightforward illustration arises in the context of considering 5-dimensional quantum gravity compactified on $S_{1}$. In this scenario, the masses of Kaluza-Klein (KK) modes are inversely proportional to the radius $r$ of $S_{1}$ as determined by the expectation value of the dilaton. The metric for the moduli space is $ds^2=\frac{dr^2}{r^2}$, and the distance between two points in the moduli space is expressed as $\text{log}\frac{r}{r_{0}}$. As $r$ tends to infinity, the distance also diverges, and the associated mass scale approaches zero, aligning with the predictions of the distance conjecture. Additional supporting examples encompass the compactification of M-theory on $R_{10}\times S_{1}$ and type IIB string theory compactified on a $K_{3}$ surface.

\textbf{dS Conjecture}:

Astrophysical observations indicate a very small yet positive cosmological constant in our universe. Given the anticipation that string theory should be reflective of the physical reality, the construction of De Sitter vacuum solutions within string compactification models becomes a crucial pursuit. Over the years, several models introducing positive vacuum energy, such as KKLT, have been developed; however, these models are not without their challenges. In my thesis, I will refrain from diving into the controversial swampland conjecture but instead focus on the Festina-Lente conjecture, as proposed in \cite{Montero:2019ekk}. This conjecture serves as a version of the Weak Gravity Conjecture tailored for the De Sitter spacetime, guiding our exploration in the quest for viable string compactification models.

In the context of the evaporation of a black hole in De Sitter space, as long as all the charged particles are heavy enough (i.e their masses satisfy the following constraint):
\begin{equation}
    m^2\geq eg M_P H.
\end{equation}

In the context of black hole evaporation, $e$ and $m$ represent the charge and mass of the evaporating particle, respectively, and $M_P$ denotes the Planck mass, with $g$ as the gauge coupling and $H$ representing the Hubble radius, adherence to a specific constraint ensures the emergence of an empty de Sitter space as a result of the evaporation process. Conversely, failure to satisfy this constraint during the evaporation of a Nariai black hole(the event horizon coincides with the cosmological horizon), leads to the formation of a singularity. Such a singularity is deemed inconsistent within the framework of quantum gravity.

\textbf{AdS Distance Conjecture}

This conjecture posits that within a quantum gravity framework situated in an Anti-de Sitter spacetime, as the cosmological constant tends towards zero, a novel tower of light states emerges, characterized by masses \cite{Lust:2019zwm}:
\begin{equation}
    m\sim|\Lambda|^\alpha,
\end{equation}
where $a$ is a positive number.
\section{Brief Introduction To AdS/CFT}
As we study holography of string vacua, it becomes imperative to provide an introduction to AdS/CFT. The integration of the Swampland Conjecture with AdS/CFT reveals an interconnection between these two previously distinct areas. This synthesis not only enhances our comprehension of string theory but also underscores the intricate relationship between the holographic properties of string vacua and the AdS/CFT correspondence.

\subsection{Top-down approach to AdS/CFT}

AdS/CFT establishes a duality between string theory and conformal field theory. In this framework, the string theory operates within the confines of a spacetime formed by the product of an asymptotically Anti-de Sitter space and a compact manifold. Simultaneously, the conformal field theory finds its definition on the boundary of the AdS spacetime. Notably, Anti-de Sitter space represents a maximally symmetric manifold characterized by constant negative curvature. Mathematically, $AdS_{p+2}$ is formally defined as a hypersurface in $(p+2)$-dimensional space. 
\begin{equation}  X_{0}^{2}+X_{p+2}^{2}-\mathop{\sum}\limits_{i=1}^{p+1}X_{i}^{2}=R^{2},
\end{equation}
embedded in a $(p+3)$-dimensional flat spacetime:
\begin{equation}
    ds^{2}=-dX_{0}^{2}-dX_{p+2}^{2}+\mathop{\sum}\limits_{i=1}^{p+1}dX_{i}^{2}.
\end{equation}
One can also use global coordinates:
\begin{equation}
    ds^{2}=R^{2}(-\text{cosh}^{2}\rho d\tau^{2}+d\rho^{2}+\text{sinh}^{2}\rho d\Omega^{2}),
\end{equation}
or Poincare coordinates:
\begin{equation}
    ds^{2}=R^{2}(\frac{du^{2}}{u^{2}}+u^{2}(-dt^{2}+\mathop{\sum}\limits_{i=1}^{p}dx_{i}^{2})).
\end{equation}

Two approaches exist for characterizing charged D-branes in the low-energy limit. The first method involves regarding them as charged solutions under R-R fields within the context of type IIB supergravity. In this perspective, D-branes are treated as sources of RR fields, and their dynamics are described accordingly. The second approach adopts a string theory viewpoint, treating D-branes as the endpoints where open strings terminate. Here, D-branes serve as the boundary conditions for open strings, and their dynamics are governed by the Dirac-Born-Infeld (DBI) action. In the low-energy limit, only massless modes of open strings are considered, leading to the emergence of gauge symmetry.

The starting point in our discussion is the extremal black Dp-brane, a solution to the equations of motion derived from the type II supergravity action. These Dp-branes carry charge, and our focus is specifically on type IIB, where the bosonic sector of the supergravity action is specified:
\begin{equation}
    S_{IIB}=\frac{1}{2k_{10}^{2}}\int dx^{10}\sqrt{g}\left(e^{-2\phi}(R+4(\partial\phi)^{2})-\frac{1}{2}|F_{p+2}|^{2}\right),
\end{equation}

The metric describing extremal black Dp-brane with electric charge $N$ is:
\begin{equation}
\begin{aligned}
    ds^{2}&=H^{-\frac{1}{2}}(r)dx_{u}dx^{u}+H^{\frac{1}{2}}(r)dy_{v}dy^{v},\ u=0,1,...,p, v=p+1,...,9\\
    &H(r)=1+\frac{(2\sqrt{\pi})^{5-p}\Gamma(\frac{7-p}{2})g_{s}N l_{s}^{7-p}}{r^{7-p}}\equiv 1+\frac{L^{7-p}}{r^{7-p}},\\
\end{aligned}   
\end{equation}
where $dx_{u}dx^{u}=\eta_{ua}dx^{u}dx^{a}$ is a $(p+1)$-dimensional Minkowski metric along the brane and  $dy_{v}dy^{v}=dr^{2}+ r^{2}d\Omega_{8-p}^{2}$ is a $(9-p)$-dimensional Euclidean metric perpendicular to the brane.

The dilaton and gauge fields read:
\begin{equation}
\begin{aligned}
    &\phi=\text{Log}(g_{s}H^{\frac{3-p}{4}}),\\
    &A_{p+1}=H(r)dx^{0}\wedge dx^{1}\wedge...\wedge dx^{p}.
\end{aligned}   
\end{equation}

One can check that the solution indeed carries electric charge $N$:
\begin{equation}
    \int_{S^{8-p}}*F_{p+2}=N.
\end{equation}
Note that in the case of $p=3$, the dilaton does not depend on $r$ and is therefore a constant field.

The effective regime of supergravity approximation is when $L\gg l_{s}$ ($l_s$ is the string scale), where $L\propto (g_{s}N)^{\frac{1}{7-p}}l_{s}$. Therefore, the supergravity description is valid for $\lambda\equiv g_{s}N\gg 1$.

The metric exhibits two noteworthy limits: Firstly, for $r \gg L$, it converges to the 10-dimensional Minkowski spacetime. Secondly, for $r \ll L$, the metric transforms into $AdS_{5} \times S_{5}$. Following the Dirac quantization condition, each D-brane carries integer multiples of charge. Consequently, an extremal black Dp-brane possessing a charge of $N$ can be equivalently interpreted as $N$ coincident branes, each carrying unit charge.

From the perspective of open strings, a single D-brane gives rise to an open string ending on it, resulting in a U(1) gauge theory in the low-energy limit. However, when N coincident D-branes are considered, the spectrum of massless open string excitations and the associated gauge groups undergo enhancement. The gauge groups become matrix-valued, signaling the emergence of a non-abelian gauge group. Notably, in the case of N coincident Dp-branes, this corresponds to a U(N) gauge symmetry. Specifically, for $p=3$, the theory describes a $d=3+1$ $\mathcal{N}=4$ U(N) super-Yang-Mills field theory.

Here, "coincident" implies that the distance $l$ between two branes satisfies $l \ll l_{s}$. This discussion involves taking the decoupling limit: 
\begin{equation}
    \alpha\to 0,\quad \frac{L}{\alpha} \quad\text{fixed},
\end{equation}

In the low-energy scenario with $E \ll \frac{1}{l_{s}}$, only approximately massless string states are activated. This condition corresponds to the near-brane limit in the supergravity description, where $r\ll L$. Notably, the energy near the horizon undergoes a redshift in this regime:
\begin{equation}
    E=E_{\infty}\frac{r^{2}}{L^{2}},
\end{equation}

This energy, significantly smaller than the string scale energy $\frac{1}{l_{s}}$, is confined to the vicinity of the brane in the supergravity regime. Importantly, these excitations remain localized near the brane and are unable to propagate to infinity due to the prevailing gravitational potential. In the decoupling limit, the bulk supergravity transmutes into a free theory set against a flat space background. Consequently, the system manifests two decoupling limits: one characterized by free bulk Type IIB supergravity, and the other governed by Type IIB string theory on $AdS_{5}\times S_{5}$ depicting the near-horizon limit geometry of the supergravity solution.

On the other hand, the open string description of D-branes is given by the following action:
\begin{equation}
    S=S_{bulk}+S_{int}+S_{brane},
\end{equation}
where $S_{bulk}$ is the action of 10d type IIB supergravity plus higher order quantum corrections, $S_{brane}$ is the 3+1 DBI action, which gives the $\mathcal{N}=4$ SYM action at leading order expansion of $\alpha$ plus higher derivative terms. $S_{int}$ is defined as the interaction between bulk and brane modes. In the decoupling limit and low energy limit, $S_{bulk}$
 becomes a free gravity theory on flat 10d spacetime, $S_{brane}$ describes $\mathcal{N}=4$ SYM , all higher derivative terms and $S_{int}$ vanishes in this limit. Then the system again consists of two decoupled theories: One is free bulk supergravity and the other is  $\mathcal{N}=4$ U(N) Super-Yang-Mills theory, which is also a conformal field theory. From the view of effective field theory, the two descriptions of the same system should be equivalent with each other. Therefore, it was conjectured that there is a duality between type IIB string theory on $AdS_{5}\times S_{5}$ and $\mathcal{N}=4$ U(N) Super-Yang-Mills theory in d=3+1 dimensions.

From two equivalent descriptions of D-branes, we know that the two coupling constants are related:
\begin{equation}
    \tau\equiv \frac{i}{g_{s}}+\frac{C_{0}}{2\pi}=\frac{4\pi i}{g_{YM}^{2}}+\frac{\theta}{2\pi},
\end{equation}
which implies the fact that both theories have a SL(2,Z) symmetry which sends $\tau$ to $\frac{a\tau+b}{c\tau+d}$( where $a,b,c,d\in Z$ with $ad-bc=1$).
\subsection{Bottom-up approach to AdS/CFT}

The known top-down constructions are quite limited and often describe gauge theories that are not that interesting from a practical standpoint. So, researchers have shifted gears, taking a bottom-up approach by crafting gravity actions that involve various additional fields. The idea is that, in principle, any brane construction, when compactified to lower dimensions, should give us standard covariant terms. While the coefficients are determined by the brane configuration, it is reasonable to consider generic coefficients in the bottom-up narrative. This strategy, while broadening our phenomenological understanding, comes with a catch: it does not provide strong predictive power because we do not know upfront if the bottom-up construction holds up consistently in the UV. Nonetheless, it has been proven useful in uncovering qualitative features.

We start from considering a massive scalar particle in the $d+1$ AdS background. We use the Poincare coordinate description of $AdS_{d+1}$ geometry, which is given by:
\begin{equation}
    ds^{2}=\frac{R_{AdS}^{2}}{z^{2}}(dz^{2}+\eta_{uv}dx^{u}dx^{v})\label{ads},
\end{equation}
where z goes from 0 to plus infinity. The Poincare patch of $AdS_{d+1}$ is conformally equivalent with the half of the d+1 Minkowski spacetime. In the following discussion we set $R_{AdS}=1$ for convenience.

The equation of motion of non-minimal coupling of massive scalar particle in curved spacetime is given by the following Klein-Gordon equation:
\begin{equation}
    \left(\Box_{d+1}+\frac{d-1}{4d}R-m^{2}\right)\phi(z,t,\vec{x})=0,
\end{equation}
where $\Box\equiv\frac{1}{\sqrt{-g}}\partial_{u}(\sqrt{-g}g^{uv}\partial_{v})$ is the d'Alembert's operator. Note that the equation of motion is invariant under conformal transformation:
\begin{equation}
    \begin{aligned}
        &g_{uv}\rightarrow \Omega^{2}g_{uv},\\
        &\phi\rightarrow \Omega^{-\frac{d-1}{2}}\phi.
    \end{aligned}
\end{equation}
Combining the facts that Eq (\ref{ads}) is conformally equivalent to Minkowski space and the equation of motion is invariant under conformal rescaling, we have:
\begin{equation}
    \left(\partial_{z}^{2}-\frac{M^{2}}{z^{2}}+\Box_{d}\right)\Tilde{\phi}(t,x,z)=0,\label{ads2}
\end{equation}
where
\begin{equation}
    \Tilde{\phi}(t,x,z)=z^{-\frac{d-1}{2}}\phi(t,x,z),\quad M^{2}=m^{2}+\frac{(d+1)(d-1)}{4},
\end{equation}
where we use the fact that $R=-d(d+1)$ for $AdS_{d+1}$. The solution of $\Tilde{\phi}$ near $z=0$ scales like:
\begin{equation}
    \Tilde{\phi}\sim z^{\alpha}+O(z^{\alpha+1}).
\end{equation}

Placing this in Eq (\ref{ads2}) leads to:
\begin{equation}
    \alpha(\alpha-1)=M^{2}.
\end{equation}

Now we define the boundary operator that is dual to the massive scalar field $\phi$:
\begin{equation}
    O= z^{-\alpha}\Tilde{\phi}=z^{-\Delta}\phi,\label{ads3}
\end{equation}
where $\Delta=\alpha+\frac{d-1}{2}$. One can check that:
\begin{equation}
    \Delta(\Delta-d)=m^{2}R_{AdS}^{2}.
\end{equation}

Classical field theory computations give the boundary to bulk operator:
\begin{equation}
    K=C_{\Delta}\left(\frac{z_{0}}{z_{0}^{2}+(\vec{z}-\vec{x})^{2}}\right)^{\Delta}.
\end{equation}

In the limit of $z_{0}\rightarrow 0$, it goes back to the two point correlation function in conformal field theory:
\begin{equation}
    \langle O(x_{1},z_{1})O(x_{2},z_{2}\rangle\sim \frac{1}{(\vec{z}-\vec{x})^{2\Delta}},
\end{equation}
and therefore the operator $O$ defined in Eq (\ref{ads3}) can be identified as the dual primary operator with conformal dimension $\Delta$ of the massive scalar field in $AdS_{d+1}$.
\section{Some General Properties of CFT}

In this section, we discuss how conformal field theories are naturally introduced in string theory and how they generalize to the superconformal field theory.

The sigma model action of string theory is written as:
\begin{equation}
    S=-T\int d^{2}\sigma \sqrt{-g}g^{\alpha\beta}\nabla_{\alpha}X_{u}\nabla_{\beta}X^{u}.
\end{equation}

One can check that, this action has two gauge symmetries: diffeomorphism invariance and Weyl invariance. In two-dimensional space, these two symmetries together are equivalent with conformal symmetry in the classical level.
Mathematically, the conformal symmetry is generated by diffeomorphisms that preserve the metric up to an overall factor $\Omega(x)^{2}$:
\begin{equation}
    g'_{ab}(x)=\Omega^{2}(x)g_{ab}(x),
\end{equation}
where the generators in $d$ dimensional spacetime are named as conformal Killing vectors. They obey the following partial differential equations:
\begin{equation}
    \nabla_{u}\xi_{v}+\nabla_{v}\xi_{u}=\frac{2}{d}g_{ab}\nabla^{m}\xi_{m}.
\end{equation}

Since the sigma model action has conformal symmetry, it is enough to discuss the physics in d-dimensional Minkowski spacetime. (As diffeomorphism and Weyl invariance can fix three degrees of freedom, which can exactly fix the metric.) Then the above equation becomes:
\begin{equation}
    \partial_{u}\xi_{v}+\partial_{v}\xi_{u}=\frac{2}{d}\eta_{ab}\partial^{m}\xi_{m}.
\end{equation}

Note that the solution of these equation depends on dimension. For $d=1,2$, there are infinitely many generators. For $d>2$, there are $\frac{1}{2}(d+1)(d+2)$ generators, which has $d$ translations, $\frac{1}{2}d(d-1)$ rotations, 1 scale transformation and $d$ special conformal transformations.

The bosonic string worldsheet is described by 2d CFT. On the string worldsheet, the coordinates are $\tau$ and $\sigma$. Using the holomorphic and antiholomorphic coordinates, the conformal transformation is written as:
\begin{equation}
    z\rightarrow f(z),\quad \overline{z}\rightarrow\overline{f}(\overline{z}),
\end{equation}
which corresponds to the following infinitesimal generators:
\begin{equation}
    l_{n}=-z^{n+1}\partial,\quad \overline{l}_{n}=-\overline{z}^{n+1}\overline{\partial}.
\end{equation}

These generators form the classical Virasoro algebra and obey the following commutation relations:
\begin{equation}
    [l_{m},l_{n}]=(m-n)l_{m+n},\quad [\overline{l}_{m},\overline{l}_{n}]=(m-n)\overline{l}_{m+n}. 
\end{equation}

Quantising the classical string sigma model gives the quantum level of conformal symmetry. Note that in the quantum case, diffeomorphism invariance is still preserved. However, the Weyl invariance does not hold anymore, which leads to the conformal anomaly. The commutation relation of Virosoro algebra in the quantum level has an extra term proportional to the central charge:
\begin{equation}
    [L_{m},L_{n}]=(m-n)L_{m+n}+\frac{c}{12}m(m^{2}-1)\delta_{m+n,0}\label{eq3}.
\end{equation}

The energy momentum tensor of 2d CFT can be expressed in terms of the Virasoro algebra:
\begin{equation}
    T(z)=\sum_{n=-\infty}^{\infty}\frac{L_{n}}{z^{n+2}}.
\end{equation}
Eq (\ref{eq3}) leads to the following OPE:
\begin{equation}
    T(z)T(w)\sim \frac{c/2}{(z-w)^{4}}+\frac{2T(w)}{(z-w)^{2}}+\frac{\partial T(w)}{z-w}.
\end{equation}

One can generalize to $\mathcal{N}=1$ Superconformal field theory, where the OPE is given as:
\begin{equation}
    \begin{aligned}
         &T(z)T(w)\sim \frac{c/2}{(z-w)^{4}}+\frac{2T(w)}{(z-w)^{2}}+\frac{\partial T(w)}{z-w},\\
         &T(z)G(w)\sim \frac{3}{2z^{2}}G(w)+\frac{1}{z}\partial G(w),\\
         &G(z)G(w)\sim \frac{2c}{3z^{3}}+\frac{2}{z}T(w).
    \end{aligned}
\end{equation}
In conformal field theory, the physical significance for anomalous dimension is the quantum correction to the scaling dimension. Most operators have a classical scaling dimension (e.g, a scalar field in 4d has mass dimension 1). Quantum corrections change classical scaling dimension. For marginal operators, by dimensional analysis, negative anomalous dimension leads to a strong coupling CFT, and positive anomalous dimension results in a free CFT. Therefore we are more interested in negative anomalous dimension.  

\section{Outline of the thesis}
People have taken a keen interest in unraveling the mysteries of string vacua through an innovative avenue known as the Swampland program. This effort aims to identify broad techniques to weed out numerous low-energy Lagrangians as potential candidates for string theory vacua. The aim is to uncover fundamental principles that absolutely apply in quantum gravity. In a somewhat akin philosophy, albeit using markedly different techniques, the Bootstrap program has been making impressive strides in setting boundaries for consistent Conformal Field Theories (CFTs). This effort relies solely on fundamental requirements like  crossing symmetry and unitarity. The Bootstrap program has made significant progress in mapping out the parameter space of consistent CFTs. As an illustration, it successfully demonstrates that the 3D Ising model lives on the fringes of the permissible regions \cite{El-Showk:2012cjh}.

The primary theme of this thesis revolves around exploring the potential connection between these two programs distinct in their principles through the AdS/CFT correspondence, as it might offer a new perspective on the swampland conjectures. This inclination gains support from the fact that certain swampland properties, such as distance conjecture \cite{Perlmutter:2020buo}, also have a dual CFT description.

In Chapter 2 and 3 we calculate the holographic dual of two specific examples: fibred Large Volume Scenario and M theory stabilizated with fluxes and non perturbative effects. In Chapter 4, we study the Wheeler-DeWitt state of a charged black hole.

\chapter{Moduli Stabilization Of Fibred Calabi Yau Scenarios}

In this section we study the moduli stabilization of fibred Calabi Yau Scenarios. This is based on the earlier work \cite{Conlon:2020wmc} where they studied holographic aspects of LVS, KKLT and racetrack moduli stabilization scenarios. These displayed interesting properties. LVS displayed a universality of conformal dimensions in the large-volume limit, and
all scenarios had an interesting negative behaviour for the mixed anomalous dimensions.
We are therefore interested in more scenarios, and in this section we extend these studies to
fibred versions of the original LVS scenario.

\section{A Brief Review of Fibred Calabi-Yau Scenarios}

One interesting variation on the original LVS is to consider models with fibred Calabi-Yau (CY) structures. In such models, there are multiple large moduli associated both to base and fibre moduli \cite{Cicoli:2008gp,Cicoli:2011it}. We shall take the form of the volume of a fibred Calabi-Yau to be given as the following (based on \cite{Cicoli:2008gp}):
\begin{equation}
\mathcal{V}=f_{\frac{3}{2}}(\tau_{j})-\sum_{i=1}^{N_{small}}\lambda_{i}\tau_{i}^{3/2},
\end{equation} 
where $f$ is a homogeneous function of degree 3/2, $j$ runs from 1 to $N_{large}$ and $h_{1,1}=N_{small}+N_{large}$, where $N_{large}$ is the number of large moduli and $N_{small}$ is the number of blow up moduli.

The fibred CY shares a similar feature with the ``Swiss Cheese'' form used in the original LVS, in terms of the blow-up moduli included in the volume. The difference lies in the function $f$, as this now involves multiple moduli: both the overall volume modulus and the fibration moduli. Different moduli play different roles. The volume modulus controls the LVS potential. The fibre modulus is a flat direction of the LVS potential, which means it will be absent in the LVS potential at leading order. However, it will appear in the kinetic terms, and also at subleading order in the string loop expansion. The blow up modulus controls the heavy mode which will subsequently be integrated out of the LVS potential within the low-energy effective field theory.

In the next two subsections, we will use two specific examples to illustrate these ideas which will involve respectively one or two fibre moduli.

\subsection{ Fibred Calabi-Yau: Example with One Fibre Modulus}

We will first look at moduli stabilisation in order to establish the AdS effective action (essentially a review of \cite{Grana:2005jc}), before considering the holographic implications of this effective action.

\subsubsection{Moduli Stabilisation}

We assume as in \cite{Cicoli:2008gp} that the volume takes the form: 
\begin{equation}
\mathcal{V}=\alpha(\sqrt{\tau_{1}}\tau_{2})-\gamma\tau_{3}^{3/2}.
\end{equation}

 Here $\tau_{1}$ controls the volume of the fibre, $\tau_{2}$ controls the volume of the base and $\tau_{3}$ describes the blow up modulus. Locally, the fibred Calabi Yau manifold is a direct product of fibre and base. Globally, it defines a fibre bundle (the math definition of a fibre bundle can be found in \cite{Nakahara:2003nw}).

The Large Volume Scenario is a specific type IIB flux compactification moduli stabilisation scenario, which stabilises all moduli in a non-SUSY AdS vacuum at an exponentially large volume $\mathcal{V}$ (as per \cite{Balasubramanian:2005zx, Conlon:2005ki,Conlon:2006gv}). Following the usual approach in LVS, after stabilising  and integrating out the complex structure moduli and dilaton,  the K\" ahler potential and superpotential read:
\begin{equation}
K=-2 \textrm{ln}(\mathcal{V}+\hat{\xi}), 
\end{equation}
\begin{equation}
W=W_{0}+A_{3}e^{-a_{3}T_{3}},
\end{equation}
\begin{equation}
\hat{\xi}=\frac{\xi}{g_{s}^{\frac{3}{2}}},
\end{equation}
\begin{equation}
\xi=-\frac{\chi(X)\zeta(3)}{2(2\pi)^{3}},
\end{equation}
Here $\chi(X)$ is the Euler number of the Calabi-Yau manifold X,  $\zeta$ is the Riemann zeta function with $\zeta(3)\approx1.2$. $T_{3}$ is the chiral multiplet for the blow-up K\" ahler modulus, $T_{3}=\tau_{3}+iC_{3}$.

The $\mathcal{N}=1$ F-term supergravity scalar potential takes the following form:
\begin{equation}
V=e^{K}(K^{i\overline{j}}D_{i}WD_{\overline{j}}\overline{W}-3|W|^{2}),
\end{equation}
where we have set $M_{P}=1$. After substituting the K\" ahler potential and superpotential into the scalar potential, the effective potential becomes:

\begin{equation}
V_{scalar}=\frac{Aa_{3}^{2}\sqrt{\tau_{3}}}{\mathcal{V}}e^{-2a_{3}\tau_{3}}-\frac{BW_{0}a_{3}A_{3}\tau_{3}}{\mathcal{V}^{2}}e^{-a_{3}\tau_{3}}+\frac{C\xi W_{0}^{2}}{g_{s}^{\frac{3}{2}}\mathcal{V}^{3}},
\end{equation}
where $A,B,C$ are numerical constants. The important point is that the fibre modulus does not appear in the effective potential at tree level, and so at this order is a flat direction of the scalar potential. Therefore there is an additional light scalar degree of freedom, which is an interesting feature of fibred Calabi-Yau models compared to the original LVS.

As with standard LVS, this scalar potential has a minimum at exponentially large volumes,
\begin{equation}
\langle \tau_{3}\rangle\sim \frac{\xi^{\frac{2}{3}}}{g_{s}},
\end{equation}
\begin{equation}
\langle \mathcal{V} \rangle\sim e^{a_{3}\langle\tau_{3}\rangle}.
\end{equation}

After integrating out the blow up modulus $\tau_{3}$, the effective potential for the LVS volume modulus reads \cite{Conlon:2018vov}:
\begin{equation}
V_{scalar}=[-A^{'}\left(\textrm{ln}(f\mathcal{V})\right)^{\frac{3}{2}}+B^{'}]\frac{W_{0}^{2}}{\mathcal{V}^{3}}.
\end{equation}
where $A^{'}$, $B^{'}$ are numerical constants.

\subsubsection{Kinetic Terms and Canonical Fields}
The effective field theory of these fibred models contains four light moduli: one overall volume modulus which is massive, plus one fibre modulus and two axions, which are approximately massless.

In order to see how they interact with each other, we write down the kinetic terms for the theory in the large volume limit using standard EFT techniques. Neglecting higher order terms, the kinetic terms read:
\begin{equation}
L_{kin}=\frac{1}{4\tau_{1}^{2}}\partial_{u}\tau_{1}\partial^{u}\tau_{1}+\frac{1}{2\tau_{2}^{2}}\partial_{u}\tau_{2}\partial^{u}\tau_{2}
+\frac{1}{4\tau_{1}^{2}}\partial_{u}a_{1}\partial^{u}a_{1}+\frac{1}{2\tau_{2}^{2}}\partial_{u}a_{2}\partial^{u}a_{2}.
\end{equation}

As a first step in diagonalising the action, we re-express the mode $\tau_{2}$  in terms of the overall LVS volume:
\begin{equation}
\tau_{2}=\frac{\mathcal{V}}{\alpha\sqrt{\tau_{1}}}.
\end{equation}

The kinetic terms then become
\begin{equation}
\mathcal{L} \supset \frac{3}{8\tau_{1}^{2}}\partial_{u}\tau_{1}\partial^{u}\tau_{1}-\frac{1}{2\tau_{1}\mathcal{V}}\partial_{u}\tau_{1}\partial^{u}\mathcal{V}+\frac{1}{2\mathcal{V}^{2}}\partial_{u}\mathcal{V}\partial^{u}\mathcal{V}
+\frac{1}{4\tau_{1}^{2}}\partial_{u}a_{1}\partial^{u}a_{1}+\frac{\alpha^{2}\tau_{1}}{2\mathcal{V}^{2}}\partial_{u}a_{2}\partial^{u}a_{2}.
\end{equation}

We next diagonalise both kinetic terms and mass terms in order to get a canonically normalised theory.
\begin{equation}
\textrm{ln}\, \tau_{1}=a\Phi_{1}+b\Phi_{2},
\end{equation}
\begin{equation}
\textrm{ln}\, \mathcal{V} =c\Phi_{1}+d\Phi_{2}.
\end{equation}

Here $\Phi_{1}$ and $\Phi_{2}$ will be the canonically normalised fields. Before we go into the details of the calculation, we make an observation. As the mass term only depends on the overall LVS volume, we must choose either $c=0$ or $d=0$ when diagonalising. Here we choose $d=0$, so our ansatz becomes:
\begin{equation}
\begin{aligned}
&\textrm{ln}\,\tau_{1}=a\Phi_{1}+b\Phi_{2},\\
&\textrm{ln}\, \mathcal{V}=c\Phi_{1}.\\
\end{aligned}
\end{equation}

We refer interested readers to appendix \ref{app:1m} for the details. Here we just state the final results:

\begin{equation}
\begin{aligned}
&\textrm{ln}\, \tau_{1}=\sqrt{\frac{2}{3}}(\Phi_{1}+\sqrt{2}\Phi_{2}),\\
&\textrm{ln}\, \mathcal{V}=\sqrt{\frac{3}{2}}\Phi_{1}.\\
\end{aligned}
\end{equation}

The kinetic terms then read:
\begin{equation}\label{eq1}
\frac{1}{2}(\partial_{u}\Phi_{1})^{2}+\frac{1}{2}(\partial_{u}\Phi_{2})^{2}+\frac{1}{2}e^{-2\sqrt{\frac{2}{3}}(\Phi_{1}+2\Phi_{2})}(\partial_{u}a_{1})^2
+\frac{1}{2}e^{\left(\sqrt{\frac{2}{3}}-2\sqrt{\frac{3}{2}}\right)\Phi_{1}+2\sqrt{\frac{2}{3}}\Phi_{2}}(\partial_{u}a_{2})^2,
\end{equation}
where the axions have also been rescaled by a constant factor to achieve canonical normalization. The potential only depends 
on $\Phi_{1}$ and so is diagonal in this basis. Having expressed the Lagrangian in canonical form,  we can now move to a discussion of the interactions and their holographic interpretation.
\subsubsection{Holographic Interpretation}
From the couplings in (\ref{eq1}) and in the scalar potential, we can easily read off the OPE coefficients, following AdS/CFT dictionary. The idea is to calculate the Witten diagram in the bulk and compare it with the 3 point correlation function at the boundary. For details see \cite{Penedones_2016}. The results are the following:
\begin{equation}
\begin{aligned}
&f_{a_{1}a_{1}}^{\Phi_{1}}= \sqrt\frac{2}{3} \frac{\pi^{\frac{d}{2}}\Gamma\left(\frac{2\Delta_{a_{1}}+\Delta_{\Phi_{1}}-d}{2}\right)}{\Gamma\left(\Delta_{\Phi_{1}}\right)\Gamma\left(\Delta_{a_{1}}\right)^{2}}\left(\Delta_{\Phi_{1}}+2\Delta_{a_{1}}-\Delta_{a_{1}}^{2}-3\right),\\
&f_{a_{1}a_{1}}^{\Phi_{2}}= 2\sqrt\frac{2}{3} \frac{\pi^{\frac{d}{2}}\Gamma\left(\frac{2\Delta_{a_{1}}+\Delta_{\Phi_{2}}-d}{2}\right)}{\Gamma\left(\Delta_{\Phi_{2}}\right)\Gamma\left(\Delta_{a_{1}}\right)^{2}}\left(\Delta_{\Phi_{2}}+2\Delta_{a_{1}}-\Delta_{a_{1}}^{2}-3\right),\\
&f_{a_{2}a_{2}}^{\Phi_{1}}=  \left( \sqrt{\frac{3}{2}} -\frac{1}{2}\sqrt{\frac{2}{3}} \right) \frac{\pi^{\frac{d}{2}}\Gamma\left(\frac{2\Delta_{a_{2}}+\Delta_{\Phi_{1}}-d}{2}\right)}{\Gamma\left(\Delta_{\Phi_{1}}\right)\Gamma\left(\Delta_{a_{2}}\right)^{2}}\left(\Delta_{\Phi_{1}}+2\Delta_{a_{2}}-\Delta_{a_{2}}^{2}-3\right),\\
&f_{a_{2}a_{2}}^{\Phi_{2}}= -\sqrt\frac{2}{3} \frac{\pi^{\frac{d}{2}}\Gamma\left(\frac{2\Delta_{a_{2}}+\Delta_{\Phi_{2}}-d}{2}\right)}{\Gamma\left(\Delta_{\Phi_{2}}\right)\Gamma\left(\Delta_{a_{2}}\right)^{2}}\left(\Delta_{\Phi_{2}}+2\Delta_{a_{2}}-\Delta_{a_{2}}^{2}-3\right),\\
&f_{\Phi_{1}\Phi_{1}}^{\Phi_{1}}=\frac{81\sqrt{6}}{2}\frac{3\pi^{\frac{d}{2}}\Gamma\left(\frac{3\Delta_{\Phi_{1}}-3}{2}\right)}{\Gamma\left(\Delta_{\Phi_{1}}\right)^{3}},
\end{aligned}
\end{equation}
where $\Delta_O$ represents the conformal dimension for the corresponding operator $O$. The key difference between the fibred Calabi-Yau LVS model and the original LVS model is that there are a total of 3 massless moduli in the effective field theory rather than just one. This leads to operator mixing in the operator product expansion. On the other hand, the volume modulus behaves exactly as in the pure LVS scenario, obtaining a conformal dimension of \cite{Conlon:2018vov,Conlon:2020wmc}
\begin{equation}
\Delta_{\Phi_1}= \frac{3(1+\sqrt{19})}{2},
\end{equation}
and with the same self-couplings.

In principle, the following double trace operators are degenerate
\begin{equation}
[a_{1}a_{1}]_{n,\ell},\quad [a_{2}a_{2}]_{n,\ell},\quad[a_{1}a_{2}]_{n,\ell},\quad [\Phi_{2}\Phi_{2}]_{n,\ell},\quad[a_{1}\Phi_{2}]_{n,\ell},\quad[a_{2}\Phi_{2}]_{n,\ell},
\end{equation}
while 
\begin{equation}
\quad[a_{1}\Phi_{1}]_{n,\ell}, \quad[a_{2}\Phi_{1}]_{n,\ell}
\end{equation}
also mix within themselves.
Each axion is invariant under an independent discrete $\mathbb{Z}_{2}$ symmetry acting as $a_i \rightarrow - a_i$, which results in selection rules for the double trace operators that can appear in a given OPE. This implies that states in different symmetry classes cannot mix with one another even though their classical dimension is the same. This is shown in Table \ref{tab:Z2}.

\begin{table}[h!]
\centering
\begin{tabular}{|c|c|c|c|}
\hline  
 &$\mathbb{Z}_{2}(a_{1})$&$\mathbb{Z}_{2}(a_{2})$\\
\hline 
$[a_{1}\Phi_{2}]_{n,\ell}$&-1&1\\
\hline 
$[a_{2}\Phi_{2}]_{n,\ell}$&1&-1\\
\hline 
$[a_{1}a_{2}]_{n,\ell}$&-1&-1\\
\hline 
$[a_{1}a_{1}]_{n,\ell}, [a_{2}a_{2}]_{n,\ell},[\Phi_{2}\Phi_{2}]_{n,\ell}$&1&1\\
\hline \hline
$[a_{1}\Phi_{1}]_{n,\ell}$&-1&1\\
\hline
$[a_{2}\Phi_{1}]_{n,\ell}$&1&-1\\
\hline
\end{tabular}
\caption{Irreducible representation of the double trace operators under the $\mathbb{Z}_{2} \times \mathbb{Z}_{2} $ symmetry. The two different blocks correspond to distinct subspaces, where the degeneracies are partly broken by the discrete symmetries.}
\label{tab:Z2}
\end{table}

For the non-degenerate operators involving different fields $ [a_{1}\Phi_{1}]_{0,\ell}, [a_{2}\Phi_{1}]_{0,\ell} 
 [a_{1}\Phi_{2}]_{0,\ell},$ $ [a_{2}\Phi_{2}]_{0,\ell}, [a_{1}a_{2}]_{0,\ell} $, we can directly  work out the anomalous dimension. In particular, since the kinematic factors are equal for the two axions
\begin{equation}
\begin{aligned}
& \gamma(0,\ell)_{a_{1}\Phi_{1}}  \propto -f_{a_{1}a_{1}\Phi_{1}}f_{\Phi_{1} \Phi_{1} \Phi_{1}} < 0,\\
&\gamma(0,\ell)_{a_{2}\Phi_{1}}  =0,\\
&\gamma(0,\ell)_{a_{1}\Phi_{2}}  =0,\\
&\gamma(0,\ell)_{a_{2}\Phi_{2}} =0.\\
&\gamma(0,\ell)_{a_{1}a_{2}}\propto -f_{a_{1}a_{1}\Phi_{2}}f_{a_{2}a_{2}\Phi_{2}} >0,\\
\end{aligned}
\end{equation}

The first two anomalous dimensions are exactly those of LVS without fibred directions, and their agreement with the original conjecture proposed in \cite{Conlon:2020wmc} should not come as a surprise. However, $\gamma(0,\ell)_{a_{1}a_{2}}$ is positive, violating the expected behaviour, which means the repulsive interaction between $a_1$ and $a_2$. The interpretation for this is that maintaining a fixed volume requires one cycle to increase in size while the other decreases, and so the coefficients reflecting the coupling of the associated axions to the modulus have opposite signs.

Let us now analyse the properly degenerate subspace. Taking perturbative corrections into account \cite{Cicoli:2007xp}, D-brane loops lift the flat potential for $\Phi_{2}$ with a scaling
\begin{equation}
V_{\rm{loop}} \sim \frac{1}{\mathcal{V}^{10/3}},
\end{equation} 
which is parametrically faster than $\mathcal{O}(1/N)$ and thus dominates over anomalous dimensions. Therefore the degeneracy between $a_{1},a_{2}$ and $\Phi_{2}$ breaks to $a_{1}$ and $a_{2}$ only. Since there are no interactions mixing $a_1$ and $a_2$ in the Lagrangian, the resulting matrix is diagonal, with
\begin{equation}
\begin{aligned}
&\gamma(0,\ell)_{a_{1}a_{1}}\propto -f_{a_{1}a_{1}\Phi_{2}}^2 <0,\\
&\gamma(0,\ell)_{a_{2}a_{2}}\propto -f_{a_{2}a_{2}\Phi_{2}}^2 <0.\\
\end{aligned}
\end{equation}
Similarly, the anomalous dimension is negative (by construction) for $ [\Phi_{2}\Phi_{1}]_{0,\ell}$.

\subsection{Fibred Calabi Yau: Example with Two Fibred Moduli}

A similar analysis can be carried out for other fibred Calabi-Yau models (here Appendix \ref{app:2m}) gives the full details). One direct extension is to consider a model with two fibred moduli, in which the volume of the fibred Calabi-Yau manifold reads: 
\begin{equation}
\mathcal{V}=\sqrt{\tau_{1}\tau_{2}\tau_{3}}-\tau_{s}^{3/2}.
\end{equation}
Here $\tau_{1},\tau_{2},\tau_{3}$ control the volumes of individual 4-cycles, with the overall volume direction corresponding in the rescaling $\tau_{i}\rightarrow \lambda\tau_{i}$. The $\sqrt{\tau_{1}\tau_{2}\tau_{3}} $ form is reminiscent of toroidal orbifolds based on $T_{2}\times T_{2}\times T_{2}$, while $\tau_{s}$ is the blow up mode necessary to realize LVS.

In the large volume limit, the K\"ahler metric reads:
$$
K_{i\overline{j}}=\left(
\begin{matrix}
\frac{1}{4\tau_{1}^{2}} & \frac{\tau_{s}^{\frac{3}{2}}}{8\tau_{1}^{\frac{3}{2}}\tau_{2}^{\frac{3}{2}}\tau_{3}^{\frac{1}{2}}}
&\frac{\tau_{s}^{\frac{3}{2}}}{8\tau_{1}^{\frac{3}{2}}\tau_{2}^{\frac{1}{2}}\tau_{3}^{\frac{3}{2}}}&-\frac{3\tau_{s}^{\frac{1}{2}}}{8\tau_{1}\mathcal{V}} \\\\
\frac{\tau_{s}^{\frac{3}{2}}}{8\tau_{1}^{\frac{3}{2}}\tau_{2}^{\frac{3}{2}}\tau_{3}^{\frac{1}{2}}}& \frac{1}{4\tau_{2}^{2}}&
\frac{\tau_{s}^{\frac{3}{2}}}{8\tau_{1}^{\frac{1}{2}}\tau_{2}^{\frac{3}{2}}\tau_{3}^{\frac{3}{2}}}& -\frac{3\tau_{s}^{\frac{1}{2}}}{8\tau_{2}\mathcal{V}}\\\\
\frac{\tau_{s}^{\frac{3}{2}}}{8\tau_{1}^{\frac{3}{2}}\tau_{2}^{\frac{1}{2}}\tau_{3}^{\frac{3}{2}}} &\frac{\tau_{s}^{\frac{3}{2}}}{8\tau_{1}^{\frac{1}{2}}\tau_{2}^{\frac{3}{2}}\tau_{3}^{\frac{3}{2}}} & \frac{1}{4\tau_{3}^{2}}&-\frac{3\tau_{s}^{\frac{1}{2}}}{8\tau_{3}\mathcal{V}}\\\\
-\frac{3\tau_{s}^{\frac{1}{2}}}{8\tau_{1}\mathcal{V}} & -\frac{3\tau_{s}^{\frac{1}{2}}}{8\tau_{2}\mathcal{V}}& -\frac{3\tau_{s}^{\frac{1}{2}}}{8\tau_{3}\mathcal{V}} &\frac{3}{8}\frac{1}{\tau_{s}^{\frac{1}{2}}\mathcal{V}}\\\\
\end{matrix}
\right),
$$
while the inverse matrix is:
$$
K^{-1}=\left(
\begin{matrix}
4\tau_{1}^{2} & -2\frac{\tau_{s}^{\frac{3}{2}}\tau_{1}^{\frac{1}{2}}\tau_{2}^{\frac{1}{2}}}{\tau_{3}^{\frac{1}{2}}}
& -2\frac{\tau_{s}^{\frac{3}{2}}\tau_{1}^{\frac{1}{2}}\tau_{3}^{\frac{1}{2}}}{\tau_{2}^{\frac{1}{2}}}&4\tau_{1}\tau_{s} \\\\
 -2\frac{\tau_{s}^{\frac{3}{2}}\tau_{1}^{\frac{1}{2}}\tau_{2}^{\frac{1}{2}}}{\tau_{3}^{\frac{1}{2}}} &4\tau_{2}^{2}&  -2\frac{\tau_{s}^{\frac{3}{2}}\tau_{2}^{\frac{1}{2}}\tau_{3}^{\frac{1}{2}}}{\tau_{1}^{\frac{1}{2}}}& 4\tau_{2}\tau_{s}\\\\
-2\frac{\tau_{s}^{\frac{3}{2}}\tau_{1}^{\frac{1}{2}}\tau_{3}^{\frac{1}{2}}}{\tau_{2}^{\frac{3}{2}}}& -2\frac{\tau_{s}^{\frac{3}{2}}\tau_{3}^{\frac{1}{2}}\tau_{2}^{\frac{1}{2}}}{\tau_{1}^{\frac{1}{2}}} &4\tau_{3}^{2}&4\tau_{3}\tau_{s}\\\\
4\tau_{1}\tau_{s} & 4\tau_{2}\tau_{s} & 4\tau_{3}\tau_{s} & \frac{8}{3}\mathcal{V}\tau_{s}^{\frac{1}{2}}\\\\
\end{matrix}
\right).
$$
The kinetic term for the moduli $\tau_{1},\tau_{2},\tau_{3}$ and their axion partners can be expressed as the following:
\begin{equation}
\begin{aligned}
\mathcal{L} \supset \, &\frac{1}{4\tau_{1}^{2}}\partial^{u}\tau_{1}\partial_{u}\tau_{1}+\frac{1}{4\tau_{1}^{2}}\partial^{u}a_{1}\partial_{u}a_{1}
+\frac{1}{4\tau_{2}^{2}}\partial^{u}\tau_{2}\partial_{u}\tau_{2}+\\
&\frac{1}{4\tau_{2}^{2}}\partial^{u}a_{2}\partial_{u}a_{2}+
\frac{1}{4\tau_{3}^{2}}\partial^{u}\tau_{3}\partial_{u}\tau_{3}+\frac{1}{4\tau_{3}^{2}}\partial^{u}a_{3}\partial_{u}a_{3}.\\
\end{aligned}
\end{equation}
Following similar techniques to the last section, we can substitute one of the flat moduli with the volume modulus, diagonalise the kinetic term and write down the scalar potential. After putting all the constituents together, the resulting effective field theory reads:

\begin{equation}
\begin{aligned}
\mathcal{L}= &\frac{1}{2}(\partial_{u}\Phi_{1})^{2}+\frac{1}{2}(\partial_{u}\Phi_{2})^{2}+\frac{1}{2}(\partial_{u}\Phi_{3})^{2}+\frac{1}{2}e^{-2\sqrt{\frac{2}{3}}\Phi_{1}-2\sqrt{\frac{1}{3}}\Phi_{2}-2\Phi_{3}}(\partial_{u}a_{1})^{2}\\
&+\frac{1}{2}e^{-2\sqrt{\frac{2}{3}}\Phi_{1}-2\sqrt{\frac{1}{3}}\Phi_{2}+2\Phi_{3}}(\partial_{u}a_{2})^{2}
+\frac{1}{2}e^{\left(2\sqrt{\frac{2}{3}}-2\sqrt{\frac{3}{2}}\right)\Phi_{1}+4\sqrt{\frac{1}{3}}\Phi_{2}}(\partial_{u}a_{3})^{2}-V_{LVS},\\
\end{aligned}
\end{equation}
where the axions have also been rescaled to give canonical kinetic terms. As before, the OPE coefficients can be read off:
\begin{equation}
\begin{aligned}
&f_{a_{1}a_{1}}^{\Phi_{1}}= \sqrt\frac{2}{3} \frac{\pi^{\frac{d}{2}}\Gamma\left(\frac{2\Delta_{a_{1}}+\Delta_{\Phi_{1}}-d}{2}\right)}{\Gamma\left(\Delta_{\Phi_{1}}\right)\Gamma\left(\Delta_{a_{1}}\right)^{2}}\left(\Delta_{\Phi_{1}}+2\Delta_{a_{1}}-\Delta_{a_{1}}^{2}-3\right),\\
&f_{a_{1}a_{1}}^{\Phi_{2}}=\sqrt\frac{1}{3} \frac{\pi^{\frac{d}{2}}\Gamma\left(\frac{2\Delta_{a_{1}}+\Delta_{\Phi_{2}}-d}{2}\right)}{\Gamma\left(\Delta_{\Phi_{2}}\right)\Gamma\left(\Delta_{a_{1}}\right)^{2}}\left(\Delta_{\Phi_{2}}+2\Delta_{a_{1}}-\Delta_{a_{1}}^{2}-3\right),\\
&f_{a_{1}a_{1}}^{\Phi_{3}}= \frac{\pi^{\frac{d}{2}}\Gamma\left(\frac{2\Delta_{a_{1}}+\Delta_{\Phi_{3}}-d}{2}\right)}{\Gamma\left(\Delta_{\Phi_{3}}\right)\Gamma\left(\Delta_{a_{1}}\right)^{2}}\left(\Delta_{\Phi_{3}}+2\Delta_{a_{1}}-\Delta_{a_{1}}^{2}-3\right),\\
&f_{a_{2}a_{2}}^{\Phi_{1}}=  \sqrt\frac{2}{3} \frac{\pi^{\frac{d}{2}}\Gamma\left(\frac{2\Delta_{a_{2}}+\Delta_{\Phi_{1}}-d}{2}\right)}{\Gamma\left(\Delta_{\Phi_{1}}\right)\Gamma\left(\Delta_{a_{2}}\right)^{2}}\left(\Delta_{\Phi_{1}}+2\Delta_{a_{2}}-\Delta_{a_{2}}^{2}-3\right),\\
&f_{a_{2}a_{2}}^{\Phi_{2}}= \sqrt\frac{1}{3} \frac{\pi^{\frac{d}{2}}\Gamma\left(\frac{2\Delta_{a_{2}}+\Delta_{\Phi_{2}}-d}{2}\right)}{\Gamma\left(\Delta_{\Phi_{2}}\right)\Gamma\left(\Delta_{a_{2}}\right)^{2}}\left(\Delta_{\Phi_{2}}+2\Delta_{a_{2}}-\Delta_{a_{2}}^{2}-3\right),\\
&f_{a_{2}a_{2}}^{\Phi_{3}}= -\frac{\pi^{\frac{d}{2}}\Gamma\left(\frac{2\Delta_{a_{2}}+\Delta_{\Phi_{3}}-d}{2}\right)}{\Gamma\left(\Delta_{\Phi_{3}}\right)\Gamma\left(\Delta_{a_{2}}\right)^{2}}\left(\Delta_{\Phi_{3}}+2\Delta_{a_{2}}-\Delta_{a_{2}}^{2}-3\right),\\
&f_{a_{3}a_{3}}^{\Phi_{1}}= \left( \sqrt\frac{3}{2}-\sqrt\frac{2}{3} \right) \frac{\pi^{\frac{d}{2}}\Gamma\left(\frac{2\Delta_{a_{3}}+\Delta_{\Phi_{1}}-d}{2}\right)}{\Gamma\left(\Delta_{\Phi_{1}}\right)\Gamma\left(\Delta_{a_{3}}\right)^{2}}\left(\Delta_{\Phi_{1}}+2\Delta_{a_{3}}-\Delta_{a_{3}}^{2}-3\right),\\
&f_{a_{3}a_{3}}^{\Phi_{2}}= -2\sqrt\frac{1}{3} \frac{\pi^{\frac{d}{2}}\Gamma\left(\frac{2\Delta_{a_{3}}+\Delta_{\Phi_{2}}-d}{2}\right)}{\Gamma\left(\Delta_{\Phi_{2}}\right)\Gamma\left(\Delta_{a_{3}}\right)^{2}}\left(\Delta_{\Phi_{2}}+2\Delta_{a_{3}}-\Delta_{a_{3}}^{2}-3\right),\\
&f_{a_{3}a_{3}}^{\Phi_{3}}= 0 ,\\
&f_{\Phi_{1}\Phi_{1}}^{\Phi_{1}}=\frac{81\sqrt{6}}{2}\frac{3\pi^{\frac{d}{2}}\Gamma\left(\frac{3\Delta_{\Phi_{1}}-3}{2}\right)}{\Gamma\left(\Delta_{\Phi_{1}}\right)^{3}}.
\end{aligned}
\end{equation}

Analogously to what happens in the previous examples, all the axions and the two fibre moduli are massless at zeroth order, leading to a large degeneracy between the double trace operators. A first simplification is again provided by the existence of a $\mathbb{Z}_2 \times \mathbb{Z}_2 \times \mathbb{Z}_2 $, symmetry, which partially breaks it down to subset of operators in the same representations. A classification is provided in Table \ref{tab:Z23}. 
\begin{table}
\begin{tabular}{|c|c|c|c|}
\hline  
 &$\mathbb{Z}_{2}(a_{1})$&$\mathbb{Z}_{2}(a_{2})$&$\mathbb{Z}_{2}(a_{3})$\\
\hline 
$[a_{1}a_{2}]_{n,\ell}$&-1&-1&1\\
\hline 
$[a_{1}a_{3}]_{n,\ell}$&-1&1&-1\\
\hline 
$[a_{2}a_{3}]_{n,\ell}$&1&-1&-1\\
\hline
$[a_{1}\Phi_{2}]_{n,\ell},[a_{1}\Phi_{3}]_{n,\ell}$&-1&1&1\\
\hline 
$[a_{2}\Phi_{2}]_{n,\ell},[a_{2}\Phi_{3}]_{n,\ell}$&1&-1&1\\
\hline 
$[a_{3}\Phi_{2}]_{n,\ell},[a_{3}\Phi_{3}]_{n,\ell}$&1&1&-1\\
\hline 
$[a_{1}a_{1}]_{n,\ell}, [a_{2}a_{2}]_{n,\ell}, [a_{3}a_{3}]_{n,\ell},[\Phi_{2}\Phi_{2}]_{n,\ell},[\Phi_{3}\Phi_{3}]_{n,\ell},,[\Phi_{2}\Phi_{3}]_{n,\ell}$&1&1&1\\
\hline 
\hline
$[a_{1}\Phi_{2}]_{n,\ell},[a_{1}\Phi_{1}]_{n,\ell}$&-1&1&1\\
\hline 
$[a_{2}\Phi_{2}]_{n,\ell},[a_{2}\Phi_{1}]_{n,\ell}$&1&-1&1\\
\hline 
$[a_{3}\Phi_{2}]_{n,\ell},[a_{3}\Phi_{1}]_{n,\ell}$&1&1&-1\\
\hline 
\end{tabular}
\caption{Irreducible representation of the double trace operators under the $\mathbb{Z}_{2} \times \mathbb{Z}_{2} \times \mathbb{Z}_{2} $ symmetry. The two different blocks correspond to distinct subspaces, where the degeneracies are partly broken by the discrete symmetries.}
\label{tab:Z23}
\end{table}

We are now ready to discuss the anomalous dimensions. For operators involving axions only,
\begin{equation}
\begin{aligned}
&\gamma_{a_{1}a_{2}}\propto -(f_{a_{1}a_{1}}^{\Phi_{2}}f_{a_{2}a_{2}}^{\Phi_{2}}+f_{a_{1}a_{1}}^{\Phi_{3}}f_{a_{2}a_{2}}^{\Phi_{3}})
>0,\\
&\gamma_{a_{1}a_{3}}\propto -(f_{a_{1}a_{1}}^{\Phi_{2}}f_{a_{3}a_{3}}^{\Phi_{2}})>0,\\
&\gamma_{a_{2}a_{3}}\propto -(f_{a_{2}a_{2}}^{\Phi_{2}}f_{a_{3}a_{3}}^{\Phi_{2}})>0.\\
\end{aligned}
\end{equation}

We have obtained the same, puzzling result encountered in the previous section - positive anomalous dimensions for operators involving two axions. Moreover, all three anomalous dimensions are identical in this case. On the other hand, for the ones involving the volume modulus
\begin{equation}
\begin{aligned}
&\gamma_{a_{1} \Phi_1}\propto -f_{a_{1}a_{1}}^{\Phi_{1}}f_{\Phi_{1} \Phi_{1}}^{\Phi_{1}} <0,\\
&\gamma_{a_{2} \Phi_1}\propto -f_{a_{2}a_{2}}^{\Phi_{1}}f_{\Phi_{1} \Phi_{1}}^{\Phi_{1}} <0,\\
&\gamma_{a_{3} \Phi_1}\propto -f_{a_{3}a_{3}}^{\Phi_{1}}f_{\Phi_{1} \Phi_{1}}^{\Phi_{1}} <0\\
\end{aligned}
\end{equation}
are all negative as in previous studies.

Again, all of the operators of the form $[a_{i}\Phi_{2}]_{0,\ell}, [a_{j}\Phi_{3}]_{0,\ell}$ decouple through a combination of the discrete symmetries and sub-leading volume correction, which lift the degeneracy between $\Phi_2$ and $\Phi_3$ (and with the axions too). In particular they are all zero at this order in perturbation theory. The same happens for $[\Phi_2\Phi_{2}]_{0,\ell},[\Phi_2\Phi_{3}]_{0,\ell}, [\Phi_3\Phi_{3}]_{0,\ell} $, which do not mix with the axion double trace operators for the same reason. Finally, 
$$[a_{1}a_{1}]_{n,\ell},\quad[a_{2}a_{2}]_{n,\ell},\quad[a_{3}a_{3}]_{n,\ell}$$
are still degenerate,
but the corresponding anomalous dimension matrix is diagonal and all entries are necessarily negative, since they are proportional to a sum of squared OPE coefficients (with a minus sign in front).

\chapter{M Theory Moduli Stabilization}
\section{Introduction}
The properties of the vacuum in string theory are important because if string theory is true, then the vacuum will be able to describe our universe. The approach to connect the original formulation of string theory and the real world is string compactification (for reviews see \cite{Conlon:2006gv, Douglas:2006es,Denef:2008wq,Denef:2007pq}), which reduces the full 10d string vacuum to a (3+1)-d spacetime, while other dimensions are compactified. However, in the process of string compactification, new massless scalar fields will emerge. The vacuum expectation values of them, called moduli, do not at first appear in the 4d effective potential, so they are not constrained. 

If moduli can take arbitrary values (for example, time dependent), then this will conflict with observations. Therefore, it is essential to stabilize the moduli. The ingredients of moduli stabilization are fluxes and non-perturbative effects (for example, world sheet instantons \cite{Dine:1986zy} and gaugino condensation) to make new terms appear in the 4d effective potential to minimize the moduli. Fluxes obey a higher dimensional generalization of Dirac quantization, so they take integer values. Therefore, the corresponding vacua are also labeled by discrete integer fluxes. The set of all vacua generated by these fluxes form the so-called "landscape".

M-Theory is another limit of string theory. Moduli stabilization is also interesting for M-Theory compactification. This has been studied in \cite{PhysRevLett.97.191601,Acharya:2002kv,Acharya:2007rc} by both flux and non-perturbative stabilization. In the flux-stablized vacua, all the saxions and a linear combination of axions are stabilized at SUSY AdS vacua under certain topological conditions. The non-perturbative stabilized vacuum is the multiple moduli version of the racetrack model.

The above method goes from top to bottom. However, in recent years an alternative approach has risen: the swampland program \cite{Palti:2019pca,Ooguri:2006in,Vafa:2005ui}. This aims at using basic principles in quantum gravity to exclude many effective field theories and work out properties that low energy effective field theory must satisfy. It is beneficial to understand the swampland program from holographic perspectives: for references see \cite{Benjamin:2016fhe,Nakayama:2015hga,Montero:2016tif,Giombi:2017mxl,Urbano:2018kax,Harlow:2018tng,Harlow:2018jwu,Montero:2018fns,Baume:2020dqd,Perlmutter:2020buo}.

A novel approach from holography was proposed \cite{Conlon:2018vov,Apers:2022zjx,Conlon:2020wmc,Conlon:2021cjk,Apers:2022tfm,Quirant:2022fpn}, which provides a new perspective to the problem. The motivation is to determine the consistency of these 4d vacua from the CFT side. So far, what has been studied are the LVS, the fibred LVS and DGKT. In particular DGKT gives an interesting spectrum leading to the integer conformal dimensions.
  
In this chapter we extend the holographic swampland story to M-Theory, both models with fluxes and without fluxes. The chapter is organised as follows. In Section \ref{Chapter3.2} we describe general aspects of M-Theory moduli stabilization on a $G_{2}$-holomony manifold and the mass matrix elements. In Section \ref{Chapter3.3} we study the holographic dual of M-Theory vacuum with flux and compare it with the previous results of DGKT \cite{Conlon:2021cjk,Apers:2022tfm}.  In Section \ref{Chapter3.4} we study the holographic dual of M-Theory vacuum with zero flux background and compare it with the previous results of KKLT and racetrack \cite{Conlon:2020wmc}. In the Appendix the detailed calculations are presented.
\section{General Aspects Of M-Theory Moduli Stabilization}\label{Chapter3.2}
In this section we give a brief description of M-Theory moduli stabilization on a $G_{2}$ holonomy manifold \cite{Acharya:2002kv}. The 11d low energy supergravity description of M-theory is given by the following action \cite{Cremmer:1978km}:
\begin{equation}
S=\frac{1}{2 \textit{k}_{11}^{2}}\left[\int d^{11}x\sqrt{-g}R-\int \left(\frac{1}{2}G_{4}\wedge \ast G_{4}-\frac{1}{6} C_{3}\wedge G_{4}\wedge G_{4}\right) \right],
\end{equation}
where the 11d metric $\textit{g}$ and a 3-form $C_{3}$ consists of the bosonic components. $G_{4}=dC_{3}$ is the field strength.
By compactifying this theory on a 7d compact manifold X with $G_{2}$ holonomy group, one obtains the 4d $\mathcal{N}=1$ supergravity theory. A covariantly constant 3-form $\phi$ always accompanies a manifold X with $G_{2}$ holonomy. Following the argument in \cite{Acharya:2002kv}, the moduli space of X has the same dimension as $H^{3}(X,R)$, therefore the 4d $\mathcal{N}=1$ supergravity theory has $b_{3}(X)$ moduli.
The corresponding chiral fields read: 
\begin{equation}
z_{i}=t_{i}+i s_{i},\quad i=1,...N,
\end{equation}
$s_{i}$ are volume of 3-cycles and $t_{i}$ are the corresponding axions. 
In the case that is consistent with $G_{2}$ holonomy \cite{Acharya:2007rc}, the K\" ahler potential is expressed by:
\begin{equation}
K=-3\textrm{Log}(\mathcal{V}).
\end{equation}
We assume, following \cite{Acharya:2007rc},  the volume of the 7d-manifold can be written as $\mathcal{V}=\prod_{i=1}^{N}s_{i}^{a_{i}}.$
The parameters have the following constraints:
\begin{equation}
\Sigma_{i=1}^{N}a_{i}=\frac{7}{3}.
\end{equation}
The K\" ahler potential satisfies the following no scale relationships:
\begin{equation}
\label{noscale}
K^{ij}K_{j}=-s^{i}, \quad K^{ij}K_{i}K_{j}=7.
\end{equation}
We now derive general expressions for the Hessians of both volume moduli and axions, assuming that we have a SUSY vacuum. Details are presented in  Appendix \ref{a1}. We start from the standard $\mathcal{N}=1$ supergravity potential form (taking $M_{P}=1$):
\begin{equation}
\textrm{V}=e^{K}(K^{i\overline{j}}D_{i}WD_{\overline{j}}\overline{W}-3|W|^{2}).
\end{equation}
The SUSY condition is:
\begin{equation}
D_{i}W=\partial_{i}W+K_{i}W=0\label{SUSY Condition}.
\end{equation}
For the volume moduli, the corresponding Hessian is expressed:
\begin{equation}
\begin{aligned}
\label{eq:mass matrix}
\partial_{b}\partial_{a}\textrm{V}&=K_{ab}\textrm{V}-3e^{K}K_{b}\partial_{a}|W|^{2}-3e^{K}\partial_{b}\partial_{a}|W|^{2}+e^{K}K^{i\overline{j}}(\partial_{a}D_{i}W)(\partial_{b}D_{\overline{j}}\overline{W})\\
&+e^{K}K^{i\overline{j}}(\partial_{b}D_{i}W)(\partial_{a}D_{\overline{j}}\overline{W}).\\
\end{aligned}
\end{equation}

Following Appendix \ref{a1} (also see \cite{Apers:2022tfm}),  putting everything together:

\begin{equation}
\label{eq:3}
\begin{aligned}
H^{V}_{ab}=\frac{\textrm{V}_{ab}}{e^{K}|W|^{2}}&=-K_{ab}+3K_{a}K_{b}+2\frac{W_{ab}}{W}+8K^{ij}\frac{W_{ia}W_{jb}}{|W|^{2}}\\
&+2s_{i}K_{b}\frac{W_{ia}}{W}+2s_{j}K_{a}\frac{W_{jb}}{W},\\
\end{aligned}
\end{equation}
where $K$, $W$ are the K\"ahler potential and superpotential evaluated at the SUSY vacuum. 

The Hessian for the axions can be obtained similarly from Eq (\ref{eq:mass matrix}), with all derivatives of the K\"ahler potential vanishing:
\begin{equation}
\begin{aligned}
\partial_{b}\partial_{a}V&=-3e^{K}\partial_{b}\partial_{a}|W|^{2}+e^{K}K^{i\overline{j}}(\partial_{a}D_{i}W)(\partial_{b}D_{\overline{j}}\overline{W})\\
&+e^{K}K^{i\overline{j}}(\partial_{b}D_{i}W)(\partial_{a}D_{\overline{j}}\overline{W}),\\
\end{aligned}
\end{equation}
where $\partial_{a}$,$\partial_{b}$ means $\partial_{t_{a}}$, $\partial_{t_{b}}$.
Again, following Appendix \ref{a1} (also see \cite{Apers:2022tfm}), we can get the axion Hessians similarly:
\begin{equation}
\begin{aligned}
\label{eq:4}
H^{A}_{ab}&=\frac{\partial_{t_{b}}\partial_{t_{a}}V}{e^{K}|W|^{2}}=2K_{a}K_{b}+6\frac{W_{ab}}{W}+8K^{ij}\frac{W_{ia}W_{jb}}{|W|^{2}}\\
&+2s_{i}K_{b}\frac{W_{ia}}{W}+2s_{j}K_{a}\frac{W_{jb}}{W},\\
\end{aligned}
\end{equation}
where  $\partial_{a}$,$\partial_{b}$ means derivatives to  $\partial_{s_{a}}$, $\partial_{s_{b}}$.\\

We now apply this formalism to two models of moduli stabilization. The first is flux-stabilization and the second is non-perturbative stabilization. In practice the superpotential will be either Eq (\ref{super1}) or Eq (\ref{super2}). 
\section{Flux-Stabilized M-Theory Vacuum}\label{Chapter3.3}
This section investigates the moduli stabilization method of \cite{Acharya:2002kv,Acharya:2005ez}, which aims at stabilizing the moduli with fluxes. This is analogous to the DGKT model in type IIA, which also uses fluxes. We give an introduction to M-Theory compactifications on $G_{2}$ manifolds with fluxes turned on.

In order to stabilize the moduli, certain topological conditions have to be imposed: The $G_{2}$ manifold X needs to have an ADE singularity along a 3-manifold Q \cite{Acharya:2002kv}. Following \cite{Beasley:2002db,Acharya:2002kv}, the superpotential induced by turning on a background flux G and background fields at the singularities is:
\begin{equation}
W=\frac{1}{8\pi^{2}}\int \left(\frac{C_{3}}{2}+i\phi \right)\wedge G+c_{1}+ic_{2}.
\end{equation}
Expanding the 4-form flux G into a harmonic basis: 
\begin{equation}
G=N^{i}\rho_{i},
\end{equation}
where $N_{i}$ are fluxes and $\rho_{i}\in H^{4}(X,\mathbb{Z})$, results in the following superpotential:
\begin{equation}
\label{super1}
W(z)=N^{i}z_{i}+c_{1}+ic_{2}.
\end{equation}
Following \cite{Acharya:2002kv},  $c_{1}+i c_{2}$ is a complex Chern-Simons invariant. This term is essential for this mechanism because without this term the moduli could not be stabilized and the cosmological constant would be zero. Without loss of generality we can take $c_{2}$ to be positive. 

The SUSY condition $D_i W=0$ implies:
\begin{equation}
\quad \frac{-3a_{i}}{2i s_{i}}(N^{j}z_{j}+c_{1}+ic_{2})+N_{i}=0.
\end{equation}
Separating the real and imaginary part of this equation, we have:
\begin{equation}
\begin{aligned}
&N^{i}t_{i}+c_{1}=0,\\
&\frac{3a_{i}}{2s_{i}}(N^{j}s_{j}+c_{2})=N^{i},\\
\end{aligned}
\end{equation}
which results in:
\begin{equation}
s_{i}\frac{3a_{i}}{2s_{i}}(N^{j}s_{j}+c_{2})=\frac{7}{2}(N^{j}s_{j}+c_{2})=N^{i}s_{i}.
\end{equation}
This leads to the following solution:
\begin{equation}
\begin{aligned}
&c_{2}=-\frac{5}{7}N^{i}s_{i},
&s_{i}=-\frac{3a_{i}}{5N_{i}}c_{2},
\end{aligned}
\end{equation}
which implies a vacuum expectation value $W=-i \frac{2c_{2}}{5}$.\\

Before we proceed, we need to check the validity of the effective field theory description. In order to realize scale separation, we need the Kaluza-Klein radius to be much smaller than the AdS radius:
\begin{equation}
\begin{aligned}
&R^{2}_{AdS}=\frac{1}{e^{K}|W|^{2}}=\frac{\mathcal{V}^{3}}{|W|^{2}}\sim \frac{c_{2}^{7}}{c_{2}^{2}}\sim c_{2}^{5},\\
&R^{2}_{KK}\sim \mathcal{V}^{\frac{2}{7}}\sim c_{2}^{\frac{2}{3}}.
\end{aligned}
\end{equation}
so $R_{KK}\ll R_{AdS}$ if $c_{2}\gg 1$. Following \cite{Acharya:2002kv}, we assume that large values of $c_{2}$ can be obtained for some special $G_{2}-$manifolds. 

This superpotential implies that the second order derivatives of the superpotential all vanish. Following Eq (\ref{eq:3}), the Hessian corresponding to the volume moduli is:
\begin{equation}
M_{ij}=-K_{ij}+3K_{i}K_{j}.
\end{equation}

The physical masses can be extracted as the eigenvalues of $m=2K^{-1}M$, where the matrix elements read:
\begin{equation}
\begin{aligned}
m_{ij}&=2K^{ia}M_{aj}\\
&=2K^{ia}(-K_{aj}+3K_{a}K_{j})\\
&=-2\delta_{ij}-6s_{i}K_{j},
\end{aligned}
\end{equation}
where we use the no scale relationship Eq (\ref{noscale}).

The eigenvalues are:
\begin{equation}
\begin{aligned}
&\lambda_{1}=...\lambda_{N-1}=-2,\\
&\lambda_{N}=-2-6s_{i}K_{i}\\
&=-2-6s_{i}\frac{-3\frac{\partial \mathcal{V}}{\partial s_{i}}}{\mathcal{V}}\\
&=-2+18\times \frac{\frac{7\mathcal{V}}{3}}{\mathcal{V}}\\
&=40,
\end{aligned}
\end{equation}
where we use the property that $\mathcal{V}$ is a homogeneous function of $s_{i}$ of degree $\frac{7}{3}$.\\

Using the standard relationship $\Delta(\Delta-3)=m^{2}R_{AdS}^{2}$, the corresponding conformal dimensions are:
\begin{equation}
\Delta_{1}=...=\Delta_{N-1}=2,
\end{equation}
\begin{equation}
\Delta_{N}=8.
\end{equation}
For the axions, the corresponding Hessian is:
\begin{equation}
M_{t_{i}t_{j}}=2K_{i}K_{j}.
\end{equation}
Similarly, the matrix elements of $m=2K^{-1}M$ are expressed as:
\begin{equation}
\begin{aligned}
m_{ij}&=2K^{ia}M_{aj}\\
&=4K^{ia} K_{a}K_{j}\\
&=-4s_{i}K_{j}.
\end{aligned}
\end{equation}
The corresponding conformal dimensions for the axions are:
\begin{equation}
\Delta_{1}=...=\Delta_{N-1}=3,
\end{equation}
\begin{equation}
\Delta_{N}=7.
\end{equation}

These are the main results for Section \ref{Chapter3.3}. This is rather interesting because all the conformal dimensions are integers. This is reminiscent of DGKT, where similar results were found for general Calabi Yau manifolds \cite{Conlon:2021cjk,Apers:2022tfm}. It was argued in \cite{Kachru:2001je} that for a special class of M-theory compactifications on 7d manifolds X with $G_{2}$ holonomy, special loci in their moduli space are well described by type IIA orientifolds. Therefore, it is natural to compare with DGKT. There may be some connections between the integers of M-Theory and the integers in DGKT, but their values are different. We give a summary of the results for general DGKT.

Following \cite{Apers:2022tfm}, for the  K\" ahler moduli and axion-dilaton sector:
\begin{equation}
\Delta_{1}=10, \quad \Delta_{2...h^{1,1}_{-}+1}=6,
\end{equation}
for the saxions and 
\begin{equation}
\Delta_{1}=11, \quad \Delta_{2...h^{1,1}_{-}+1}=5.
\end{equation}
for the corresponding axions. For the complex structure moduli sector:
\begin{equation}
\Delta_{u_{a}}=2,\quad \Delta_{a_{a}}=3, \quad a=1,...h^{2,1}.
\end{equation}
In DGKT, the K\" ahler potential and superpotential can be expressed as \cite{DeWolfe:2005uu}:
\begin{equation}
K=K^{K}+K^{Q}=-\textrm{Log}(\mathcal{V})+4D,
\end{equation}
\begin{equation}
W=W^{K}+W^{Q},
\end{equation}
where
\begin{equation}
W^{K}=e_{0}+e_{a}z^{a}+\frac{1}{2}k_{abc}m_{a}z_{b}z_{c}-\frac{m_{0}}{6}k_{abc}z_{a}z_{b}z_{c},
\end{equation}
and
\begin{equation}
W^{Q}=-2p_{k}N_{k}-iq_{\lambda}T_{\lambda}.
\end{equation}

Scale separation and geometric limit rely on large $c_{2}\gg 1$. Note that in DGKT fluxes can be chosen arbitrarily large for any Calabi Yau 3-folds. However, in the M-Theory flux-stabilization, the value of $c_{2}$ depends on the topological conditions of the 7d $G_{2}$ manifold. An example of large $c_{2}$ is given in \cite{Acharya:2002kv}, where $Q=H^{3}/\Gamma$.

Note that the complex structure moduli sector in DGKT is similar to M-theory flux vacuum. This is because $W_{Q}$ is also a linear combination of complex structure moduli, which is along the same line as Eq (\ref{super1}). The resemblance has been predicted in \cite{DeWolfe:2005uu}. However, the volume moduli and axion-dilaton sector seems to be different because there is no quadratic and cubic terms in Eq (\ref{super1}).

\section{Stabilization By Non-Perturbative Effects}\label{Chapter3.4}
Up to now we have focused on scenarios arising from M-theory flux vacua. In this section, we study a different model based on M-theory. The authors took a further step in \cite{PhysRevLett.97.191601,Acharya:2007rc}: They considered the M-theory vacuum with zero flux background. We will revisit this model and focus on the detailed properties of its holographic dual in this section. 

We give a quick summary to the model. The  K\" ahler potential remains the same in this case:
\begin{equation}
K=-3\textrm{Log}(\mathcal{V}),
\end{equation}
where $\mathcal{V}=\prod_{i=1}^{N}s_{i}^{a_{i}}$ with $\Sigma_{i=1}^{N}a_{i}=\frac{7}{3}.$
The superpotential is generated by the non-perturbative effects \cite{PhysRevLett.97.191601,Acharya:2007rc}:
\begin{equation}
\label{super2}
W=A_{1}e^{i b_{1}\Sigma_{i=1}^{N}N_{i}^{1}z_{i}}+A_{2}e^{i b_{2}\Sigma_{i=1}^{N}N_{i}^{2}z_{i}},
\end{equation}
where $A_{k}$ are numerical constants. $b_{1}=\frac{2\pi}{P}$ and $b_{2}=\frac{2\pi}{Q}$ with $\textit{P}, \textit{Q}$ being the rank of the gauge group for gauge condensation. The sets of $N_{i}^{1}, N_{i}^{2}$ are all integers. Therefore, the M-theory vacuum is fully determined by the constants $(a_{i}, b_{1}, b_{2}, N_{i}^{1}, N_{i}^{2}, A_{1}, A_{2})$. In this paper, without loss of generality, we take positive $A_{1}, A_{2}$.

The SUSY condition is:
\begin{equation}
\label{eq:susy}
D_{i}W=\partial_{i}W+K_{i}W=0.
\end{equation}
The vacuum solution to Eq (\ref{eq:susy}) is:
\begin{equation}
\frac{A_{1}}{A_{2}}=-\textrm{cos}[(b_{1}\Sigma_{i=1}^{N}N^{1}_{i}s_{i}-b_{2}\Sigma_{i=1}^{N}N^{2}_{i}s_{i})]\frac{2b_{2}N^{2}_{i}s_{i}+3a_{i}}{2b_{1}N^{1}_{i}s_{i}+3a_{i}}e^{(b_{1}\Sigma_{i=1}^{N}N^{1}_{i}s_{i}-b_{2}\Sigma_{i=1}^{N}N^{2}_{i}s_{i})},
\end{equation}
\begin{equation}
\textrm{sin}[(b_{1}\Sigma_{i=1}^{N}N^{1}_{i}s_{i}-b_{2}\Sigma_{i=1}^{N}N^{2}_{i}s_{i})]=0.
\end{equation}
To prove them, we can start by taking the expression of $W$ and $K$ into SUSY condition :

\begin{equation}
i(A_{1}b_{1}N^{1}_{i}e^{i b_{1}\Sigma_{i=1}^{N}N^{1}_{i}z_{i}}+A_{2}b_{2}N^{2}_{i}e^{i b_{2}\Sigma_{i=1}^{N}N^{2}_{i}z_{i}})-\frac{3a_{i}}{2is_{i}}(A_{1}e^{i b_{1}\Sigma_{i=1}^{N}N^{1}_{i}z_{i}}+A_{2}e^{i b_{2}\Sigma_{i=1}^{N}N^{2}_{i}z_{i}})=0.
\end{equation}
which implies:
\begin{equation}
\begin{aligned}
\frac{A_{1}}{A_{2}}&=-\frac{2b_{2}N^{2}_{i}s_{i}+3a_{i}}{2b_{1}N^{1}_{i}s_{i}+3a_{i}}e^{ib_{2}\Sigma_{i=1}^{N}N^{2}_{i}z_{i}-i b_{1}\Sigma_{i=1}^{N}N^{1}_{i}z_{i}}\\
&=-\frac{2b_{2}N^{2}_{i}s_{i}+3a_{i}}{2b_{1}N^{1}_{i}s_{i}+3a_{i}}e^{-b_{2}\Sigma_{i=1}^{N}N^{2}_{i}s_{i}+ b_{1}\Sigma_{i=1}^{N}N^{1}_{i}s_{i}}\\
&(\textrm{cos}[-b_{2}\Sigma_{i=1}^{N}N^{2}_{i}t_{i}+ b_{1}\Sigma_{i=1}^{N}N^{1}_{i}t_{i}]-i\textrm{sin}[-b_{2}\Sigma_{i=1}^{N}N^{2}_{i}t_{i}+ b_{1}\Sigma_{i=1}^{N}N^{1}_{i}t_{i}]).\\
\end{aligned}
\end{equation}
So we prove the SUSY condition.

The overall phase of $\textit{W}$,
$e^{ib_{1}\Sigma_{i=1}^{N}N^{1}_{i}t_{i}}$,
does not have physical meaning, since it is only relative phase factors that matter. If we have $A_{1}, A_{2}$ real and positive, the SUSY condition implies:
\begin{equation}
\label{av}
\textrm{cos}[(-b_{2}\Sigma_{i=1}^{N}N^{2}_{i}t_{i}+ b_{1}\Sigma_{i=1}^{N}N^{1}_{i}t_{i})]=-1,
\end{equation}
which also gives:
\begin{equation}
\label{eq:1}
\frac{A_{1}}{A_{2}}=\frac{2b_{2}N^{2}_{i}s_{i}+3a_{i}}{2b_{1}N^{1}_{i}s_{i}+3a_{i}}e^{-b_{2}\Sigma_{i=1}^{N}N^{2}_{i}s_{i}+ b_{1}\Sigma_{i=1}^{N}N^{1}_{i}s_{i}},
\end{equation}
which results in the following equations:
\begin{equation}
\label{16}
\frac{A_{2}}{A_{1}}=\frac{1}{\alpha} e^{-b_{1}\sum_{i=1}^{N}N^{1}_{i}s_{i}+b_{2}\sum_{i=1}^{N}N^{2}_{i}s_{i}},
\end{equation}
\begin{equation}
\label{11}
s_{i}=-\frac{3a_{i}(\alpha-1)}{2(b_{1}N^{1}_{i}\alpha-b_{2}N^{2}_{i})},
\end{equation}
where one can solve $\alpha$ and $s_{i}$ numerically.

The Hessians of the volume moduli and the axions in AdS units read:
\begin{equation}
\begin{aligned}
H^{V}_{ab}=R_{AdS}^{2}\textrm{V}_{ab}=\frac{\textrm{V}_{ab}}{e^{K}|W|^{2}}&=-K_{ab}+3K_{a}K_{b}+2\frac{W_{ab}}{W}+8K^{ij}\frac{W_{ia}W_{jb}}{|W|^{2}}\\
&+2s_{i}K_{b}\frac{W_{ia}}{W}+2s_{j}K_{a}\frac{W_{jb}}{W},\\
\end{aligned}
\end{equation}
and
\begin{equation}
\begin{aligned}
H^{A}_{ab}=R_{AdS}^{2}\textrm{V}_{ab}=\frac{\textrm{V}_{ab}}{e^{K}|W|^{2}}&=2K_{a}K_{b}+6\frac{W_{ab}}{W}+8K^{ij}\frac{W_{ia}W_{jb}}{|W|^{2}}\\
&+2s_{i}K_{b}\frac{W_{ia}}{W}+2s_{j}K_{a}\frac{W_{jb}}{W}.\\
\end{aligned}
\end{equation}
The physical masses can be acquired as the eigenvalues of $M=2K^{-1}H$. Using the explicit expressions for the Hessians in the Appendix, we have:
\begin{equation}
\begin{aligned}
M^{V}_{ab}=2K^{ai}H^{V}_{ib}=&=-2\delta_{ab}+\alpha\frac{s_{a}^{2}}{3a_{a}}\left(b_{1}N^{1}_{a}+\frac{3a_{a}}{2s_{a}}\right)\\
&\left(b_{1}N^{1}_{b}+\frac{3a_{b}}{2s_{b}}\right)\left(\Sigma_{i=1}^{N}\frac{8}{3a_{i}}\alpha\left(b_{1}N^{1}_{i}s_{i}+\frac{3a_{i}}{2}\right)^{2}-2\right),\\
\end{aligned}
\end{equation}
\begin{equation}
\begin{aligned}
M^{A}_{ab}=2K^{ai}H^{A}_{ib}=&=\alpha\frac{s_{a}^{2}}{3a_{a}}\left(b_{1}N^{1}_{a}+\frac{3a_{a}}{2s_{a}}\right)\left(b_{1}N^{1}_{b}+\frac{3a_{b}}{2s_{b}}\right)\\
&\left(\Sigma_{i=1}^{N}\frac{8}{3a_{i}}\alpha\left(b_{1}N^{1}_{i}s_{i}+\frac{3a_{i}}{2}\right)^{2}-6\right).\\
\end{aligned}
\end{equation}
The eigenvalues are:
\begin{equation}
\lambda^{V}_{i}=-2, \quad i=1,...N-1,
\end{equation}
\begin{equation}
\begin{aligned}
\lambda^{V}_{N}&=-2+\alpha\Sigma_{a=1}^{N}\frac{s_{a}^{2}}{3a_{a}}\left(b_{1}N^{1}_{a}+\frac{3a_{a}}{2s_{a}}\right)^{2}\left(\Sigma_{i=1}^{N}\frac{8}{3a_{i}}\alpha\left(b_{1}N^{1}_{i}s_{i}+\frac{3a_{i}}{2}\right)^{2}-2\right)\\
&=-2+\alpha\Sigma_{a=1}^{N}\frac{1}{3a_{a}}\left(b_{1}N^{1}_{a}s_{a}+\frac{3a_{a}}{2}\right)^{2}\left(\Sigma_{i=1}^{N}\frac{8}{3a_{i}}\alpha\left(b_{1}N^{1}_{i}s_{i}+\frac{3a_{i}}{2}\right)^{2}-2\right),\\
\end{aligned}
\end{equation}
and
\begin{equation}
\lambda^{A}_{i}=0,\quad i=1,...N-1,
\end{equation}
\begin{equation}
\begin{aligned}
\lambda^{A}_{N}&=\alpha\Sigma_{a=1}^{N}\frac{s_{a}^{2}}{3a_{a}}\left(b_{1}N^{1}_{a}+\frac{3a_{a}}{2s_{a}}\right)^{2}\left(\Sigma_{i=1}^{N}\frac{8}{3a_{i}}\alpha\left(b_{1}N^{1}_{i}s_{i}+\frac{3a_{i}}{2}\right)^{2}-6\right)\\
&=\alpha\Sigma_{a=1}^{N}\frac{1}{3a_{a}}\left(b_{1}N^{1}_{a}s_{a}+\frac{3a_{a}}{2}\right)^{2}\left(\Sigma_{i=1}^{N}\frac{8}{3a_{i}}\alpha\left(b_{1}N^{1}_{i}s_{i}+\frac{3a_{i}}{2}\right)^{2}-6\right).\\
\end{aligned}
\end{equation}
respectively.\\
Using the standard relationship $\Delta(\Delta-3)=m^{2}R^{2}_{AdS}$,
the corresponding conformal dimensions are:
\begin{equation}
\Delta^{V}_{i}=2, \quad i=1,...N-1,
\end{equation}
\begin{equation}
\label{cd}
\Delta^{V}_{N}=\frac{3+\sqrt{1+4\alpha\Sigma_{a=1}^{N}\frac{1}{3a_{a}}\left(b_{1}N^{1}_{a}s_{a}+\frac{3a_{a}}{2}\right)^{2}\left(\Sigma_{i=1}^{N}\frac{8}{3a_{i}}\alpha\left(b_{1}N^{1}_{i}s_{i}+\frac{3a_{i}}{2}\right)^{2}-2\right)}}{2},
\end{equation}
and
\begin{equation}
\Delta^{A}_{i}=3, \quad i=1,...N-1,
\end{equation}
\begin{equation}
\Delta^{A}_{N}=\frac{3+\sqrt{9+4\alpha\Sigma_{a=1}^{N}\frac{1}{3a_{a}}\left(b_{1}N^{1}_{a}s_{a}+\frac{3a_{a}}{2}\right)^{2}\left(\Sigma_{i=1}^{N}\frac{8}{3a_{i}}\alpha\left(b_{1}N^{1}_{i}s_{i}+\frac{3a_{i}}{2}\right)^{2}-6\right)}}{2}.
\end{equation}
These are the main results for Section \ref{Chapter3.4}. Note that the option $\Delta^{V}_{i}=1$ is excluded by $N=1$ supersymmetry, as the volume moduli and the axions are in the same three dimensional boundary $N=1$ supermultiplet \cite{Cordova:2016emh}.

Following \cite{Acharya:2007rc}, we work in the following two branches in which the supergravity description is meaningful:
\begin{equation}
\begin{aligned}
&a)\frac{A_{2}}{A_{1}}>1, \textrm{min}{\frac{b_{2}N_{i}^{2}}{b_{1}N_{i}^{1}};i=1...N}>\alpha>\textrm{max}{\frac{b_{2}N_{i}^{2}+\frac{3a_{i}}{2}}{b_{1}N_{i}^{1}+\frac{3a_{i}}{2}};i=1...N}\\
&b)\frac{A_{2}}{A_{1}}<1, \textrm{max}{\frac{b_{2}N_{i}^{2}}{b_{1}N_{i}^{1}};i=1...N}<\alpha<\textrm{min}{\frac{b_{2}N_{i}^{2}+\frac{3a_{i}}{2}}{b_{1}N_{i}^{1}+\frac{3a_{i}}{2}};i=1...N}\\
\end{aligned}
\end{equation}

It remains to show how the conformal dimension (\ref{cd}) grows with the $R_{AdS}$ in the scale seperation limit where $R_{AdS}$ goes to infinity. Since we have a supersymmetric vacuum, it follows that $V=-3e^{K}|W|^{2}$, therefore
$\frac{1}{R_{AdS}^{2}}=e^{K}|W|^{2}$,
which implies:
\begin{equation}
\textrm{Log}  R_{AdS}=-\frac{1}{2}K- \textrm{Log} W.
\end{equation}
Following the definition of  K\" ahler potential, superpotential, Eq (\ref{16}), Eq (\ref{18}) and combining them, we have:
\begin{equation}
\textrm{Log}   R_{AdS}=\Sigma_{i=1}^{N}\frac{3}{2}a_{i}\textrm{Log} s_{i}+b_{1}\Sigma N_{i}^{1}s_{i}+\textrm{Log}\left(A_{1}\left(1-\frac{1}{\alpha}\right)\right).
\end{equation}
In the scale separation limit, the leading order contribution of the conformal dimension and the $\textrm{Log}R_{AdS}$ is:
\begin{equation}
\Delta_{N}\propto \alpha\Sigma_{a=1}^{N}\frac{1}{3a_{a}}\left(b_{1}N^{1}_{a}s_{a}+\frac{3a_{a}}{2}\right)^{2}\propto b_{1}(\Sigma_{i=1}^{N} N_{i}^{1}s_{i})^{2},
\end{equation}
\begin{equation}
\textrm{Log}^{2}R_{AdS}\propto b_{1}(\Sigma_{i=1}^{N} N_{i}^{1}s_{i})^{2},
\end{equation}
Therefore, in the scale separation limit:
\begin{equation}
\Delta_{Heavy}\propto \textrm{Log}^{2}R_{AdS}.
\end{equation}

We end this section by comparing the results with other scenarios like KKLT \cite{Kachru:2003aw} and Racetrack \cite{Escoda:2003fa}, as they are both non-perturbative stabilized. KKLT is a type IIB flux compactification model in which both fluxes and non-perturbative effects are used to stabilize the moduli supersymmetrically. The AdS minimum is found after moduli stabilization then lifted to dS. The 4d effective field theory is described by:
\begin{equation}
\begin{aligned}
&K=-3\textrm{Log}(-i(\rho-\overline{\rho})),\\
&W=W_{0}+Ae^{i\alpha \rho}.\\
\end{aligned}
\end{equation}
In the limit when $|W_{0}|\ll 1$,  one obtains $\Delta\propto \textrm{Log} R_{AdS}$ \cite{Conlon:2018vov,Choi:2005ge}.

Racetrack still occurs in the context of both type IIB and heterotic string compactification but with only non-perturbative effects. The K\" ahler potential is the same as KKLT while the superpotential are dominated by two different non-perturbative effects:
\begin{equation}
W=Ae^{i\alpha \rho}-Be^{i\beta \rho}.
\end{equation}
Following \cite{Conlon:2020wmc}, one can show  $\Delta\propto \textrm{Log}^{2} R_{AdS}$ in the scale separation limit. 

It is natural that M-theory gives similar results to IIB racetrack because racetrack model corresponds to the one modulus case of M-theory with zero flux background.

\chapter{Wheeler DeWitt State of Charged Black Hole}
\section{Introduction}
In recent decades, the holographic principle has made it possible to study the bulk dynamics of gravitational theories via processes in a dual quantum theory. This dual quantum theory lives on some slice, usually the boundary, of the spacetime. For asymptotically AdS (anti de Sitter) spacetimes, an important part of the holographic toolbox has been the holographic renormalization group flow. This tells us that events further from the boundary correspond to lower energy processes in the dual theory \cite{susskind_holographic_1998,skenderis_lecture_2002}. The location in the bulk can be quantified by the metric function $g_{tt}$, which is equivalent to imposing a canonical transformation from the usual coordinates. Specifically, the AdS boundary lives at $g_{tt} \rightarrow -\infty$ and corresponds to the ultraviolet limit of the dual theory, and moving to lower energy processes corresponds to increasing $g_{tt}$. Reaching a horizon in the spacetime corresponds to $g_{tt} \rightarrow 0$, or equivalently the far infrared of the renormalization group. At this point, the usual interpretation from the dual theory perspective is that there are no more modes to integrate out.

However, it was recently emphasised in \cite{Frenkel:2020ysx} that it is natural to extend the renormalization group flow in the bulk through the horizon, including as far as the singularity. Subsequent work has further explored extending the renormalizaton group flow through the horizon in various setups \cite{Wang:2020nkd,Caceres:2021fuw,Mansoori:2021wxf,Bhattacharya:2021nqj,Das:2021vjf,Caceres:2022smh}, including when the black hole has charge \cite{Hartnoll:2020fhc,Hartnoll:2020rwq,Sword:2021pfm}. The upshot of these developments is that the renormalization group flow might be added to our current list of tools \cite{fidkowski_black_2004-1,hartman_time_2013-1,stanford_complexity_2014-1,brown_holographic_2016} for probing black hole interiors. What remains to be achieved is using the flow to explicitly construct the interior from the exterior.

A hint on how to do this is offered by the observation that the natural language for describing the renormalization group flow is via the Hamilton-Jacobi equation of classical mechanics \cite{deBoer:1999tgo,Faulkner:2010jy,Heemskerk:2010hk}. The Hamilton-Jacobi equation arises in the context of quantum cosmology as the classical limit of the Wheeler DeWitt equation \cite{Halliwell:1987eu,Halliwell:1989myn}. Furthermore, the Wheeler DeWitt equation arises from the canonical quantization of the Einstein-Hilbert action so is a natural tool for studying quantum aspects of spacetime. This point has been emphasised recently in \cite{chowdhury_holography_2022,witten_note_2022}. Therefore, Wheeler DeWitt states constructed from Hamilton-Jacobi functions are a probe of semiclassical aspects of spacetime that naturally connect with the holographic renormalization group flow. They may therefore offer a window into the black hole interior.

This was indeed the approach of \cite{Hartnoll:2022snh}, which studied how the (semiclassical) quantum state of the interior of an AdS-Schwarzschild black hole is prepared by the AdS boundary. In that work, the interior wavefunction was identified with the boundary partition function extended to $g_{tt} > 0$. This was achieved by constructing Wheeler DeWitt states from solutions to the Hamilton-Jacobi equation for a number of different clocks. The most convenient of these solutions was linear in $g_{tt}$, so it was natural to extend from $g_{tt} < 0$ to $g_{tt} >0$ when passing through the horizon. Indeed, in the setup of \cite{Hartnoll:2022snh} $g_{tt}$ was monotonic and a well-defined clock from the boundary to the singularity. When a non-monotonic clock such as volume is used, solutions to the Wheeler-DeWitt equation involve, for example, superpositions of spacetimes close to the singularity and close to the horizon -- both of which have small spatial volumes -- and are hence harder to interpret. Non-monotonic clocks are therefore a less convenient choice for probing the full space time.

What happens, then, if $g_{tt}$ is not monotonic? This arises in spacetimes with more than one horizon. It is natural to ask what choice of clock would be best for applying the framework of \cite{Hartnoll:2022snh} to these spacetimes. One such case is the RN AdS spacetime, where the charge of the black hole introduces a second (Cauchy) horizon in the black hole interior. \begin{figure}[h!]
\centering
\includegraphics[width=.45\textwidth]{Figure1}
\caption{The regions of the Penrose diagram of the RN-AdS spacetime that will play a role in our discussion. The black dotted lines indicate the black hole and cauchy horizons, where $g_{xx}= g_{xx,+}$ and $g_{xx} = g_{xx,-}$ respectively. At the AdS Boundary $g_{xx} = \infty$ and at the singularity $g_{xx} = 0$. The value of $g_{tt}$ at each of these points is indicated to emphasise that it is not monotonic. We solve the Wheeler DeWitt equation in the black hole interior on constant $g_{xx}$ slices, and extend these to the cauchy interior in Section \ref{SectionClocks} and the exterior in Section \ref{SecPartitionFunction}.\label{FigPenrose}}
\end{figure}

Exploring the procedure of \cite{Hartnoll:2022snh} in the (planar) RN AdS spacetime is the aim of this work. As is emphasised in Figure \ref{FigPenrose}, in the RN AdS spacetime $g_{tt}$ is not monotonic from the AdS boundary to the singularity, whereas $g_{xx}$ is. We therefore solve the Wheeler DeWitt equation in the black hole interior, and use $g_{xx}$ as a clock to extend these solutions to the cauchy interior and the exterior. Specifically, we extend our solutions through the left hand horizon, as illustrated in Figure \ref{FigPenrose}, to obtain a positive definite norm. In the cauchy interior, we study whether these states can probe the singularity. In the exterior, we identify the Wheeler DeWitt states with the partition function of the dual theory living on the AdS boundary via a renormalization group flow in $g_{xx}$.

This chapter is organised as follows. In Section \ref{SecRNAdSHJ}, we recover the RN-AdS black hole from Hamilton-Jacobi equation. Using the corresponding Hamilton-Jacobi function, we construct semiclassical Wheeler DeWitt states in Section \ref{SecSemiClassWdW}. We then show three main results. Firstly, in Section \ref{SectionClocks} we use $g_{xx}$ as a clock to show that the variance in $\langle g_{tt} \rangle$ blows up as the singularity is approached. This result indicates that minisuperspace classicality breaks down. Secondly, in Section \ref{SecPartitionFunction} we show that the Wheeler DeWitt states are the continuation of the boundary partition function along $g_{xx}$ slices. That is, the Wheeler DeWitt states are part of the renormalization group flow of a dual theory living on the AdS boundary. The dual theory is specified by an energy and a charge. Finally, in Section \ref{SecThermodynamics} we show that the Wheeler DeWitt states know about the thermodynamics of the black hole horizon. We do so by averaging over states of fixed extrinsic curvature at the black hole horizon in the spirit of \cite{Blacker:2023oan}, to recover the grand canonical thermodynamic potential.
\section{Reissner-Nordstrom-AdS from the Hamilton-Jacobi equation}
\label{SecRNAdSHJ}
As will be elaborated upon in Section \ref{SecSemiClassWdW}, the classical limit of the Wheeler-DeWitt equation is the Hamilton-Jacobi equation for the Einstein-Hilbert action. These classical solutions can then be used to construct semi-classical wavepackets. We will thus begin by recovering the RN-AdS black hole using the Hamilton-Jacobi formulation of gravity, following the approach of \cite{Blacker:2023oan,Hartnoll:2022snh}.

The action for four dimensional gravity with a negative cosmological constant and a Maxwell field is
\begin{align}
    I\left[ g, A \right]= \int d^4x\sqrt{-g}\left( R+ \frac{6}{L^2}\right)
    -\frac{1}{4}\int d^4x \sqrt{-g} F^2 + 2 \int d^3x \sqrt{h}K .
\label{EqAction}
\end{align}

Here, we work in units $16 \pi G_N = 1$. The second term in \eqref{EqAction} is the action of a Maxwell field of strength $F = dA$ for a vector potential $A$, and the final term is the Gibbons-Hawking boundary term. Throughout this paper, we will consider the ans\"{a}tze
    \begin{align}
    ds^2 = - N^2 dr^2 + g_{tt}dt^2 + g_{xx} \left( dx^2 + dy^2 \right), \text{ and } A = \phi_t dt,
\label{EqAnsatz}
\end{align}
where $N$, $g_{tt}$, $g_{xx}$ and $\phi_t$ are functions of $r$ only.

We are interested classically in the black hole interior, as depicted in Figure \ref{FigPenrose}. On the ansatz \eqref{EqAnsatz}, the action is
\begin{align}
    I [N, g_{tt}, g_{xx}, \phi_t] = \int dr \mathcal{L},
\end{align}
where the Lagrangian density is
\begin{align}
    \mathcal{L} = \frac{g_{xx}^2 \left(\left(\partial_r \phi_t \right)^2+(12/L^2) g_{tt} N^2\right)-2 g_{xx} \partial_r g_{tt} \partial_r g_{xx}-g_{tt} \left( \partial_r g_{xx} \right)^2}{2 \sqrt{g_{tt}} g_{xx} N}.
\label{EqEulerLagrange}
\end{align}

From this action we define the momenta $\pi_i = \lbrace \pi_{tt}, \pi_{xx}, \pi_{\phi} \rbrace$ as conjugate to $g_i = \lbrace g_{tt}, g_{xx}, \phi_t \rbrace$, and construct a Hamiltonian in the usual way. $N$ plays the role of a Lagrange multiplier, and imposes the Hamiltonian constraint 
\begin{align}
    -\frac{1}{2} g_{tt} \pi_{tt}^2 +g_{xx} \pi_{tt} \pi_{xx}+ \frac{6}{L^2} g_{xx}^2-\frac{\pi_{\phi}^2}{2} = 0.
\label{EqHamiltonConstraint}
\end{align}

To reconstruct the Hamilton-Jacobi equation, we introduce the Hamilton-Jacobi function $S(g_i)$ such that $\pi_{i} = \partial_{g_i} S(g_i)$. From \eqref{EqHamiltonConstraint} we have
\begin{align}
    g_{xx} \left( \partial_{g_{xx}} S \right) \left( \partial_{g_{tt}} S \right) -\frac{1}{2} \left( \partial_{\phi_t} S \right)^2-\frac{1}{2} g_{tt} \left( \partial_{g_{tt}} S \right)^2+ \frac{6}{L^2} g_{xx}^2 = 0.
\label{EqHJEquation}
\end{align}

The classical equations of motion are obtained from the Hamilton-Jacobi equation by first finding one member of the family of solutions to \eqref{EqHJEquation}. As there are three variables $g_i$, we expect the solution to have three constants of integration. Two of these are non-trivial, and the third leads to an overall shift in in $S$ which does not contribute to the equations of motion and we therefore do not consider. The first such solution with which we will be concerned is
\begin{align}
    S_1 \left( g_{tt}, g_{xx}, \phi_t; k_0, c_0 \right) = -\frac{e^{-k_0} \left(c_0^2+4 g_{xx}^2/L^2\right)}{2 \sqrt{g_{xx}}}+c_0 \phi_t+2 g_{tt} \sqrt{g_{xx}} e^{k_0},
\label{solution}
\end{align}
where $\lbrace k_0, c_0 \rbrace$ are the non-trivial constants of integration. The classical solution (i.e. the solution to Euler-Lagrange equations corresponding to \eqref{EqEulerLagrange}) is obtained by introducing another pair of constants $\lbrace \epsilon, \mu \rbrace$ such that
\begin{align}
    \partial_{k_0} S_1 = \epsilon, \text{ and } \partial_{c_0} S_1 = \mu,
\label{EqClassicalconstraints}
\end{align}
where $k_o$ is related to boundary time normalization, $\epsilon$ is the ADM mass of black hole, $c_o$ is the charge density and $\mu$ denotes chemical potential. Solving this pair of equations we can rewrite two of our coordinates $\lbrace g_{tt}, \phi_t \rbrace$ in terms of the other and the constants of integration, as
\begin{align}
    g_{tt} = \frac{e^{-2 k_0} \left(-c_0^2 L^2 -4 g_{xx}^2+2 \sqrt{g_{xx}} L^2 e^{k_0} \epsilon \right)}{4 L^2 g_{xx}}, \text{ and } \phi_t - \mu = \frac{c_0 e^{-k_0}}{\sqrt{g_{xx}}}.
\label{EqClassicalsol}
\end{align}
The upshot of \eqref{EqClassicalsol} is that we can now recover the RN-AdS solution. Substituting \eqref{EqClassicalsol} into the equation of motion for $N$ from \eqref{EqEulerLagrange}, we obtain
\begin{align}
    N^2 dr^2 = - \frac{dg_{xx}^2}{c_0^2 - 2e^{k_0}\epsilon \sqrt{g_{xx}} + 4 g_{xx}^2/L^2}.
\label{EqNsquaredonshell}
\end{align}
Finally, it will be convenient to define $z = 1/\sqrt{g_{xx}}$, so that when substituting \eqref{EqNsquaredonshell} into \eqref{EqAnsatz} we obtain
\begin{align}
    ds^{2}=\frac{1}{z^2} \left(-f(z)e^{-2k_0} 
dt^2 +\frac{dz^2}{f(z)}+\left( dx^2+dy^2 
 \right)  \right),\text{ and } A=\Phi(z)dt,
\label{EqRNAdS}
\end{align}
where
\begin{align}
    f(z)=\left( \frac{1}{L^2} -\frac{z^3}{2}\epsilon e^{k_0}+\frac{c_0^2}{4}z^4\right),\text{ and } \Phi(z)=\mu+c_0 e^{-k_0}z.
\label{EqFZ}
\end{align}

We recognise \eqref{EqRNAdS} as the RN-AdS solution. The solution has a Cauchy and black hole horizon where $f(z_h) = 0$. The function $f(z) > 0$ in the black hole exterior where $t$ is the timelike direction, changes sign to $f(z) < 0$ after crossing the black hole horizon, and changes sign again to $f(z) > 0$ in the interior of the Cauchy horizon. In the coordinate $z$, we have that (on-shell)
\begin{align}
    g_{tt} = - \frac{f(z)}{z^2} e^{-2k_0}, \text{ and } \phi_t - \mu = c_0e^{-k_0} z.
\label{EqClassicalsolinz}
\end{align}

Therefore, a consequence of having two horizons is that $g_{tt}$ is only monotonic when restricted to either $z > z^*$ or $z < z^*$, where $z^* = 3\epsilon e^{k_0}/2c_0^2$. This means that $g_{tt}$ is not a suitable candidate for a clock to define a relational notion of time. However, $g_{xx}$ is trivially monotonic, and from \eqref{EqClassicalsolinz} we can see that $\phi_t$ is also monotonic, and so are both alternative choices of clock. This point will be discussed further in Section \ref{SectionClocks}.

The radial functions should satisfy the well-known asymptotic properties of charged black hole solutions to Einstein-Maxwell-AdS theory \cite{hartnoll_holographic_2018}. To recognise our solution with these, it is instructive for a moment to consider the re-scaled time coordinate $t' = e^{-k_0} t$ where
\begin{align}
    ds^{2}=\frac{1}{z^2} \left(-f(z)dt'^2+\frac{dz^2}{f(z)}+\left( dx^2+dy^2 
 \right)\right),\text{ and } A=\Phi(z)e^{k_0} dt'. 
\label{EqRNAdSRescaled}
\end{align}

We therefore expect to recover from the electromagnetic potential the boundary chemical potential $\mu_B$ 
\begin{align}
    \lim_{z\rightarrow 0} \Phi(z) e^{k_0} = \mu_B,
\end{align}
and the boundary charge density $\rho$
\begin{align}
    \rho = - \lim_{z\rightarrow 0} \partial_z \Phi(z) e^{k_0}.
\end{align}

In particular, for the classical solution \eqref{EqClassicalsolinz} we recover
\begin{align}
    \mu_B = \mu e^{k_0} \text{ and } \rho = -c_0.
\label{EqMuandRho}
\end{align}

We can compute the associated charge $Q$ via the four-current $J_{\mu}$ and associated four vector $n^{\mu}$ on an $r$ slice
\begin{align}
    Q = \int d^3 x \sqrt{-g} J_a n^a = \rho e^{-k} = -c_0 e^{-k_0}.
\label{EqChargeDefinition}
\end{align}

We can additionally fix the value of $\mu$ (or equivalently $\mu_B$). In the RN-AdS geometry, if one rotates to Euclidean signature only the outer horizon remains and the thermal circle shrinks to zero there \cite{hartnoll_holographic_2018}. Thus, for a Wilson loop of the Maxwell field to be regular, we require $\Phi(z_+) = 0$, where $z_+$ is the outer horizon. This regularity condition in the bulk fixes the chemical potential of the boundary theory to
\begin{align}
    \mu_B = - c_0 z_{+},
\label{EqFixMuB}
\end{align}
where $z_+$ is the value of $z = 1/\sqrt{g_{xx}}$ at the black hole horizon.

So far, we have only considered one member of the family of solutions to \eqref{EqHJEquation}. We could have, of course, constructed a classically equivalent solution where $\lbrace \epsilon, \mu \rbrace$ are the constants of integration. In fact, provided that we do not use both elements of a conjugate pair $\lbrace k_0, \epsilon \rbrace$ or $\lbrace c_0, \mu \rbrace$, we could construct a solution using any combination of these constants. In particular, we could construct the three following solutions
\begin{subequations}
    \begin{align}
        S_{\epsilon,c_0} = & c_0 \phi_t + \frac{F_1 - \epsilon^2 + \epsilon \sqrt{\epsilon^2 - F_1}}{\sqrt{\epsilon^2 - F_1} - \epsilon} - \epsilon \log \frac{\epsilon - \sqrt{\epsilon^2 - F_1}}{4g_{tt}\sqrt{g_{xx}}}, \label{EqSEpsc} \\
        S_{\epsilon,\mu} = & - F_2 - \epsilon \log \frac{4g_{xx}^{3/2}}{L^2(F_2 + \epsilon)}, \label{EqSEpsF} \\
        S_{k_0,\mu} = & \frac{e^{-k_0} \sqrt{g_{xx}}}{2} \left( e^{2k_0} \left( \left( \mu - \phi_t \right)^2 + 4g_{tt} \right) - \frac{4 g_{xx}}{L^2} \right). \label{EqSkf}
    \end{align}
\label{EqOtherHJsolutions}
\end{subequations}
Here, it has been convenient to define
\begin{subequations}
    \begin{align}
        F_1 \left(g_{tt},g_{xx};c\right) = & 4 g_{tt} \left( c^2 + 4 g_{xx}^2/L^2 \right), \\
        F_2 \left(g_{tt},g_{xx},\phi_t; \epsilon, f \right) = & \sqrt{\epsilon^2 - 4 \left( \left( \mu - \phi_t \right)^2 + 4g_{tt} \right) g_{xx}^2/L^2}.
    \end{align}
\end{subequations}

We have labelled each solution by the constants of integration it depends on. We wish to emphasise again here that each of \eqref{EqSEpsc}, \eqref{EqSEpsF}, and \eqref{EqSkf} is classically equivalent to $S_1$, in the sense that they lead to the same general solutions to the equations of motion. We have chosen to first focus on $S_1$ because in Section \ref{SecSemiClassWdW} it can be used to construct exact solutions to the Wheeler DeWitt equation. However, the relevance of each of \eqref{EqOtherHJsolutions} will become apparent as we proceed.

\section{Semiclassical Wheeler-DeWitt States}
\label{SecSemiClassWdW}
\subsection{From classical to semiclassical}
\label{SecClasstoSemiClass}

We can use these classical results to construct semiclassical quantum states, following a procedure we briefly review below. For a more detailed introduction to the procedure which follows, the interested reader should consult \cite{Halliwell:1987eu,Halliwell:1989myn}.

For the gravitational degrees of freedom, we have an action of the form
\begin{align}
    I[g] = \int dr \mathcal{L} = \int dr N \left[ \frac{1}{2N^2} G_{ab} \dot{g}^a \dot{g}^b - V(q) \right],
\end{align}
where $G_{ab}(g)$ are the minisuperspace metric and $V(g)$ an effective potential respectively, and are functions of some metric functions $g$. In the quantum theory, the Hamiltonian constraint is promoted to the Wheeler-DeWitt equation for a wavefunction $\Psi(g)$
\begin{align}
    \left( - \frac{1}{2} \nabla^2 + V(g) \right) \Psi \left( g \right) = 0,
\end{align}
where $\nabla^2$ is the Laplacian on the inverse DeWitt metric $ \sqrt{-G} \nabla^2 = \partial_{g^a} \left( \sqrt{-G} G^{ab} \partial_{g^b} \right)$. Strictly, there is an ordering ambiguity in quantizing the Hamiltonian constraint, which we discuss momentarily. If we restore units and consider solutions of the form $\Psi = \exp \left( i S(q)/\hbar \right)$, we find 
\begin{subequations}
    \begin{align}
        \mathcal{O} \left( \hbar^0 \right): & \left( \frac{1}{2} \left( \nabla S \right)^2 + V \right) \Psi = 0, \\
        \mathcal{O} \left( \hbar^1 \right): & \nabla^2 S \Psi = 0.
    \end{align}
\end{subequations}

Observe that the $\mathcal{O} \left( \hbar^0 \right)$ constraint is nothing more than the Hamilton-Jacobi equation. Moreover, the ordering ambiguity in defining the Laplacian only arises at order $\mathcal{O} \left( \hbar^1 \right)$. Thus, if we are only concerned with the leading order semiclassical physics, we can form a basis of states by exponentiating the Hamilton-Jacobi function.

In addition to the gravitational degrees of freedom, we also have electromagnetic terms, which amount to an additional term proportional to $\left( \partial_r \phi_t \right)^2$ in our action. In this case the above reasoning still holds, with the $\mathcal{O} \left( \hbar^0 \right)$ equation of motion still being the Hamilton-Jacobi equation and the $\mathcal{O} \left( \hbar^1 \right)$ equation containing an additional $\partial^2_{\phi_t} S$ term.
\subsection{Constructing states of the charged black hole}
\label{SecConstructingStates}
With the above discussion in mind, in our setup the Wheeler-DeWitt equation becomes
\begin{align}
    \frac{\partial}{\partial g_{tt}} \left( g_{tt} \frac{\partial \Psi}{\partial g_{tt}} - 2 g_{xx} \frac{\partial \Psi}{\partial g_{xx}} \right) + \frac{\partial^2 \Psi}{\partial \phi_t^2} + \frac{12}{L^2} g_{xx}^2 \Psi = 0.
\label{EqWdWEquation}
\end{align}

From section \ref{SecClasstoSemiClass}, we know that $e^{\pm i S}$ form a basis of semiclassical solutions to \eqref{EqWdWEquation} if $S$ is a solution to \eqref{EqHJEquation}. In fact, for $S_1$ \eqref{solution}, the solution
\begin{align}
    \Psi \left( g_{tt},g_{xx},\phi_t; k,c \right) = e^{iS_1\left( g_{tt},g_{xx},\phi_t; k,c \right)},
\label{EqBasisS1}
\end{align}
is an exact solution to the Wheeler DeWitt equation \eqref{EqWdWEquation}. Note that for the other Hamilton-Jacobi functions in \eqref{EqOtherHJsolutions}, $e^{\pm i S}$ are only solutions to leading order. We will thus first consider the basis of exact solutions to \eqref{EqWdWEquation} constructed from $S_1$. From the basis \eqref{EqBasisS1}, we construct a general solution
\begin{align}
    \Psi \left( g_{tt}, g_{xx}, \phi_t \right) = \int_{-\infty}^{\infty} \frac{dk}{2\pi} \int_{-\infty}^{\infty} \frac{dc}{2\pi} \beta \left( k,c \right) e^{iS_1\left( g_{tt},g_{xx},\phi_t; k,c \right)}.
\label{EqS1Solution}
\end{align}

Here $\beta \left( k, c \right)$ is an arbitrary function. Note here that we only include $e^{+iS_1}$ solutions in our superposition (as opposed to also including the $e^{-iS_1}$ solutions). This choice of sign determines the sign of momentum operators, and is necessary to define a positive definite norm when we pick a clock in Section \ref{SectionClocks}. Physically, this corresponds to evolving the wavefunction through only the left horizons (see Figure \ref{FigPenrose}). A superposition of the $e^{\pm i S_1}$ solutions would lead to decoherence in the wavefunction, a phenomenon we are not concerned with in this work but has been widely discussed in quantum cosmology (see \cite{halliwell_decoherence_1989,padmanabhan_decoherence_1989}).

As in \cite{Blacker:2023oan,Hartnoll:2022snh}, we can obtain another set of semiclassical solutions by fourier transforming to the basis corresponding to the classical constants $\lbrace \epsilon, \mu \rbrace$. We define
\begin{align}
    \beta \left( k, c \right) = \int \int \frac{d\epsilon}{\sqrt{2\pi}} \frac{d\mu}{\sqrt{2\pi}} \alpha \left( \epsilon, \mu \right) e^{- i \epsilon k } e^{-i \mu c},
\end{align}
where $\alpha \left( \epsilon, \mu \right)$ is an arbitrary function. The solution \eqref{EqS1Solution} then becomes
\begin{align}
    \begin{split}
        \Psi \left( g_{tt}, g_{xx}, \phi_t \right) = \int \frac{d\epsilon}{2\pi} \int\frac{d\mu}{2\pi} \alpha \left( \epsilon, \mu \right) & \frac{2\left( - g_{xx}/L^2 \right)^{1/2 + i\epsilon/2}}{\sqrt{2\pi}} \left( \frac{\left( \mu + \phi_t \right)^2 + 4 g_{tt}}{4} \right)^{-\frac{1}{2} \left( \frac{1}{2} + i \epsilon \right)} \\
        & \times K_{\frac{1}{2}+i \epsilon} \left( \frac{2 \sqrt{(\mu+\phi_t)^2 + 4g_{tt}} g_{xx}}{L} \right),
    \end{split}
\label{EqFourierSolution}
\end{align}
where $K_{1/2+i\epsilon}$ is a modified bessel function of the second kind. To make use of this Fourier transform in the semiclassical limit, we want to relate it to a solution of the Hamilton-Jacobi equation. To do so, we evaluate the Fourier transform using a stationary phase approximation and obtain the sum of two solutions
\begin{align}
    \Psi \left( g_{tt}, g_{xx}, \phi_t \right) = \int \frac{d\epsilon}{2\pi} \int\frac{d\mu}{2\pi} \alpha \left( \epsilon, f \right) \left[ \psi_{+} \left( g_{tt}, g_{xx}, \phi_t; \epsilon, \mu \right) + \psi_{-} \left( g_{tt}, g_{xx}, \phi_t; \epsilon, \mu \right) \right].
\label{EqPsiinEF}
\end{align}
These solutions are
\begin{align}
    \psi_{\pm} \left( g_{tt}, g_{xx}, \phi_t; \epsilon, \mu \right) = \frac{2g_{xx}}{L \sqrt{ F_2^2 \pm \epsilon F_2 }} \exp \left \lbrace i \left( \pm S_{\pm \epsilon,\mu} + \frac{\pi}{2} \right) \right \rbrace,
\end{align}
where $S_{\pm \epsilon,f}$ is the solution to the Hamilton-Jacobi equation introduced in \eqref{EqSEpsF}. As expected, taking the limit of no charge (i.e. the AdS-Schwarzschild solution), one finds that $\psi_{\pm}$ recover the solutions obtained in \cite{Hartnoll:2022snh}. As in \cite{Hartnoll:2022snh,Blacker:2023oan}, we only consider $e^{iS_{+\epsilon,\mu}}$ to ensure we have positive norm solutions in the interior. That is, in the semiclassical regime, we will find it useful to consider states
\begin{align}
     \Psi_{+} \left( g_{tt}, g_{xx}, \phi_t \right) = \int \frac{d\epsilon}{2\pi} \int\frac{d\mu}{2\pi} \alpha \left( \epsilon, \mu \right) e^{iS_{+ \epsilon, \mu}}.
\end{align}
\subsection{Gaussian wavepackets}

We are yet to consider a particular form for $\beta \left( k, c \right)$. As in \cite{Blacker:2023oan,Hartnoll:2022snh}, it is natural to consider Gaussian wavepackets, because these are strongly supported on the classical solution. In this work, we consider a Gaussian wavepacket
\begin{align}
    \begin{split}
        \beta \left( k, c \right) = & N_{\beta} \exp \left \lbrace -i \epsilon_0 \left( k - k_0 \right) - \frac{\Delta_k^2}{2} \left( k - k_0 \right)^2 \right \rbrace \\
        & \times \exp \left \lbrace -i \mu_0 \left( c - c_0 \right) - \frac{\Delta_c^2}{2} \left( c - c_0 \right)^2 \right \rbrace,
    \end{split}
\label{EqGaussianWavepacket}
\end{align}
where $N_{\beta} = \Delta_c \Delta_k/4\pi$. These wavepackets are strongly peaked around $\lbrace k = k_0, c = c_0 \rbrace$ if $\Delta_c, \Delta_k \gg 1$. In that case, we can then perform the $k$ and $c$ integrals in \eqref{EqS1Solution} by a 2D stationary phase approximation. The wavefunction is strongly peaked on values of the metric function where
\begin{align}
    \left. \frac{\partial S_1}{\partial k} \right|_{k=k_0} = \epsilon_0, \text{ and } \left. \frac{\partial S_1}{\partial c}\right|_{c=c_0} = \mu_0.
\end{align}

That is, the wavefunction is strongly supported on the classical solution \eqref{EqClassicalconstraints} with $\epsilon = \epsilon_0$ and $\mu = \mu_0$. As when computing the fourier transform, we have a contribution from two stationary points (corresponding to $S_{\pm \epsilon, \mu}$). As noted above, we only consider the branch corresponding to $S_{+\epsilon,\mu}$ to ensure positivity of the norm. Therefore, to leading semiclassical order the wavefunction \eqref{EqS1Solution} becomes
\begin{align}
    \begin{split}
        \Psi \left( g_{tt}, z, \phi_t; k_0, c_0, \epsilon_0, \mu_0 \right) = & \delta \left( g_{tt} = - \frac{f(z)}{z^2} e^{-2k_0} \right) \delta \left( \phi_t - \mu_0 = c_0e^{-k_0} z \right) \\
        & \times \exp \left \lbrace - i \left( \epsilon_0 - 4 e^{-k_0} /L^2 z^3  \right) + i c_0 \mu_0 \right \rbrace,  
    \end{split}
\label{EqClassicalWavefunction}
\end{align}
where
\begin{align}
    f(z) = \frac{1}{L^2} - \frac{z^3}{2} \epsilon_0 e^{k_0} + \frac{c_0^2}{4}z^4.
\end{align}

Here, as in \cite{Blacker:2023oan}, we have taken the limit in which the strongly peaked Gaussian that arises after integration becomes a delta function, localising our state on the classical solution. We have used $z=1/\sqrt{g_{xx}}$ to make explicit the connection with \eqref{EqClassicalsol}. The phase in \eqref{EqClassicalWavefunction} is equal to the onshell action, up to an additional $c_0\mu_0$ term which could be incorporated into our choice of $\beta(c)$ without affecting the classical dynamics.

One could complete a similar computation in the basis $\lbrace \epsilon, \mu \rbrace$, by evaluating \eqref{EqFourierSolution} on the Fourier transform of \eqref{EqGaussianWavepacket}
\begin{align}
    \alpha \left( \epsilon, f \right) = N_{\alpha} \exp \left \lbrace i k_0 \epsilon - \frac{\left( \epsilon - \epsilon_0 \right)^2}{2 \Delta_k^2} \right \rbrace \exp \left \lbrace i c_0 \mu - \frac{\left( \mu - \mu_0 \right)^2}{2 \Delta_c^2} \right \rbrace,
\end{align}
where $N_{\alpha} = 1/\left( 4\pi \Delta_c \Delta_k \right)$. To obtain the classical solution \eqref{EqClassicalWavefunction}, we require these $\alpha \left( \epsilon,\mu \right)$ wavepackets to be strongly peaked on $\lbrace \epsilon = \epsilon_0, \mu = \mu_0 \rbrace$, and thus that $\epsilon_0 \gg \Delta_k$ and $\mu_0 \gg \Delta_c$. That is, the semiclassical regime is obtained by requiring 
\begin{align}
    \epsilon_0 \gg \Delta_k \gg 1, \text{ and } \mu_0 \gg \Delta_c \gg 1.
\label{EqSemiclassicalConstraint}
\end{align}

Physically, this is because if the wavepacket is to strongly localised on any coordinate (i.e. $k$ or $c$) the variance in its conjugate momentum (i.e. $\epsilon$ or $\mu$) will become large. The constraint \eqref{EqSemiclassicalConstraint} is therefore necessary to preserve the classicality of the minisuperspace, as will be discussed in Section \ref{SectionClocks}.
\section{Clocks and Expectation Values}\label{1}
\label{SectionClocks}
The Wheeler DeWitt equation is timeless in the sense that it contains no explicit time parameter, but provides a relation between metric functions on any slicing of spacetime. To specify our slicing is to treat some coordinate (or some combination of coordinates) as constant on each slice. Moving between slices is equivalent to evolving that coordinate. That is, that coordinate acts as a clock, and the Wheeler DeWitt equation can then be used to compute probabilities which are conditional upon the choice of clock. These probabilities can be used to compute expectation values to validate the stability of our semiclassical theory.

A number of different possible clocks where considered for the AdS-Schwarzschild black hole in \cite{Hartnoll:2022snh}. The $g_{tt}$ clock turned out to be convenient because a Hamilton-Jacobi function linear in $g_{tt}$ was obtained. In that work, evolving $g_{tt}$ from $-\infty$ to $\infty$ was equivalent to evolving the wavefunction from the AdS boundary ($g_{tt} = - \infty$), through the horizon ($g_{tt} = 0$), and to the singularity ($g_{tt} = \infty$). However, as discussed in Section \ref{SecRNAdSHJ}, when the black hole is imbued with a charge, $g_{tt}$ is no longer monotonic from the boundary to the singularity. Therefore, although \eqref{solution} is also linear in $g_{tt}$, $g_{tt}$ may not be the most convenient choice of clock. The fact that $g_{tt}$ is not monotonic corresponds to the presence of two horizons, which also occurs in the dS-Schwarzschild black hole studied in \cite{Blacker:2023oan}. In that work, constant $R$ slices were used to move from the cosmological to the black hole horizons.

The analogous choice here, as advertised in Section \ref{SecRNAdSHJ}, is to choose $g_{xx}$ as our clock. This is because $g_{xx}$ is monotonic from the AdS Boundary ($g_{xx} \rightarrow \infty$) to the singularity ($g_{xx} \rightarrow 0$). It is straightforward to show that $g_{xx}$ is a null direction in the minisuperspace we are considering here. This means to take $g_{xx}$ as a clock is to define the conserved norm via a limiting sequence of spacelike slices. That is, we can compute conditional probabilities for any given $g_{xx}$, by defining a conserved norm as in \cite{PhysRev.160.1113}. For $g_{xx}$, this norm is
\begin{align}
    \left| \Psi \right|_{g_{xx}}^2 = - \frac{i}{2} \int d\phi_t \int dg_{tt} \left( \Psi^* \partial_{g_{tt}} \Psi - \Psi \partial_{g_{tt}} \Psi^* \right). 
\label{EqGXXNorm}
\end{align}

One can show using the Wheeler DeWitt equation \eqref{EqWdWEquation} that $\partial_{g_{xx}} \left| \Psi \right|_{g_{xx}}^2 = 0$, provided $\Psi$ decays at large and small $\lbrace \phi_t, g_{tt} \rbrace$ on each $g_{xx}$ slice. That is, the norm is conserved under evolution between different slices of fixed $g_{xx}$. The norm \eqref{EqGXXNorm} is defined to be positive definite on an $e^{+iS_1}$ state, and evaluating it on a general semiclassical state of the form \eqref{EqS1Solution}, we find that
\begin{align}
    \left| \Psi \right|_{g_{xx}}^2 = \int \frac{dkdc}{(2\pi)^2} \left| \beta(k,c) \right|^2.
\end{align}

One could alternatively compute the norm on the Fourier transformed state \eqref{EqFourierSolution}, however the $\lbrace k,c \rbrace$ basis is a more computationally convenient choice.

An upshot of having defined a norm is that we can now compute expectation values, which can be used to examine the stability of minisuperspace classicality in the semi-classical regime. The simplest, non-trivial expectation values are those of the unfixed metric functions $\langle \phi_t \rangle$ and $\langle g_{tt} \rangle$, which we evaluate on \eqref{EqS1Solution} to obtain respectively
\begin{align}
    \begin{split}
        \langle \phi_t \rangle = & - \frac{i}{2} \int d\phi_t \phi_t \int dg_{tt} \left( \Psi^* \partial_{g_{tt}} \Psi - \Psi \partial_{g_{tt}} \Psi^* \right) \\
    = & \frac{i}{2} \int \frac{dk dc}{(2\pi)^2} \left[ \beta^* \partial_c \beta - \beta \partial_c \beta^* \right] + \int \frac{dkdc}{(2\pi)^2} \frac{e^{-k}c}{\sqrt{g_{xx}}} \left| \beta \right|^2,
    \end{split}
\label{EqGAexpection}
\end{align}
and
\begin{align}
    \begin{split}
        \langle g_{tt} \rangle = & - \frac{i}{2} \int d\phi_t \int dg_{tt} g_{tt} \left( \Psi^* \partial_{g_{tt}} \Psi - \Psi \partial_{g_{tt}} \Psi^* \right) \\
        = & -\int \frac{dkdc}{(2\pi)^2} \left( \frac{i}{2} \left[ \beta \partial_k \beta^* - \beta^* \partial_k \beta \right] \frac{e^{-k}}{2\sqrt{g_{xx}}} + \left| \beta \right|^2 \frac{c^2+4g_{xx}^2/L^2}{4g_{xx}} e^{-2k} \right).
    \end{split}
\label{EqGTTexpectation}
\end{align}
Note here we have left the $\lbrace k,c \rbrace$ dependence implicit. If we evaluate \eqref{EqGAexpection} on the Gaussian wavepacket \eqref{EqGaussianWavepacket}, we obtain 
\begin{align}
    \langle \phi_t \rangle - \mu_0 = \frac{c_0 e^{-k_0}}{\sqrt{g_{xx}}} \exp \left( \frac{1}{4\Delta_k^2} \right) = \frac{c_0 e^{-k_0}}{\sqrt{g_{xx}}} + O \left( 1 / \Delta_k^2 \right),
\label{EqGAsemiclassical}
\end{align}
where expanding in $\Delta_k \gg 1$ we obtain the classical solution \eqref{EqClassicalsol} as the leading contribution. Here, we have included the full quantum correction, to make explicit that the $\mu_0$ contribution is independent of the quantum fluctuations. Thus, the freedom to redefine $\phi_t \rightarrow \phi_t - \mu_0$ is preserved. We can similarly evaluate \eqref{EqGTTexpectation} on the Gaussian wavepacket \eqref{EqGaussianWavepacket}, expanding in $\Delta_k, \Delta_c \gg 1$ to obtain
\begin{align}
    \langle g_{tt} \rangle = -\frac{e^{-2k_0}}{g_{xx}} \left( g_{xx}^2 - \frac{1}{2} e^{k_0} \sqrt{g_{xx}} \epsilon_0  + \frac{c_0^2}{4} \right) + O \left( \frac{1}{\Delta_c^2}, \frac{1}{\Delta_k^2} \right), 
\end{align}
again obtaining to leading order the classical solution \eqref{EqClassicalsol}.

We can also compute the expectation values of momenta. For example, if $g_{xx}$ is our clock defining our notion of time, $\pi_{xx} = - i \partial_{g_{xx}}$ generates `time translations' and therefore defines the `Hamiltonian' for this clock. We compute the expectation value on the state \eqref{EqS1Solution} to obtain
\begin{align}
    \begin{split}
        \langle \pi_{xx} \rangle = & -\frac{1}{2} \int d\phi_t \int dg_{tt} \left( \Psi^* \partial_{g_{xx}} \left( \partial_{g_{tt}} \Psi \right) - \partial_{g_{xx}} \Psi \partial_{g_{tt}} \Psi^* \right) \\
        = & \int d\phi_t \int dg_{tt} \left[ \frac{g_{tt}}{2 g_{xx}} \left| \frac{\partial \Psi}{\partial g_{tt}} \right|^2 + \frac{1}{2g_{xx}} \left| \frac{\partial \Psi}{\partial \phi_t} \right|^2 - \frac{6}{L^2} g_{xx} \left| \Psi \right|^2 \right] \\
        = & \int \frac{dk dc}{(2\pi)^2} \left( \frac{i}{4g_{xx}} \left( \beta^* \partial_k \beta - \beta^* \partial_k \beta' \right) - 4 e^{-k} \sqrt{g_{xx}}/L^2 \left| \beta \right|^2 \right).
    \end{split}
\label{EqPiXXGeneral}
\end{align}
To obtain the second line we used the Wheeler DeWitt equation \eqref{EqWdWEquation}. Evaluating \eqref{EqPiXXGeneral} on the Guassian wavepacket we obtain
\begin{align}
    \langle \pi_{xx} \rangle = \frac{\epsilon_0}{2g_{xx}} - 4 e^{-k_0} \sqrt{g_{xx}}/L^2 \exp \left( \frac{1}{\Delta_k^2} \right) = \frac{\epsilon_0}{2g_{xx}} - 4 e^{-k_0} \sqrt{g_{xx}}/L^2 + O \left( 1/ \Delta_k^2 \right),
\end{align}
where expanding in $\Delta_k \gg 1$ we obtain the classical solution as the leading contribution, as from \eqref{solution} $\pi_{xx} = \partial S_1/\partial g_{xx} = \epsilon_0/(2g_{xx}) - 4 e^{-k_0} \sqrt{g_{xx}}/L^2$. For $\epsilon_0 >0$, $\langle \pi_{xx} \rangle$ is monotonic. That the expansion does not break down anywhere indicates we have picked a good clock for extending our Wheeler DeWitt solutions through the spacetime.

Recall that the motivation for introducing these expectation values was to study the stability of minisuperspace classicality as the singularity is approached as $g_{xx} \rightarrow 0$. A natural quantity to test that is the fluctuations of the metric function $g_{tt}$. Computing the variance of $g_{tt}$ we obtain
\begin{align}
    \langle g_{tt}^2 \rangle - \langle g_{tt} \rangle^2 = \frac{\Delta_k^2}{e^{2k_0}8g_{xx}} + O \left( \frac{1}{\Delta_k^2}, \frac{1}{\Delta_c^2} \right),
\label{EqgttVariance}
\end{align}
where we have used $\Delta_k, \Delta_c \gg 1$. One may recognise the denominator as (to leading order) $\langle \pi_{tt} \rangle^2 = 4e^{2k_0}g_{xx}$, and indeed it is straightforward recover the uncertainty relation
\begin{align}
    \left( \langle \pi_{tt}^2 \rangle - \langle \pi_{tt} \rangle^2 \right) \left( \langle g_{tt}^2 \rangle - \langle g_{tt} \rangle^2 \right) = \frac{1}{4}  + O \left( \frac{1}{\Delta_k^2}, \frac{1}{\Delta_c^2} \right).
\end{align}
To interpret \eqref{EqgttVariance} near the singularity it will be useful to remove the $g_{xx}$ dependence. The leading order dependence of $g_{tt}$ on $g_{xx}$ depends on whether or not charge is present. In the neutral ($c_0 = 0$) case
\begin{align}
    \left. \lim_{g_{xx}\rightarrow 0} \langle g_{tt} \rangle \right|_{c_0 = 0} =  \frac{\epsilon_0 e^{-k_0}}{2\sqrt{g_{xx}}} + O \left( g_{xx}, \frac{1}{\Delta_k^2}, \frac{1}{\Delta_c^2} \right),
\label{Eqgttlimitneutral}
\end{align}
so the singularity is spacelike as $\langle g_{tt} \rangle > 0$. Using \eqref{Eqgttlimitneutral} we find that in the neutral case
\begin{align}
    \left. \lim_{g_{xx}\rightarrow 0} \langle g_{tt}^2 \rangle - \langle g_{tt} \rangle^2 \right|_{c_0 = 0} = \frac{\Delta_k^2}{2 \epsilon_0^2} \langle g_{tt} \rangle^2 + + O \left( \frac{1}{\Delta_k^2}, \frac{1}{\Delta_c^2} \right).
\label{EqVarianceNoCharge}
\end{align}
Recall that in the semiclassical regime \eqref{EqSemiclassicalConstraint} $\epsilon_0 \gg \Delta_k$, so the variance is small compared to the expectation value and the Gaussian wavepacket is able to probe the singularity, as noted in \cite{Hartnoll:2022snh}.\\\\
When the black hole has a charge (i.e. $c_0 \not = 0$), the leading order dependence of $g_{tt}$ on $g_{xx}$ becomes
\begin{align}
    \lim_{g_{xx}\rightarrow 0} \langle g_{tt} \rangle = - \frac{c_0^2 e^{-2k_0}}{4g_{xx}} + O \left( \frac{1}{\sqrt{g_{xx}}}, \frac{1}{\Delta_k^2}, \frac{1}{\Delta_c^2} \right).
\label{Eqgttlimitcharge}
\end{align}

As is to be expected from the classical solution \eqref{EqClassicalsolinz}, we see that in the charged case the singularity is timelike as $\langle g_{tt} \rangle < 0$. Note that by taking the $g_{xx} \rightarrow \infty$ limit, we have continued our wavepacket through the inner Cauchy horizon. This inner horizon is believed to be unstable \cite{papadodimas_simple_2020,kehle_blowup_2021}, however this instability does not arise in any of the quantities we have computed in our minisuperspace description. Hence, it appears safe to continue the wavepacket through the Cauchy horizon.\\\\
Using \eqref{Eqgttlimitcharge}, we compute
\begin{align}
    \lim_{g_{xx}\rightarrow 0} \langle g_{tt}^2 \rangle - \langle g_{tt} \rangle^2 = -\frac{\Delta_k^2}{2 c_0^2} \langle g_{tt} \rangle + O \left( \frac{1}{\Delta_k^2}, \frac{1}{\Delta_c^2} \right).
\label{Eqgxxvariancecharge}
\end{align}

In \eqref{Eqgxxvariancecharge} the sign of the variance is still positive, as from \eqref{Eqgttlimitcharge} we have that $\langle g_{tt} \rangle < 0$ in the limit $g_{xx} \rightarrow 0$. The significance of \eqref{Eqgxxvariancecharge} is that the variance in $g_{tt}$ is linear in $g_{tt}$ rather than quadratic when $\Delta_k, \Delta_c \gg 1$. If we divide by $\langle g_{tt} \rangle^2$, we see that for any choice of $\Delta_k$ and $c_0$ the variance will go to zero at the singularity. That means the quantum fluctuations are not enough to prevent this direction from collapsing as $g_{xx} \rightarrow 0$, and the classicality of the minisuperspace approximation breaks down. Therefore, the Wheeler DeWitt state is not able to resolve the singularity.

One possible interpretation is that this result is because the singularity is timelike. With that in mind, it is useful to recall the strong cosmic censorship conjecture: that from physically reasonable generic initial conditions, only spacelike or null singularities can form classically \cite{penrose_notitle_1979}. There has been some evidence \cite{balasubramanian_holographic_2020} via holography that quantum effects enforce strong cosmic censorship in charged AdS black holes. The result \eqref{Eqgxxvariancecharge} may therefore indicate the quantum corrections accounted for by our semiclassical picture are similarly enforcing strong cosmic censorship. It would be interesting to perform computations analogous to \eqref{Eqgxxvariancecharge} for Wheeler DeWitt states in other spacetimes which contain a timelike singularity, to see if a similar result is obtained.

Throughout these calculations, we have considered $g_{xx}$ as a monotonic function from which to define a clock. As noted in Section \ref{SecRNAdSHJ}, $\phi_t \propto 1/\sqrt{g_{xx}}$ is also monotonic from the AdS boundary ($\phi_t \rightarrow 0$) to the singularity ($\phi_t \rightarrow \infty$). Indeed, one may have instead chosen to define $\phi_t$ as a clock, with an associated conserved norm
\begin{align}
    \left| \Psi \right|_{\phi_t}^2 = \frac{i}{2} \int dg_{xx} \int dg_{tt} \frac{1}{g_{xx}} \left( \Psi^* \partial_{\phi_t} \Psi - \Psi \partial_{\phi_t} \Psi^* \right).
\end{align}
In that case, one can show that on states of the form \eqref{EqS1Solution}
\begin{align}
    \left| \Psi \right|_{\phi_t}^2 = \int \frac{dk dc}{(2\pi)^2} \left| \beta (k,c) \right|^2.
\end{align}
That is, the $g_{xx}$ and $\phi_t$ clocks lead to the same norm. This is to be expected, given the inverse proportionality of the functions. However, the integrals used to compute expectation values in the $\phi_t$ norm are not as easy to evaluate as when using the $g_{xx}$ clock, hence our preference for that choice in this work.
\section{The Boundary Partition Function}
\label{SecPartitionFunction}

From the AdS/CFT correspondence, we expect our theory of the RN-AdS bulk to be dual to some quantum theory living on the AdS boundary. In particular, the semiclassical bulk wavefunctions \eqref{EqS1Solution} constructed from \eqref{solution} should be related to some partition function living on the AdS boundary $g_{xx} \rightarrow \infty$. As noted in the introduction, the Hamilton-Jacobi function $S$ should control the holographic renormalization group flow, with the arguments of $S$ playing the role of coupling in the dual quantum theory. Here, we make that link explicit by studying the energy flow of the dual theory, and show that by running the coupling $g_{xx}$ we move our partition function into the bulk and obtain a Wheeler DeWitt state. Equivalently, by evolving the Wheeler DeWitt state along $g_{xx}$ slices to the boundary, we recover the partition function.

The holographic framework we consider is analogous to \cite{Hartnoll:2022snh}. Specifically, the standard holographic relation tells us that the partition function of the Lorentzian quantum field theory (QFT) on the AdS boundary is
\begin{align}
    Z_{\text{QFT}} \left[ \gamma, A_{\gamma} \right] = \int \mathcal{D}g \mathcal{D}A e^{iI[g,A]+i S_{\text{ct}}[\gamma]},
\label{EqDictionary}
\end{align}
where the path integrals are over bulk metrics $g$ that are asymptotically AdS with conformal boundary metric $\gamma$, and over vector potentials $A$. $A_{\gamma}$ is the value of the Maxwell field at the boundary. In our minisuperspace approximation, $\gamma_{tt}= g_{tt}$ and $\gamma_{xx} = \gamma_{yy} = g_{xx}$. Here, $I[g,A]$ is the bulk action and $S_{\text{ct}} [\gamma]$ a boundary counterterm action to cancel the divergence in the action at the boundary \cite{Balasubramanian:1999re,Emparan:1999pm}
\begin{align}
    S_{\text{ct}}[\gamma] = 4 \int dx^3 \frac{\sqrt{-\gamma}}{\Delta t \left( \Delta x \right)^2 L} = \frac{4 \sqrt{-g_{tt}}g_{xx}}{L}.
\label{Eqcounterterms}
\end{align}

A difference between our setup and that of \cite{Hartnoll:2022snh} is, as motivated by Section \ref{SectionClocks}, we will use $g_{xx}$ as our clock. The AdS boundary lives at $g_{xx} \rightarrow \infty$, and we will consider the holographic renormalization group flow that evolves as a function of $g_{xx}$. That is, we start with a QFT for a finite $g_{xx}$, and then obtain a conformal field theory (CFT) in the limit that $g_{xx} \rightarrow \infty$. As in \cite{Hartnoll:2022snh}, there is no need to remove a conformal factor from the metric, as conformal invariance at the AdS boundary constrains observables here.

If we wish to make more explicit our definition of the boundary partition function in terms of a trace over states in the boundary theory, we should first understand how imbuing the black hole with charge affects the dual QFT. The leading order contribution to \eqref{EqDictionary} is the classical solution, upon which $I[g,A]$ is precisely the Hamilton-Jacobi function. In Section \ref{SecRNAdSHJ}, we found several classically equivalent Hamilton-Jacobi solutions, which are each described by a different choice of constants from each pair $\lbrace k_0, \epsilon \rbrace$ and $\lbrace c_0 , \mu \rbrace$. To determine which of these will be most convenient for describing our dual theory, we will take advantage of the AdS/CFT correspondence and draw some intuition from the bulk. Classically, the RN-AdS bulk is described by its mass and charge, which would suggest that $\epsilon$ and $c_0$ should specify the dual quantum theory. We therefore want their conjugate variables $k_0$ and $\mu$ to be variables in our dual partition function, so \eqref{EqSkf} is a useful form of the Hamilton-Jacobi function for us to use. Thus, with the addition of the counterterms \eqref{Eqcounterterms}, we obtain 
\begin{align}
    \begin{split}
        \log Z_{\text{QFT}} \left[ g_{tt}, g_{xx}, \phi_t; k_0,\mu_0 \right] = & - 2i e^{-k_0} \sqrt{g_{xx}}\left( e^{k_0} \sqrt{-g_{tt}}- \sqrt{g_{xx}}/L \right)^2 \\
        & - \frac{i}{2} e^{k_0} \left( \mu_0^2 - \phi_t^2 \right) \sqrt{g_{xx}}.
\label{EqPartionfunctionf0}
    \end{split}
\end{align}

We see explicitly that this partition funciton depends not only on the gauge field $\phi_t$ and the boundary data $g_{tt}$ and $g_{xx}$, but additionally the classical parameters $k_0$ and $\mu_0$. It obeys the scale invariance
\begin{align}
    \log Z_{\text{QFT}} [g_{tt}e^{-\frac{4}{3}k},g_{xx}e^{\frac{2}{3}k},\phi_te^{-\frac{2}{3}k};\mu_0 e^{-\frac{2}{3}k},k_0+k] = \log Z_{\text{QFT}} [g_{tt},g_{xx},\phi_t;\mu_0,k_0].
\label{EqZInvariancef0}
\end{align}

The energy density of the dual field theory is defined in the usual way, by the momentum conjugate to $g_{tt}$
\begin{align}
    \sqrt{-\gamma}\langle T^t_t\rangle_{\text{QFT}}=-2i\gamma_{tt}\frac{\partial \log Z_{\text{QFT}}}{\partial \gamma_{tt}}=4\sqrt{-g_{tt}g_{xx}}(\sqrt{g_{xx}}/L-e^{k_0}\sqrt{-g_{tt}}).
\end{align}

To obtain the CFT limit, we eliminate $g_{tt}$ using \eqref{EqClassicalsol}, and expand in $g_{xx} \rightarrow \infty$ to obtain
\begin{align}
    \lim_{g_{xx} \rightarrow \infty}\sqrt{-\gamma}\langle T^t_t\rangle_{\text{QFT}} = \epsilon.
\label{EqEnergyDensityBoundary}
\end{align}

We recover, as in \cite{Hartnoll:2022snh}, that the energy density of the black hole $\epsilon$ is the energy density of the dual theory. Indeed, from \cite{Hartnoll:2022snh} we expect the conjugate variable $k_0$ to play a role anaologous to time in the CFT limit. To learn more about our dual theory, it will be instructive to study the dependence of $\log Z_{\text{QFT}}$ on $k_0$. In particular, observe that
\begin{align}
    \begin{split}
        - i \frac{\partial \log Z_{\text{QFT}}}{\partial k_0} = &  \frac{1}{2} \sqrt{g_{xx}} \left(e^{k_{0}} \left(-f_{0}^2+\phi_t^2+4 g_{tt}\right)+4 g_{xx} e^{-k_0}/L^2 \right) \\
        = & \epsilon + c_0 \mu_0 \\
        = & \epsilon - \mu_B Q.
    \end{split}
\label{EqDLogDk0boudnary}
\end{align}

In the second line we have used the classical solutions for $g_{tt}$ and $\phi_t$. We recognise from \eqref{EqEnergyDensityBoundary} that $\epsilon$ is the energy density in the $g_{xx} \rightarrow \infty$. In the third line we replaced $c_0 \mu_0$ with the boundary charge and chemical potential from \eqref{EqMuandRho} and \eqref{EqChargeDefinition}. From \eqref{EqDLogDk0boudnary} it appears that our dual theory is characterised by an energy and a charge. That is, beyond the classical regime we expect
\begin{align}
    - i \frac{\partial \log Z_{\text{QFT}}}{\partial k_0} = H_{\text{QFT}} - \mu_B Q_{\text{QFT}}.
\label{EqDlokDk0semi}
\end{align}
Here, $H_{QFT} = \epsilon$ and $Q_{\text{QFT}} = Q$ in the classical limit. We therefore expect
\begin{align}
    Z_{\text{QFT}} [g_{tt},g_{xx},\phi_t;k_0,f_0] = \text{Tr} \left( e^{i k_0 \left( H_{QFT}[g_{tt},g_{xx},\phi_t]- \mu_B Q_{QFT} \right)} \right),
\label{EqClaim}
\end{align}
to be a suitable partition function of the boundary theory.

All of our discussion so far has only considered the classical limit of \eqref{EqDictionary}.

To further justify our claim \eqref{EqClaim}, we should consider \eqref{EqDictionary} beyond the classical regime. Recall that the path integral over $e^{iI[g,A]}$ is usually associated with the wavefunction of the gravititational bulk \cite{PhysRevD.28.2960}. Therefore, when the boundary counterterms are subtracted out, in the semiclassical regime the partition function \eqref{EqDictionary} should obey the Wheeler DeWitt equation \eqref{EqWdWEquation}. In particular, this means that in the semiclassical regime we can approximate the partition function as 
\begin{align}
    Z_{\text{QFT}} [g_{tt},g_{xx},\phi_t; \beta] = e^{4i\sqrt{-g_{tt}}g_{xx}/L} \Psi \left(g_{tt},g_{xx},\phi_t; \beta \right),
\label{EqPartitionfunctionSemiclass1}
\end{align}
where $\Psi$ is a solution to \eqref{EqWdWEquation}. Note that if we rearrange \eqref{EqPartitionfunctionSemiclass1} for $\Psi$, that is
\begin{align}
    \Psi \left(g_{tt},g_{xx},\phi_t; \beta \right) = e^{-4i\sqrt{-g_{tt}}g_{xx}/L} Z_{\text{QFT}} [g_{tt},g_{xx},\phi_t; \beta],
\label{EqZdefromed}
\end{align}
we could reinterpret the Wheeler DeWitt solution as a deformation of a QFT partition function. Or indeed, treating $g_{xx}$ as a clock, as a deformation of a QFT living on the AdS boundary as we evolve away from the $g_{xx} \rightarrow \infty$ limit. This interpretation could be linked to studies of $T^2$ deformations of the boundary theory in AdS/CFT \cite{Belin:2020oib,LeFloch:2019rut,McGough:2016lol}, or indeed the explicit specification of a bulk state in terms of a boundary theory in \cite{Araujo-Regado:2022gvw}.

To explicitly compute \eqref{EqPartitionfunctionSemiclass1}, we need to use a Wheeler DeWitt state $\Psi$. In Section \ref{SecConstructingStates}, we discussed how to construct a Wheeler DeWitt state \eqref{EqS1Solution} from a basis of solutions \eqref{EqBasisS1} weighted by some function $\beta(k,c)$. What we want to do here is slightly different; as discussed earlier, \eqref{EqSkf} is a more natural representation Hamilton-Jacobi function to identify with our partition function. We will therefore perform a Fourier transform to the $c$ integral in \eqref{EqS1Solution} to recast our states in the semiclassical basis $e^{iS_{k,\mu}}$. To do so, we will assume that our wavepackets are separable. That is, $\beta(k,c) = \beta_k(k) \beta_c(c)$ and $\alpha (\epsilon,\mu) = \alpha_{\epsilon} (\epsilon) \alpha_{\mu} (\mu)$. Performing the fourier transform by stationary phase, we recover
\begin{align}
    \Psi \left( g_{tt}, g_{xx}, \phi_t \right) \sim \int dk d\mu \beta_k(k) \alpha_{\mu}(\mu) e^{iS_{g_{tt},g_{xx},\phi_t;k,\mu}}.
\label{EqinKF}
\end{align}
Here, $\sim$ means up to some prefactor. Therefore, in the semiclassical regime, from \eqref{EqPartitionfunctionSemiclass1} and \eqref{EqinKF} our claim \eqref{EqClaim} is that the partition function becomes
\begin{align}
    Z_{\text{QFT}} \left[ g_{tt}, g_{xx}, \phi_t; \beta \right] \sim \int dk df \beta_k(k) \alpha_{\mu}(\mu) \text{Tr} \left( e^{i k \left( H_{\text{QFT}}- \mu \rho_{\text{QFT}} \right)} \right).
\label{EqClaim2}
\end{align}
Note that here we use the charge density $\rho_{\text{QFT}} = e^{k_0} Q_{\text{QFT}}$. This is because it is more natural to build our wavepackets in $f$ than $\mu$, given the arguments of the Hamilton-Jacobi function. For the Gaussian wavepacket \eqref{EqGaussianWavepacket}, \eqref{EqClaim2} becomes
\begin{align}
    Z_{\text{QFT}} \sim \text{Tr}_{\epsilon,c}\left(e^{-(H_{\text{QFT}}-\epsilon_0)^2/(2\Delta_k^2)-\Delta_c^2(c-c_0)^2/2}e^{i k_0 \left( H_{QFT}- \mu_0 \rho_{QFT} \right)}\right). 
\label{EqClaim3}
\end{align}
We can now assess the validity of the claim \eqref{EqClaim} by explicitly evaluating \eqref{EqPartitionfunctionSemiclass1} and comparing it to \eqref{EqClaim3}. We evaluate \eqref{EqPartitionfunctionSemiclass1} by making a stationary phase approximation and taking the $g_{xx} \rightarrow \infty$ limit to obtain
\begin{align}
    \begin{split}
        \lim_{g_{xx} \rightarrow \infty} Z_{\text{QFT}} \sim \int d\epsilon dc & \exp \left \lbrace - \frac{\Delta_c^2}{2} \left( c- c_0 \right)^2 - \frac{\left( \epsilon - \epsilon_0\right)^2}{2\Delta_k^2} + i \epsilon \log \left( e^{k_0} L \frac{\sqrt{-g_{tt}}}{\sqrt{g_{xx}}} \right) \right \rbrace \\
        & \times \exp \left \lbrace i \left( c \phi_t + \frac{\epsilon^2 L}{8 \sqrt{-g_{tt}}g_{xx}} + \frac{c^2 L \sqrt{-g_{tt}}}{2g_{xx}}  \right)  \right \rbrace,
    \end{split}
\label{EqPartitionfunctionSemiclass3}
\end{align}
up to a prefactor which is not important here. The integrals in \eqref{EqPartitionfunctionSemiclass3} are identified with the trace in \eqref{EqClaim3}, and the Gaussian terms are identical. The third term in the first line of \eqref{EqPartitionfunctionSemiclass3} is due to the smearing of $k$ around $k_0$ and changes the sign of the $\epsilon^2$ term, as noted in the neutral case \cite{Hartnoll:2022snh}. The term linear in $\phi_t$ arises from an analogous blurring of $\mu$ around $\mu_0$, and also changes the sign of the $c^2$ term. As in the neutral case, the $\epsilon^2$ term is the scaling for the density of states in a $2+1$ dimensional CFT. To interpret the $c^2$ term, we compute the current associated with the electromagnetic potential $A$. Classically, in the $dt$ direction this is
\begin{align}
    J_t = \frac{\pi_{\phi}^2}{2g_{xx}^2},
\end{align}
so the $c^2$ term in \eqref{EqPartitionfunctionSemiclass3} is identified with the energy density $\sqrt{-\gamma} \langle J_t \rangle$ of the Maxwell field. Finally, to interpret the term linear in $\phi_t$, we compute 
\begin{align}
    -i \lim_{g_{xx}\rightarrow \infty} \frac{\partial \log Z_{\text{QFT}}}{\partial \phi_t} = \langle c \rangle = - \langle \rho \rangle.
\label{EqSemiZCharge}
\end{align}

Here, we have an expectation value because we are taking a trace of the boundary theory ensemble. The implication of \eqref{EqSemiZCharge} is that the bulk gauge field is dual to the charge density in the bulk theory. This recovers the standard result of the holographic dictionary \cite{hartnoll_holographic_2018}. We have therefore identified each term in \eqref{EqPartitionfunctionSemiclass3} with those in \eqref{EqClaim3}, validating our proposition for the dual theory ensemble.

To summarise, what we have done here is demonstrated at a classical and semiclassical level that Wheeler DeWitt states of the RN-AdS interior can be constructed from the holographic renormalization group flow of $g_{xx}$ of a partition function that lives on the AdS Boundary. The gaussian wavepacket localises to a fixed energy and charge, suggesting that the partition function of the dual theory is specified by those quantities.
\section{Black hole thermodynamics}
\label{SecThermodynamics}
Having identified the dual theory as being specified by a fixed energy and charge, we expect to be able to construct a similar thermodynamic picture for the bulk theory. In particular, we expect that at the black hole horizon, Wheeler DeWitt states should know about the black hole thermodynamics. That quantum states of a gravitational bulk should know about horizon thermodynamics has long been appreciated \cite{gibbons_action_1977}; what we do here is explicitly recover these thermodynamics from our semiclassical states.

To recover the bulk thermodynamics, we follow a similar procedure to \cite{Blacker:2023oan}. There, it was observed that it is natural to average over a Wheeler DeWitt state such as \eqref{EqClassicalWavefunction} by integrating over a metric function. This averaging has the practical benefit of removing a delta function. Physically, this averaging corresponds to fixing the gravitational gauge redundancy represented by the Wheeler DeWitt equation. In particular, in \cite{Blacker:2023oan} the redundancy was fixed by fixing the trace of the extrinsic curvature $K$, motivated by the observations of \cite{york_role_1972,witten_note_2021,witten_note_2022}. This is achieved by fourier transforming the wavefunction to a basis of extrinsic curvatures rather than metric functions, and then averaging. The trace $K$ is proportional to $\pi_{v}$, the momentum conjugate to the volume $v = \sqrt{-g_{tt}}g_{xx}$. The fourier transform is therefore taking us from wavefunctions $\Psi \left( v, [h], \phi_t \right)$ to $\tilde{\Psi} \left( \pi_v, [h], \phi_t \right)$, where $[h] = g_{tt}/g_{xx}$ is the conformal class of the induced metric. The transform is implemented by adding a term proportional to $i v \pi_v$ in the exponent of \eqref{EqClassicalWavefunction}. As in \cite{Blacker:2023oan}, we will add the Gibbons-Hawking boundary term
\begin{align}
    2 \int d^3x \sqrt{h} K = g_{tt} \pi_{tt} + g_{xx} \pi_{xx},
\end{align}
and consider the partial integration
\begin{align}
    \tilde{\Psi} \left( g_{xx}, \phi_t \right) = \int dg_{tt} e^{-i \left( g_{tt} \langle \pi_{tt} \rangle + g_{xx} \langle \pi_{xx} \rangle \right)} \Psi \left( g_{tt}, g_{xx}, \phi_t \right).
\label{EqPartialIntegrationStep1}
\end{align}

We integrate over $g_{tt}$ because $g_{xx}$ is monotonic from the AdS boundary to the singularity. We will in particular be interested in evaluating \eqref{EqPartialIntegrationStep1} at the black hole horizon $g_{xx} (z_h)$. It is worth noting that in this case, \eqref{EqPartialIntegrationStep1} is equivalent to averaging over a state in a gravitational theory with
\begin{align}
    I_{\text{GR}}\left[ g, A \right]=\int d^4x\sqrt{-g}(R+6/L^2)+2\int d^3x\sqrt{h} \left( K_I - K_h \right),
\end{align}
where $K_I$ and $K_h$ are the trace of the extrinsic curvature at the infinite boundary and the horizon respectively. This is in the spirit of the original Gibbons-Hawking procedure \cite{gibbons_action_1977}.

Integrating over $g_{tt}$ only removes one of the delta functions from \eqref{EqClassicalWavefunction}. That is, we still have a remaining distribution in $\phi_t$. We proceed by integrating out $\phi_t$ to obtain an effective field theory on a $g_{xx}$ slice to obtain
\begin{align}
    \tilde{\Psi} \left( g_{xx} \right) = \int d\phi_t \int dg_{tt} e^{-i \left( g_{tt} \langle \pi_{tt} \rangle + g_{xx} \langle \pi_{xx} \rangle \right)} \Psi \left( g_{tt}, g_{xx}, \phi_t \right).
\label{EqPartialIntegrationStep2}
\end{align}

Note that we are interested in $\tilde{\Psi}$ at the black hole horizon $g_{xx} = g_{xx,+}$, where $\phi_t = 0$ as per \eqref{EqFixMuB}. We therefore don't lose any information by performing the integral over $\phi_t$. Evaluating \eqref{EqPartialIntegrationStep2} on the solution \eqref{EqClassicalWavefunction} we obtain
\begin{align}
    \begin{split}
        \tilde{\Psi} \left( g_{xx,+} \right) = & \exp \left \lbrace i \left[ e^{-k_0} S T_h - \left( \epsilon + c_0 \mu_0 \right) \right] \right \rbrace \\
         = & \exp \left \lbrace i \left[ e^{-k_0} S_h T_h - \epsilon + \mu_B Q \right] \right \rbrace.
    \end{split}
\label{EqAveragedPsi}
\end{align}
Here we defined the Bekenstein-Hawking entropy
\begin{align}
    S_h = A_h , \text{ where } A_h = \frac{4\pi}{z_h^2} = 4\pi g_{xx,+},  
\end{align}
and the black hole temperature
\begin{align}
    T_h = \frac{1}{4\pi} \left| \frac{\partial f(z)}{\partial z} \right|_{z=z_+} = \frac{1}{4\pi} e^{2k_0} \left| 2\sqrt{g_{xx}} \frac{\partial g_{tt}}{\partial g_{xx}} \right|_{g_{xx}=g_{xx,+}}.
\end{align}

Note that, as in Section \eqref{SecPartitionFunction}, we identified $c_0 \mu_0 = - \mu_B Q$ as before. To proceed further, firstly observe from the metric \eqref{EqFZ} that $\epsilon = M e^{-k_0}$ where $M$ is the mass of the black hole. Additionally, whilst the charge $Q$ in the bulk and boundary is the same, the chemical potential is redshifted when moving from the boundary to the horizon. The redshift is explicitly computed following \cite{bardeen_four_1973}
\begin{align}
    \mu_B = \sqrt{-n_a n^a} \mu_h = e^{-k_0} \mu_h,
\end{align}
where $n^{\mu}$ is the four-vector introduced above \eqref{EqChargeDefinition}. Our averaged wavefunction \eqref{EqAveragedPsi} can therefore be expressed as
\begin{align}
    \Psi \left( g_{xx,+} \right) = & \exp \left \lbrace i e^{-k_0}  \left[ S_h T_h - M + \mu_h Q \right] \right \rbrace = \exp \left \lbrace - i e^{-k_0} W \right \rbrace,
\label{EqAveragedPsi2}
\end{align}
where $W$ is the usual grand canonical thermodynamic potential. In the spirit of \cite{Blacker:2023oan,Blacker:2024rje}, if we set $e^{-k_0} = -i/T_h$, the argument of the exponential is the usual thermodynamic argument for Euclidean quantum gravity \cite{gibbons_action_1977}. Setting $e^{-k_0} = -i/T_h$ is equivalent to performing a Wick rotation of the metric \eqref{EqFZ} to Euclidean signature. In Euclidean siganture, the radial coordinate runs from the AdS boundary to the black hole horizon, where the Euclidean time circle shrinks to zero \cite{hartnoll_holographic_2018}.

It is worth comparing the result \eqref{EqAveragedPsi2} to that in \cite{Blacker:2023oan}, in which a de Sitter-Schwarzschild black hole was considered. There, the averaged and fourier transformed wavepacket recovered exactly the horizon entropy $\tilde{\Psi} = e^{S_h}$ under an analogous Wick rotation. This is because for a closed universe, such as an asymptotically de Sitter one, $W = -TS$, and so $-WT^{-1}$ is the entropy $S$ exactly. We make this comparison to emphasise that it is the thermodynamic potential $W$ which the Wheeler DeWitt state knows about.
\chapter{Conclusion}
In Chapter 2, we demonstrated that operators involving both an axion and a modulus exhibit negative anomalous dimensions. However, an interesting twist comes in when we consider double trace operators formed by two distinct axions; in this case, the anomalous dimensions turn out to be positive.

In a prior study by \cite{Conlon:2020wmc}, they found a consistent negativity in the anomalous dimensions for double-trace operators in the moduli sector. This negativity was intriguingly linked to certain signs in the low-energy effective field theory, crucial for satisfying swampland considerations. For instance, a change in sign implies a divergent axion decay constant in the large volume limit.

Hence, a compelling question remains: what determines the sign of these anomalous dimensions? Understanding this aspect holds significance in unraveling the broader implications within the realm of swampland considerations.

In Chapter 3, we study two different models of M-theory moduli stabilization. First, for the flux-stabilized M-Theory vacuum, we have shown that the spectrum of
the dual $CFT_3$ is chacterized by a set of integer conformal dimensions. The presence of
integer conformal dimensions dual to the moduli and axions in M-Theory flux vacuum is
quite intriguing. Similar results also appear in DGKT type IIA string compactification
scenarios \cite{Conlon:2021cjk, Apers:2022tfm}. Therefore we compared them and found the flux-stabilized M-Theory
vacuum resembles the complex structure sector of DGKT, which is not surprising because
their superpotential has similar forms, as predicted in \cite{DeWolfe:2005uu}. The validity of the results heavily
rely on the existence of large $c_2$.
Second, for non-perturbatively stabilized M-Theory vacuum, we prove that
there is one heavy operator with $\Delta \propto \text{Log}R^2_
{AdS}$ , the rest are the same as flux-stabilized
M-Theory vacuum. Then we compare the results with other scenarios like KKLT \cite{Kachru:2003aw} and
racetrack \cite{Escoda:2003fa}. It is different with KKLT because there are two different non-perturbative
effects in the superpotential, while KKLT only has one non-perturbative term. Racetrack is
quite similar because it is the one modulus version of the non-perturbative effects stabilized
M-Theory vacuum.
An extremely interesting question is: What is the origin of these integer conformal
dimensions? Currently we have no explanation for this, but we believe there must be some
deeper structures behind this. Another interesting open question is: Can we prove the
existence of $G_2$ manifold with large $c_2$? It would certainly be of interest to explore these
questions in the future.

In Chapter 4, we have studied Wheeler DeWitt solutions as a probe of the interior of charged black holes. In particular, we have applied the framework of \cite{Hartnoll:2022snh} to the RN-AdS black hole, constructing Wheeler DeWitt states in the black hole interior and extending them to the cauchy interior and the exterior. We found that the close to the singularity, quantum fluctuations become significant and the state is unable to probe the singularity. Moving to the exterior and considering the AdS/CFT correspondence, we then found that our Wheeler DeWitt states were part of the renormalization group flow of a quantum theory living on the AdS boundary. The partition function of this dual theory was specified by an energy and charge. We then returned to the bulk description to recover this thermodynamic picture there. In particular, by applying the averaging procedure of \cite{Blacker:2023oan} and making an appropriate Wick rotation, we recovered the grand canonical thermodynamic potential at the black hole horizon.

Having developed this framework, it is natural to consider what other rich dynamics of charged black holes it will allow us to explore. For example, it was shown in \cite{Hartnoll:2020rwq} that the inner Cauchy horizon of the RN-AdS black hole vanishes upon coupling a scalar field to the bulk theory. Whilst directly coupling a scalar field to our theory has not yet yielded a tractable Hamilton-Jacobi solution, a simpler starting point might be to treat the scalar field as a perturbation of the Hamilton-Jacobi solution as in \cite{halliwell_origin_1985}. Such a treatment would introduce sources in the dual theory living on the AdS boundary and thus affect the renormalization group flow.

Coupling to a scalar field could equally be used to introduce inhomogeneities around our minisuperspace approximation, which would be a first step to going beyond the semiclassical regime \cite{pimentel_inflationary_2014,Ning:2023ybc}. Going beyond minisuperspace may be useful to better understand the result \eqref{Eqgxxvariancecharge}. That is, it would be interesting to learn whether petrurbations around minisuperspace make it possible for a Wheeler DeWitt state to resolve the timelike singularity. 
\begin{appendices}
\chapter{Appendix}
\section{Fibred Calabi-Yau Effective Lagrangians}
\subsection{One Fibre Modulus}\label{app:1m}
From equation(\ref{eq1}), the Lagrangian for the fibred one-modulus Calabi Yau scenario is :
\begin{equation}
L_{kin}=\frac{3}{8\tau_{1}^{2}}\partial_{u}\tau_{1}\partial^{u}\tau_{1}-\frac{1}{2\tau_{1}\mathcal{V}}\partial_{u}\tau_{1}\partial^{u}\mathcal{V}+\frac{1}{2\mathcal{V}^{2}}\partial_{u}\mathcal{V}\partial^{u}\mathcal{V}
+\frac{1}{4\tau_{1}^{2}}\partial_{u}a_{1}\partial^{u}a_{1}+\frac{\alpha^{2}\tau_{1}}{2\mathcal{V}^{2}}\partial_{u}a_{2}\partial^{u}a_{2}.
\end{equation}
To put this in a canonically normalised form, we use the ansatz:
\begin{equation}
\begin{aligned}
&\textrm{ln}\tau_{1}=a\Phi_{1}+b\Phi_{2},\\
&\textrm{ln}\mathcal{V}=c\Phi_{1}.\\
\end{aligned}
\end{equation}
We subsitute this ansatz into the Lagrangian:
\begin{equation}
\begin{aligned}
\frac{3}{8\tau_{1}^{2}}\partial_{u}\tau_{1}\partial^{u}\tau_{1}&=\frac{3}{8}(\partial_{u}\textrm{ln}\tau_{1})^{2}
=\frac{3}{8}(a\partial_{u}\Phi_{1}+b\partial_{u}\Phi_{2})^{2}\\
&=\frac{3}{8}a^{2}(\partial_{u}\Phi_{1})^{2}+\frac{3}{8}b^{2}(\partial_{u}\Phi_{2})^{2}+\frac{3}{4}ab\partial_{u}\Phi_{1}\partial^{u}\Phi_{2},\\
\end{aligned}
\end{equation}
\begin{equation}
\begin{aligned}
-\frac{1}{2\tau_{1}\mathcal{V}}\partial_{u}\tau_{1}\partial^{u}\mathcal{V}&=
-\frac{1}{2}\partial_{u}\textrm{ln}\tau_{1}\partial^{u}\textrm{ln}\mathcal{V}
=-\frac{1}{2}(a\partial_{u}\Phi_{1}+b\partial_{u}\Phi_{2})c\partial^{u}\Phi_{1}\\
&=-\frac{1}{2}ac(\partial_{u}\Phi_{1})^{2}-\frac{1}{2}bc\partial_{u}\Phi_{2}\partial^{u}\Phi_{1},\\
\end{aligned}
\end{equation}
\begin{equation}
\begin{aligned}
\frac{1}{2\mathcal{V}^{2}}\partial_{u}\mathcal{V}\partial^{u}\mathcal{V}=
\frac{1}{2}(\partial_{u}\textrm{ln}\mathcal{V})^{2}
=\frac{1}{2}c^{2}(\partial_{u}\Phi_{1})^{2}.
\end{aligned}
\end{equation}
Putting them together, the kinetic term reads:
\begin{equation}
L_{kin}=\left(\frac{3}{8}a^{2}-\frac{1}{2}ac+\frac{1}{2}c^{2}\right)(\partial_{u}\Phi_{1})^{2}
+\left(\frac{3}{4}ab-\frac{1}{2}bc\right)\partial_{u}\Phi_{1}\partial^{u}\Phi_{2}
+\frac{3}{8}b^{2}(\partial_{u}\Phi_{2})^{2}
\end{equation}
Canonical normalisation then requires satisfying the following conditions:
\begin{equation}
\frac{3}{8}a^{2}-\frac{1}{2}ac+\frac{1}{2}c^{2}=\frac{1}{2},
\end{equation}
\begin{equation}
\frac{3}{4}ab-\frac{1}{2}bc=0,
\end{equation}
\begin{equation}
\frac{3}{8}b^{2}=\frac{1}{2}.
\end{equation}
A solution is:
\begin{equation}
\begin{aligned}
&a=\sqrt{\frac{2}{3}},\\
&b=\frac{2}{\sqrt{3}},\\
&c=\sqrt{\frac{3}{2}}.\\
\end{aligned}
\end{equation}
After substituting this into the kinetic term, we have:
\begin{equation}
\frac{3}{8\tau_{1}^{2}}\partial_{u}\tau_{1}\partial^{u}\tau_{1}
=\frac{1}{4}(\partial_{u}\Phi_{1})^{2}+\frac{1}{2}(\partial_{u}\Phi_{2})^{2}+\frac{\sqrt{2}}{2}\partial_{u}\Phi_{1}\partial^{u}\Phi_{2},
\end{equation}
\begin{equation}
-\frac{1}{2\tau_{1}\mathcal{V}}\partial_{u}\tau_{1}\partial^{u}\mathcal{V}
=-\frac{1}{2}(\partial_{u}\Phi_{1})^{2}-\frac{\sqrt{2}}{2}\partial_{u}\Phi_{1}\partial^{u}\Phi_{2},
\end{equation}
\begin{equation}
\frac{1}{2\mathcal{V}^{2}}\partial_{u}\mathcal{V}\partial^{u}\mathcal{V}
=\frac{1}{2}(\partial_{u}\ln \mathcal{V})^{2}
=\frac{3}{4}(\partial_{u}\Phi_{1})^{2}.
\end{equation}
Adding these together, we have both diagonal kinetic terms and diagonal mass terms
\begin{equation}
\frac{1}{2}(\partial_{u}\Phi_{1})^{2}+\frac{1}{2}(\partial_{u}\Phi_{2})^{2}+V(\Phi).
\end{equation}

Where the canonical  fields $\Phi_{1}, \Phi_{2}$ relate to the original fields as:
\begin{equation}
\mathcal{V}=e^{\sqrt{\frac{3}{2}}\Phi_{1}},
\end{equation}
\begin{equation}
\tau_{1}=e^{\sqrt{\frac{2}{3}}(\Phi_{1}+2\Phi_{2})}.
\end{equation}

\subsection{Two Fibre Moduli}\label{app:2m}
We again start from the kinetic terms in the Lagrangian:
\begin{equation}
\begin{aligned}
L_{kin}=&\frac{1}{4\tau_{1}^{2}}\partial^{u}\tau_{1}\partial_{u}\tau_{1}+\frac{1}{4\tau_{1}^{2}}\partial^{u}a_{1}\partial_{u}a_{1}
+\frac{1}{4\tau_{2}^{2}}\partial^{u}\tau_{2}\partial_{u}\tau_{2}+\\
&\frac{1}{4\tau_{2}^{2}}\partial^{u}a_{2}\partial_{u}a_{2}+
\frac{1}{4\tau_{3}^{2}}\partial^{u}\tau_{3}\partial_{u}\tau_{3}+\frac{1}{4\tau_{3}^{2}}\partial^{u}a_{3}\partial_{u}a_{3}\\
\end{aligned}
\end{equation}

To keep the mass term diagonal, we must have:
\begin{equation}
\ln \mathcal{V}=a\Phi_{1},
\end{equation}
and we also anticipate:
\begin{equation}
\begin{aligned}
&\textrm{ln}\tau_{1}=b\Phi_{1}+c\Phi_{2}+d\Phi_{3},\\
&\textrm{ln}\tau_{2}=b\Phi_{1}+e\Phi_{2}+f\Phi_{3}.\\
\end{aligned}
\end{equation}

Similarly to the previous subsection, canonical normalisation leads to the following conditions:

\begin{equation}
\begin{aligned}
&a^{2}+\frac{3}{2}b^{2}-2ab=\frac{1}{2},\\
&c^{2}+e^{2}+ce=1,\\
&d^{2}+f^{2}+df=1,\\
&a(c+e)=\frac{3}{2}b(c+e),\\
&a(d+f)=\frac{3}{2}b(d+f),\\
&c(d+\frac{1}{2}f)+e(\frac{1}{2}d+f)=0.\\
\end{aligned}
\end{equation}

A solution is:
\begin{equation}
\begin{aligned}
&a=\sqrt{\frac{3}{2}},\\
&b=\sqrt{\frac{2}{3}},\\
&c=e=\sqrt{\frac{1}{3}},\\
&d=-f=1.\\
\end{aligned}
\end{equation}
The Lagrangian then reads:
\begin{equation}
\begin{aligned}
L_{eff}=&\frac{1}{2}(\partial_{u}\Phi_{1})^{2}+\frac{1}{2}(\partial_{u}\Phi_{2})^{2}+\frac{1}{2}(\partial_{u}\Phi_{3})^{2}
+\frac{1}{4}e^{-2\sqrt{\frac{2}{3}}\Phi_{1}-2\Phi_{3}-2\sqrt{\frac{1}{3}}\Phi_{2}}(\partial_{u}a_{1})^{2}\\
&+\frac{1}{4}e^{-2\sqrt{\frac{2}{3}}\Phi_{1}+2\Phi_{3}-2\sqrt{\frac{1}{3}}\Phi_{2}}(\partial_{u}a_{2})^{2}
+\frac{1}{4}e^{\left(2\sqrt{\frac{2}{3}}-2\sqrt{\frac{3}{2}}\right)\Phi_{1}+4\sqrt{\frac{1}{3}}\Phi_{2}}(\partial_{u}a_{3})^{2}-V_{lvs}.\\
\end{aligned}
\end{equation}

Other solutions are also possible but are physically equivalent under field redefinitions.
\section{General Expressions For Mass Matrices}\label{a1}
In this section we derive Eq (\ref{eq:3}) and Eq (\ref{eq:4}).
The general expressions of moduli and axions Hessians are:
\begin{equation}
\begin{aligned}
\label{eq:mass matrix1}
H_{ab}^{V}=\partial_{b}\partial_{a}\textrm{V}&=K_{ab}\textrm{V}-3e^{K}K_{b}\partial_{a}|W|^{2}-3e^{K}\partial_{b}\partial_{a}|W|^{2}\\
&+e^{K}K^{i\overline{j}}(\partial_{a}D_{i}W)(\partial_{b}D_{\overline{j}}\overline{W})+e^{K}K^{i\overline{j}}(\partial_{b}D_{i}W)(\partial_{a}D_{\overline{j}}\overline{W}),\\
\end{aligned}
\end{equation}
where $\partial_{a}$,$\partial_{b}$ means $\partial_{s_{a}}$, $\partial_{s_{b}}$ and $\partial_{i}$,$\partial_{j}$ means $\partial_{z_{i}}$, $\partial_{z_{j}}$.
\begin{equation}
\begin{aligned}
\label{eq:mass matrix2}
H_{ab}^{A}=\partial_{b}\partial_{a}V&=-3e^{K}\partial_{b}\partial_{a}|W|^{2}+e^{K}K^{i\overline{j}}(\partial_{a}D_{i}W)(\partial_{b}D_{\overline{j}}\overline{W})\\
&+e^{K}K^{i\overline{j}}(\partial_{b}D_{i}W)(\partial_{a}D_{\overline{j}}\overline{W}),\\
\end{aligned}
\end{equation}
where $\partial_{a}$,$\partial_{b}$ means $\partial_{t_{a}}$, $\partial_{t_{b}}$ and $\partial_{i}$,$\partial_{j}$ means $\partial_{z_{i}}$, $\partial_{z_{j}}$.

For the first line of (\ref{eq:mass matrix1}), the SUSY condition implies that:
\begin{equation}
\label{app1}
\partial_{a}|W|^{2}=2\partial_{a}W\overline{W}=-K_{a}|W|^{2},
\end{equation}
and
\begin{equation}
\begin{aligned}
\label{app2}
&\partial_{b}\partial_{a}|W|^{2}=2\partial_{b}\partial_{a}W\overline{W}+2\partial_{a}W\partial_{b}\overline{W}\\
&=(2\frac{W_{ab}}{W}+\frac{1}{2}K_{a}K_{b})|W|^{2}.\\
\end{aligned}
\end{equation}
Following \cite{Apers:2022tfm}, for the second line of (\ref{eq:mass matrix1}), we have:
\begin{equation}
\begin{aligned}
&e^{K}K^{i\overline{j}}(\partial_{a}D_{i}W)(\partial_{b}D_{\overline{j}}\overline{W})\\
&=e^{K}(K^{i\overline{j}}W_{ia}\overline{W}_{\overline{j}b}+4W_{ab}\overline{W}-\frac{1}{2}K^{i\overline{j}}W_{ia}K_{\overline{j}}K_{b}\overline{W}\\
&-\frac{1}{2}K^{i\overline{j}}W_{ib}K_{\overline{j}}K_{a}\overline{W}+(K_{ab}+\frac{3}{4}K_{a}K_{b})|W|^{2}).\\
\end{aligned}
\end{equation}
The fact that $K=K(z_{i}+\overline{z}_{i})$ and the superpotential is holomorphic implies:
\begin{equation}
\partial_{z_{i}}K=\frac{\partial_{s_{i}}K}{2i}, \partial_{\overline{z}_{i}}K=-\frac{\partial_{s_{i}}K}{2i},
\end{equation}
and
\begin{equation}
\partial_{z_{i}}W=\frac{1}{i}\partial_{s_{i}}W, \partial_{\overline{z}_{i}}\overline{W}=-\frac{1}{i}\partial_{s_{i}}\overline{W},
\end{equation}
which results in:
\begin{equation}
\begin{aligned}
\label{app3}
&e^{K}K^{i\overline{j}}(\partial_{a}D_{i}W)(\partial_{b}D_{\overline{j}}\overline{W})\\
&=e^{K}(4K^{ij}W_{ia}\overline{W}_{jb}+4W_{ab}\overline{W}+s_{i}W_{ia}K_{b}\overline{W}\\
&+s_{j}\overline{W}_{jb}K_{a}W+(K_{ab}+\frac{3}{4}K_{a}K_{b})|W|^{2}).\\
\end{aligned}
\end{equation}
Combining (\ref{app1}),(\ref{app2}) and (\ref{app3}) leads to:
\begin{equation}
\begin{aligned}
M^{V}_{ab}=\frac{\textrm{V}_{ab}}{e^{K}|W|^{2}}&=-K_{ab}+3K_{a}K_{b}+2\frac{W_{ab}}{W}+8K^{ij}\frac{W_{ia}W_{jb}}{|W|^{2}}\\
&+2s_{i}K_{b}\frac{W_{ia}}{W}+2s_{j}K_{a}\frac{W_{jb}}{W},\\
\end{aligned}
\end{equation}
The axion Hessians (\ref{eq:mass matrix2}) can be obtained similarly, note that:
\begin{equation}
\label{m2}
\frac{\partial W}{\partial t_{a}}=\frac{\partial W}{\partial z_{a}}=\frac{1}{i}\frac{\partial W}{\partial s_{a}},
\end{equation}
and
\begin{equation}
\begin{aligned}
\label{m23}
\partial_{a}D_{i}W&=\partial_{a}(\partial_{z_{i}}W+K_{i}W)\\
&=\partial_{z_{i}}\partial_{z_{a}}W+K_{z_{i}}\partial_{z_{a}}W\\
&=W_{z_{i}z_{a}}-K_{z_{i}}K_{z_{a}}W\\
&=-W_{ia}+\frac{1}{4}K_{i}K_{a}W
\end{aligned}
\end{equation}
which results in:
\begin{equation}
\label{ax1}
\begin{aligned}
\partial_{b}\partial_{a}|W|^{2}&=2\partial_{t_{b}}\partial_{t_{a}}W\overline{W}+2\partial_{t_{a}}W\partial_{t_{b}}\overline{W}\\
&=-2W_{ab}\overline{W}+\frac{1}{2}K_{a}K_{b}|W|^{2}.\\
\end{aligned}
\end{equation}
For the second term of (\ref{eq:mass matrix2}), using (\ref{m2}), (\ref{m23}) we have:
\begin{equation}
\label{ax2}
\begin{aligned}
&e^{K}K^{i\overline{j}}(\partial_{a}D_{i}W)(\partial_{b}D_{\overline{j}}\overline{W})\\
&=e^{K}4K^{ij}(-W_{ia}+\frac{1}{4}K_{i}K_{a}W)(-\overline{W}_{jb}+\frac{1}{4}K_{j}K_{b}\overline{W})\\
&=e^{K}(4K^{ij}W_{ia}W_{jb}+s_{i}W_{ia}K_{b}W+s_{j}W_{jb}K_{a}W+\frac{7}{4}K_{a}K_{b}|W|^{2}).\\
\end{aligned}
\end{equation}
Putting (\ref{ax1}), (\ref{ax2}) together into (\ref{eq:mass matrix2}):
\begin{equation}
\begin{aligned}
M^{A}_{ab}=\frac{\textrm{V}_{ab}}{e^{K}|W|^{2}}&=2K_{a}K_{b}+6\frac{W_{ab}}{W}+8K^{ij}\frac{W_{ia}W_{jb}}{|W|^{2}}\\
&+2s_{i}K_{b}\frac{W_{ia}}{W}+2s_{j}K_{a}\frac{W_{jb}}{W}.\\
\end{aligned}
\end{equation}
\section{Hessians Of Zero Fluxes Background} 
In this appendix we give the explicit expressions for the Hessians of the moduli and axions.\\
We can subsititute the vaccum expectation value of axions Eqs(\ref{av}) into the superpotential because the Hessians for moduli and axions factorise:
\begin{equation}
\label{18}
W=A_{1}e^{- b_{1}\Sigma_{i=1}^{N}N_{i}^{1}s_{i}}-A_{2}e^{-b_{2}\Sigma_{i=1}^{N}N_{i}^{2}s_{i}}.
\end{equation}
The first and second derivatives for $\textit{W}$ are:
\begin{equation}
W_{a}=-b_{1}N^{1}_{a}A_{1}e^{-b_{1}\sum_{i=1}^{N}N^{1}_{i}s_{i}}+b_{2}N^{2}_{a}A_{2}e^{-b_{2}\sum_{i=1}^{N}N^{2}_{i}s_{i}},
\end{equation}
\begin{equation}
\label{81}
W_{ab}=b_{1}^{2}N^{1}_{a}N^{1}_{b}A_{1}e^{-b_{1}\sum_{i=1}^{N}N^{1}_{i}s_{i}}-b_{2}^{2}N^{2}_{a}N^{2}_{b}A_{2}e^{-b_{2}\sum_{i=1}^{N}N^{2}_{i}s_{i}}.
\end{equation}
We substitute the SUSY condition (\ref{eq:1}) into (\ref{81}):
\begin{equation}
\frac{W_{ab}}{W}=\frac{\alpha b_{1}^{2}N^{1}_{a}N^{1}_{b}-b_{2}^{2}N^{2}_{a}N^{2}_{b}}{\alpha-1}.
\end{equation}
Eqs(\ref{11}) implies:
\begin{equation}
N_{a}^{2}=\frac{b_{1}N_{a}^{1}}{b_{2}}\alpha+\frac{3a_{a}(\alpha-1)}{2s_{a}b_{2}}.
\end{equation}
Putting them together, the second order derivatives of the superpotential are expressed as:
\begin{equation}
\begin{aligned}
\frac{W_{ab}}{W}&=-\alpha b_{1}^{2}N_{a}^{1}N_{b}^{1}-\frac{3}{2}b_{1}\alpha\left(\frac{a_{b}}{s_{b}}N_{a}^{1}+\frac{a_{a}}{s_{a}}N_{b}^{1}\right)+\frac{9a_{a}a_{b}}{4s_{a}s_{b}}\left(1-\alpha\right)\\
&=-\alpha\left(b_{1}N^{1}_{a}+\frac{3a_{a}}{2s_{a}}\right)\left(b_{1}N^{1}_{b}+\frac{3a_{b}}{2s_{b}}\right)+\frac{9a_{a}a_{b}}{4s_{a}s_{b}},\\
\end{aligned}
\end{equation}

The Hessians of the moduli and axions can be expressed in AdS units:
\begin{equation}
\begin{aligned}
\label{15}
H^{V}_{ab}=R_{AdS}^{2}\textrm{V}_{ab}=\frac{\textrm{V}_{ab}}{e^{K}|W|^{2}}&=-K_{ab}+3K_{a}K_{b}+2\frac{W_{ab}}{W}+8K^{ij}\frac{W_{ia}W_{jb}}{|W|^{2}}\\
&+2s_{i}K_{b}\frac{W_{ia}}{W}+2s_{j}K_{a}\frac{W_{jb}}{W}.\\
\end{aligned}
\end{equation}
\begin{equation}
\begin{aligned}
H^{A}_{ab}=R_{AdS}^{2}\textrm{V}_{ab}=\frac{\textrm{V}_{ab}}{e^{K}|W|^{2}}&=2K_{a}K_{b}+6\frac{W_{ab}}{W}+8K^{ij}\frac{W_{ia}W_{jb}}{|W|^{2}}\\
&+2s_{i}K_{b}\frac{W_{ia}}{W}+2s_{j}K_{a}\frac{W_{jb}}{W}.\\
\end{aligned}
\end{equation}
The derivatives of K\" ahler potential and superpotential are given by:
\begin{equation}
-K_{ab}=-\frac{s_{a}^{2}}{3a_{a}}\delta_{ab},
\end{equation}
\begin{equation}
K_{a}K_{b}=\frac{9a_{a}a_{b}}{s_{a}s_{b}},
\end{equation}
\begin{equation}
2\frac{W_{ab}}{W}=-2\alpha\left(b_{1}N^{1}_{a}+\frac{3a_{a}}{2s_{a}}\right)\left(b_{1}N^{1}_{b}+\frac{3a_{b}}{2s_{b}}\right)+\frac{9a_{a}a_{b}}{2s_{a}s_{b}}.
\end{equation}
These are used to evaluate the last term in the first line in (\ref{15}):
\begin{equation}
\begin{aligned}
8K^{ij}\frac{W_{ia}W_{jb}}{|W|^{2}}&=8\Sigma_{i=1}^{N}\frac{s_{i}^{2}}{3a_{i}}\left(-\alpha\left(b_{1}N^{1}_{a}+\frac{3a_{a}}{2s_{a}}\right)+\frac{9a_{i}a_{a}}{4s_{i}s_{a}}\right)\\
&\left(-\alpha\left(b_{1}N^{1}_{i}+\frac{3a_{i}}{2s_{i}}\right)\left(b_{1}N^{1}_{b}+\frac{3a_{b}}{2s_{b}}\right)+\frac{9a_{i}a_{b}}{4s_{i}s_{b}}\right)\\
&=\frac{8}{3a_{i}}\alpha^{2}\left(b_{1}N^{1}_{i}s_{i}+\frac{3a_{i}}{2}\right)^{2}\left(b_{1}N^{1}_{a}+\frac{3a_{a}}{2s_{a}}\right)\left(b_{1}N^{1}_{b}+\frac{3a_{b}}{2s_{b}}\right)\\
&-\frac{8s_{i}^{2}}{3a_{i}}\alpha \left(b_{1}N_{i}^{1}+\frac{3a_{i}}{2s_{i}}\right)\frac{9a_{i}}{4s_{i}}((b_{1}N^{1}_{a}\\
&+\frac{3a_{a}}{2s_{a}})\frac{a_{b}}{s_{b}}+\left(b_{1}N^{1}_{b}
+\frac{3a_{b}}{2s_{b}}\right)\frac{a_{a}}{s_{a}})+\frac{8s_{i}^{2}}{3a_{i}}\frac{81a_{i}^{2}a_{a}a_{b}}{16s_{i}^{2}s_{a}s_{b}}\\
=&\frac{8}{3a_{i}}\alpha^{2}\left(b_{1}N^{1}_{i}s_{i}+\frac{3a_{i}}{2}\right)^{2}\left(b_{1}N^{1}_{a}+\frac{3a_{a}}{2s_{a}}\right)\left(b_{1}N^{1}_{b}+\frac{3a_{b}}{2s_{b}}\right)\\
&-6s_{i}\alpha \left(b_{1}N_{i}^{1}+\frac{3a_{i}}{2s_{i}}\right)\left(b_{1}N^{1}_{a}\frac{a_{b}}{s_{b}}+b_{1}N^{1}_{b}\frac{a_{a}}{s_{a}}
+\frac{3a_{b}a_{a}}{s_{b}s_{a}}\right)+\frac{63}{2}\frac{a_{a}a_{b}}{s_{a}s_{b}}\\
\end{aligned}
\end{equation}
The second line of (\ref{15}) can be obtained similarly:
\begin{equation}
\begin{aligned}
2s_{i}K_{b}\frac{W_{ia}}{W}&=2s_{i}\frac{-3a_{b}}{s_{b}}\frac{W_{ia}}{W}\\
&=-\frac{6a_{b}}{s_{b}}\Sigma_{i=1}^{N}\left(-\alpha\left(b_{1}N^{1}_{i}s_{i}+\frac{3}{2}a_{i}\right)\left(b_{1}N^{1}_{a}+\frac{3a_{a}}{2s_{a}}\right)+\frac{9a_{a}a_{i}}{4s_{a}}\right)\\
&=\frac{6a_{b}}{s_{b}}\Sigma_{i=1}^{N}\alpha\left(b_{1}N^{1}_{i}s_{i}+\frac{3}{2}a_{i}\right)\left(b_{1}N^{1}_{a}+\frac{3a_{a}}{2s_{a}}\right)-\frac{63a_{a}a_{b}}{2s_{a}s_{b}},
\end{aligned}
\end{equation}
\begin{equation}
\begin{aligned}
2s_{i}K_{a}\frac{W_{ib}}{W}&=2s_{i}\frac{-3a_{a}}{s_{a}}\frac{W_{ib}}{W}\\
&=\frac{6a_{a}}{s_{a}}\Sigma_{i=1}^{N}\alpha\left(b_{1}N^{1}_{i}s_{i}+\frac{3}{2}a_{i}\right)\left(b_{1}N^{1}_{b}+\frac{3a_{b}}{2s_{b}}\right)-\frac{63a_{a}a_{b}}{2s_{a}s_{b}},
\end{aligned}
\end{equation}
Combining these results in:

\begin{equation}
\begin{aligned}
H^{V}_{ab}&=-K_{ab}+3K_{a}K_{b}+2\frac{W_{ab}}{W}+8K^{ij}\frac{W_{ia}W_{jb}}{|W|^{2}}\\
&+2s_{i}K_{b}\frac{W_{ia}}{W}+2s_{j}K_{a}\frac{W_{jb}}{W}.\\
&=-\frac{s_{a}^{2}}{3a_{a}}\delta_{ab}+\frac{8}{3a_{i}}\alpha^{2}\left(b_{1}N^{1}_{i}s_{i}+\frac{3a_{i}}{2}\right)^{2}\left(b_{1}N^{1}_{a}+\frac{3a_{a}}{2s_{a}}\right)\left(b_{1}N^{1}_{b}+\frac{3a_{b}}{2s_{b}}\right)\\
&-2\alpha\left(b_{1}N^{1}_{a}+\frac{3a_{a}}{2s_{a}}\right)\left(b_{1}N^{1}_{b}+\frac{3a_{b}}{2s_{b}}\right)\\
&=-\frac{s_{a}^{2}}{3a_{a}}\delta_{ab}+\alpha\left(b_{1}N^{1}_{a}+\frac{3a_{a}}{2s_{a}}\right)\left(b_{1}N^{1}_{b}+\frac{3a_{b}}{2s_{b}}\right)\left(\frac{8}{3a_{i}}\alpha\left(b_{1}N^{1}_{i}s_{i}+\frac{3a_{i}}{2}\right)^{2}-2\right),\\
\end{aligned}
\end{equation}
and
\begin{equation}
\begin{aligned}
H^{A}_{ab}&=2K_{a}K_{b}+6\frac{W_{ab}}{W}+8K^{ij}\frac{W_{ia}W_{jb}}{|W|^{2}}\\
&+2s_{i}K_{b}\frac{W_{ia}}{W}+2s_{j}K_{a}\frac{W_{jb}}{W}.\\
&=\frac{8}{3a_{i}}\alpha^{2}\left(b_{1}N^{1}_{i}s_{i}+\frac{3a_{i}}{2}\right)^{2}\left(b_{1}N^{1}_{a}+\frac{3a_{a}}{2s_{a}}\right)\left(b_{1}N^{1}_{b}+\frac{3a_{b}}{2s_{b}}\right)\\
&-6\alpha\left(b_{1}N^{1}_{a}+\frac{3a_{a}}{2s_{a}}\right)\left(b_{1}N^{1}_{b}+\frac{3a_{b}}{2s_{b}}\right)\\
&=\alpha\left(b_{1}N^{1}_{a}+\frac{3a_{a}}{2s_{a}}\right)\left(b_{1}N^{1}_{b}+\frac{3a_{b}}{2s_{b}}\right)\left(\frac{8}{3a_{i}}\alpha\left(b_{1}N^{1}_{i}s_{i}+\frac{3a_{i}}{2}\right)^{2}-6\right)\\
\end{aligned}
\end{equation}
\end{appendices}
\renewcommand{\bibname}{Bibliography}
\bibliographystyle{apacite}
\bibliography{Bibliography.bib}

\end{document}